\documentclass{aa}

\usepackage{graphicx}
\usepackage{psfig}
\usepackage{txfonts}
\usepackage[round]{natbib}
\usepackage{amssymb}

\bibpunct{(}{)}{;}{a}{}{,}
\def\ion#1#2{{\rm #1}\,{\sc #2}}

\begin{document}
\title{A new comprehensive set of elemental abundances in DLAs \\ 
II. Data analysis and chemical variation studies
\thanks{Based on UVES observations made with the European Southern Observatory 
VLT/Kueyen telescope, Paranal, Chile, collected during the programme ID 
No.~70.B--0258(A).}}


\author{M. Dessauges-Zavadsky\inst{1},
        J. X. Prochaska\inst{2},
	S. D'Odorico\inst{3},
	F. Calura\inst{4},        
	\and
        F. Matteucci\inst{4,5}
        }

\offprints{M. Dessauges-Zavadsky, \\ \email{miroslava.dessauges@obs.unige.ch}}

\institute{Observatoire de Gen\`eve, 51 Ch. des Maillettes, 1290 Sauverny, 
           Switzerland 
           \and
	   UCO/Lick Observatory, University of California at Santa Cruz, Santa 
	   Cruz, CA 95064, USA 
	   \and
	   European Southern Observatory, Karl-Schwarzschildstr. 2, 85748 
	   Garching bei M\" unchen, Germany 
	   \and
	   Dipartimento di Astronomia-Universit\'a di Trieste, Via G. B. 
	   Tiepolo 11, 34131 Trieste, Italy 
           \and
	   INAF, Osservatorio Astronomico di Trieste, Via G. B. Tiepolo 11, 
	   34131 Trieste, Italy 
           }

\date{Received; accepted}

\authorrunning{M. Dessauges-Zavadsky et al.}

\titlerunning{Data analysis and chemical variation studies}

\abstract{
We present new, comprehensive sets of elemental abundances for seven damped 
Ly$\alpha$ systems (DLAs) in the redshift range $z_{\rm abs} = 1.8-2.5$. These
were derived from UVES/VLT spectra combined with existing HIRES/Keck spectra. We 
detected 54 metal-line transitions, and obtained the column density measurements 
of 30 ions from 22 elements, $-$~B, C, N, O, Mg, Al, Si, P, S, Cl, Ar, Ti, Cr, 
Mn, Fe, Co, Ni, Cu, Zn, Ge, As, Kr. Together with the four DLAs analyzed in 
\citet{dessauges04}, we have a sample of eleven DLA galaxies with uniquely 
comprehensive and homogeneous abundance measurements. These observations allow 
one to study in detail the abundance patterns of a wide range of elements and 
the chemical variations in the interstellar medium of galaxies outside the 
Local Group. Comparing the gas-phase abundance ratios of these high redshift 
galaxies, we found that they show low RMS dispersions, reaching only up 2--3 
times the statistical errors for the majority of elements. This uniformity is 
remarkable given that the quasar sightlines cross gaseous regions with 
\ion{H}{i} column densities spanning over one order of magnitude and 
metallicities ranging from 1/55 to 1/5 solar. The uniformity is also remarkable 
since DLAs are expected (and observed at low redshift) to be associated 
with a wide range of galaxy types. This implies the respective star formation 
histories seem to have conspired to yield one set of relative abundances. 
We examined the gas-phase abundance patterns of interstellar medium ``clouds'' 
within the DLA galaxies detected along the velocity profiles. By considering 
all the clouds of all the DLAs studied together, we observe a high 
dispersion in several abundance ratios, indicating that chemical variations 
seem to be more confined to individual clouds within the DLA galaxies than to 
integrated profiles. We found unambiguous correlations between [Si/Fe], [S/Fe] 
and [S/Si] versus [Zn/Fe], and anti-correlations between [Si/Zn] and [S/Zn] 
versus [Zn/Fe]. These trends are primarily the result of differential dust 
depletion effects, which also explain the cloud abundance ratio dispersion. The 
signature of the nucleosynthesis enrichment contribution is observed in the 
[$\alpha$/Fe,Zn] ratios at low dust depletion levels, $0\leq {\rm [Zn/Fe]}\leq 
0.2$, and is characterized by an $\alpha$-enhancement in individual clouds. 
Quite surprisingly, however, while the [Si/Fe] ratios are supersolar in clouds 
with low depletion level, the [S/Zn] ratios remain almost solar, suggesting 
that [S/Zn] may not be a reliable tracer of nucleosynthesis enrichment. 
Analysis of the cloud-to-cloud chemical variations within seven individual DLA 
systems reveals that five of them show statistically significant variations, 
higher than 0.2~dex at more than 3\,$\sigma$, but only two DLAs show extreme 
variations. The sources of these variations are both the differential dust 
depletion and/or ionization effects; however, no evidence for variations due to 
different star formation histories could be highlighted. These observations 
place large constraints on the mixing timescales of protogalaxies and on 
scenarios of galaxy formation within the CDM hierarchical theory. Finally, we 
provide an astrophysical determination of the oscillator strength of the 
\ion{Ni}{ii}\,$\lambda$\,1317 transition. 

}

\maketitle
%


\section{Introduction}
\label{introduction}

One of the most exciting developments in observational cosmology over the
last decade has been the ability to extend studies of elemental abundances from
the local Universe to high redshift. This is a fundamental step toward a better 
understanding of the formation and evolution of galaxies. It has been made 
possible by the studies of absorption line systems detected in optical spectra 
of quasars (QSOs), and specifically via the damped Ly$\alpha$ systems 
\citep[DLA; e.g.][]{wolfe86}. These systems with \ion{H}{i} column densities 
higher than $2\times 10^{20}$ cm$^{-2}$ dominate the neutral gas content of the 
Universe available for star formation and are, therefore, widely believed to 
be progenitors of present-day galaxies 
\citep[][ and references therein]{storrie00}. The DLA systems provide the best 
opportunity to accurately measure the gas-phase chemical abundances of many 
elements for a variety of galactic systems spanning a wide redshift interval. 
These objects are by far the best laboratories for studying galaxies at high 
redshifts, in their early stages of evolution.

In Paper~I \citep{dessauges04}, we demonstrated the efficiency and power of 
this approach. Indeed, we obtained comprehensive sets of elemental abundances 
for four DLA systems, namely the abundance measurements of 15 elements $-$ N, O, 
Mg, Al, Si, P, S, Cl, Ar, Ti, Cr, Mn, Fe, Ni, and Zn~$-$ and the column density 
measurements of up to 21 ions. This large chemical dataset allowed us to study 
each galaxy individually, while previously the DLA galaxy population had been 
analyzed as a whole. We also provided the first constraints on their star 
formation history, age, and star formation rate through a detailed comparison 
with a grid of chemical evolution models for spiral and dwarf irregular 
galaxies by \citet{calura03}. These exciting results encouraged us to further 
extend our sample. We obtained high quality spectra of seven additional DLAs, 
bringing the full sample of DLA galaxies to eleven systems.

Three main sources influence the chemical gas-phase abundance composition of 
these high redshift galaxies: (i)~the star formation history (SFH) through
its implications for nucleosynthesis processes; (ii)~differential dust 
depletion effects, since part of elements may be removed from the gas to the 
solid phase \citep{savage96}; and (iii)~photoionization effects, because we 
are only able to measure one or a few ionization states of a given element
\citep{viegas95,vladilo01}. The contribution of each of these three sources can 
theoretically be disentangled through specific abundance ratios and is best 
constrained through comprehensive sets of elemental abundances within a DLA. 
There is a degeneracy between these effects that is particularly high for the 
routinely detected elements (Fe, Si, Cr, Ni, Al, sometimes Zn). The specific 
abundance ratios which should allow to break the degeneracy are: for (i) the 
ratios involving two elements released in the interstellar medium on different 
timescales, in particular the $\alpha$ over iron-peak element ratios; for (ii) 
the ratios of two elements with the same nucleosynthesis origin, but with 
different dust depletion levels, like [Zn/Fe]; and for (iii) the ratios of two 
elements preferentially with the same nucleosynthesis origin, but with one 
element being more sensitive to ionization than the other, like [O/Si].

Low redshift ($z_{\rm abs} < 1$) deep imaging reveals that the DLA galaxies have 
a variety of morphological types \citep[e.g.][]{lebrun97,nestor02,rao03,chen03}. 
If the high redshift DLA galaxy population also samples galaxies with different 
star formation histories, different stages of chemical evolution, and different 
interstellar medium (ISM) conditions, we could expect to see the impact of 
these differences in the chemical abundances, and highlight them through the 
study of variations in the specific abundance ratios described above. A range 
of at least 0.3~dex is, for example, observed in the $\alpha$ over iron-peak 
element ratios, when comparing the abundance measurements of stars in the Milky 
Way with those of the Small and Large Magellanic Clouds (SMC, LMC) and dwarf 
spheroidal galaxies and when comparing the abundance measurements of stars 
within a given galaxy, due to different star formation histories 
\citep{venn99,shetrone03,tolstoy03}. Similarly there is a greater than
0.5~dex range in the gas-phase abundances of different lines of sight crossing 
the Milky Way and Magellanic Clouds due to different physical conditions, e.g. 
various dust-to-gas ratios, volume densities and ionization states within the 
ISM \citep{savage96,welty99,welty01}.

Even though DLA galaxies have \ion{H}{i} column densities spanning an 
order of magnitude and metallicities spread over two orders of magnitude, 
previous studies show that DLAs have similar abundance ratios suggesting that 
these protogalaxies have common enrichment histories \citep{prochaska02c}. 
In the same way, the abundance patterns of DLA ``clouds'' for a given system 
also show uniformity suggesting that the gas of these protogalaxies has a 
similar enrichment history and uniform differential dust depletion 
\citep{prochaska96,lopez02,prochaska03a}. If confirmed, these results have 
important implications for the ISM of high redshift galaxies and the enrichment 
of gas in the early Universe.

In this paper, we further investigate the issue of chemical variations in the 
abundance ratios of both DLA galaxies and DLA gas along the velocity profiles. 
We focus on our sample of eleven DLAs with comprehensive sets of elemental 
abundances. This sample comprises the four DLAs studied in Paper~I and the seven 
new DLAs presented here (Sects.~\ref{observations} and \ref{data-analysis}). We 
discuss the global abundance patterns of DLAs in Sect.~\ref{DLA-trends}, and in 
Sects.~\ref{indiv-clouds} and \ref{cloud-to-cloud} we study the abundance 
patterns of clouds within the DLA galaxies, which we regard as individual 
entities. First, we consider all the clouds of all the DLAs from our sample 
together (Sect.~\ref{indiv-clouds}), and second we focus on the cloud-to-cloud 
chemical analysis within a single DLA (Sect.~\ref{cloud-to-cloud}). In 
Sect.~\ref{conclusions} we summarize our results. Finally, we present an 
astrophysical determination of the oscillator strength of the 
\ion{Ni}{ii}\,$\lambda$\,1317 transition in Appendix~\ref{appendix}. In two
future papers, Paper~III and Paper~IV of this series, we will further analyze 
this sample of DLAs. In Paper~III, we will study their star formation history,
age, and star formation rate through a detailed comparison with chemical 
evolution models, and in Paper~IV we plan to analyze in detail their ionization 
state through comparisons of the observed ionic ratios against photoionization 
models. Two DLAs of our sample are likely to have among the highest ionization 
levels of all previously studied DLA galaxies. 

%

\begin{table*}[t]
\begin{center}
\caption{Journal of observations} 
\label{Journal-obs}
\begin{tabular}{l c c c l c c c}
\hline \hline
\\[-0.3cm]
Quasar  & V     & $z_{\rm em}$ & $z_{\rm abs}$ & Observing date & UVES Setting & Wavelength coverage & Exposure time \\
        & [mag] &              &               &                &              & [nm]                & [s]     
\smallskip 
\\     
\hline  
\\[-0.3cm]
Q0450$-$13   & 16.5 & 2.30 & 2.067       & November 2002          & dic1\,(B346+R580) & 306$-$387/478$-$681  & 6000 \\
             &      &      &             & November 2002          & dic2\,(B346+R860) & 306$-$387/671$-$1000 & 9000 \\
Q0841+129    & 18.5 & 2.50 & 2.375/2.476 & December 2002          & dic2\,(B390+R860) & 330$-$452/671$-$1000 & 10800 \\
Q1157+014 $^{\dagger}$& 17.0 & 1.99 & 1.944 & April 2000          & dic1\,(B346+R550) & 310$-$387/448$-$651  & 7200 \\
             &      &      &             & June 2001/January 2002 & dic1\,(B380+R580) & 319$-$441/478$-$681  & 5400 \\
	     &      &      &             & June 2001              & dic2\,(B380+R750) & 319$-$441/558$-$939  & 7200 \\
Q1210+17     & 17.4 & 2.54 & 1.892       & February 2003          & dic2\,(B346+R860) & 325$-$381/671$-$1000 & 30500 \\
Q2230+02     & 18.0 & 2.15 & 1.864       & October 2003           & dic2\,(B346+R860) & 310$-$387/671$-$1000 & 13500 \\
Q2348$-$1444 & 16.9 & 2.94 & 2.279       & November 2002          & dic2\,(B390+R800) & 344$-$452/609$-$990  & 10350 \\
\hline
\end{tabular}
\begin{minipage}{170mm}
\smallskip
$^{\dagger}$ Based on UVES observations made with the European Southern Observatory VLT/Kueyen telescope obtained 
from the ESO/ST-ECF \\ \phantom{$^{\dagger}$} Science Archive Facility (programme ID No.~65.O--0063, 67.A--0078, 
and 68.A--0461).
\end{minipage}
\end{center}
\end{table*}
%

\section{Observations and data reduction}
\label{observations}

The selected quasars Q0450$-$13, Q0841+129, Q1157+014, Q1210+17, Q2230+02, and
Q2348$-$1444 with seven intervening DLAs in the redshift range $z_{\rm abs} =
1.8-2.5$ are relatively bright with $V = 16.5-18.5$. Their spectra have already
been reported in the literature, and four of them have been studied by 
\citet{prochaska99} and \citet{prochaska01} using high resolution, high quality 
spectra obtained with the HIRES echelle spectrograph on the Keck~I 10\,m 
telescope at Mauna Kea, Hawaii. More details on these DLA systems and the 
existing data can be found in Sect.~\ref{data-analysis}. This sample of six 
QSOs completes our first sample of QSOs, $-$ Q0100+13, Q1331+17, Q2231$-$00, 
and Q2343+12~$-$, with the same characteristics in redshift and magnitude as 
studied in Paper~I. The combination of these two sets of observations provides 
an impressive sample of eleven damped Ly$\alpha$ systems with comprehensive 
elemental abundances.

We used the unique capability of the Ultraviolet-Visual Echelle Spectrograph
UVES \citep{dodorico00} on the VLT 8.2\,m Kueyen ESO telescope at Cerro
Paranal, Chile, to obtain high resolution, high signal-to-noise ratio spectra
of our QSOs in the total optical spectral range, or to complete the existing 
HIRES/Keck spectra in the blue $\lambda = 3060-4500$ \AA\ and in the far-red 
$\lambda = 6700-10\,000$ \AA. The observations were performed in service mode
in period 70 from October 2002 to April 2003 (programme ID No.~70.B--0258) 
under good seeing conditions (between 0.5\arcsec\ and 1\arcsec). For one 
object, Q1157+014, we used spectra from the ESO UVES/VLT archive (programme ID
No.~65.O--063, 67.A--0078, and 68.A--0461). For each science exposure, slit 
widths of 1\arcsec\ in the blue and of 0.9\arcsec\ in red were used with a CCD 
binning of $2\times 2$, resulting in a resolution $FWHM \simeq 6.9$ km~s$^{-1}$ 
and 6.4 km~s$^{-1}$ on average, respectively. Relevant details on the
observations are collected in Table~\ref{Journal-obs}. The total exposure times
of each QSO were split in multiple exposures of 3600 or 4500~s.

The spectra were reduced using the ESO data reduction package {\tt MIDAS} and
the UVES pipeline in an interactive mode available as a {\tt MIDAS} context. A
detailed description of the pipeline can be found in \citet{ballester00}. To
optimize the results, we made a systematic check of each step of the pipeline
reduction. Once reduced, the wavelengths of the resulting one-dimensional 
spectra were converted to a vacuum-heliocentric scale. For QSOs with multiple 
exposures, the individual spectra were co-added using their signal-to-noise 
ratio as weights. The spectra were normalized by smoothly connecting regions 
free from absorption features with a spline function. In the Ly$\alpha$ forest, 
the continuum was fitted by using small regions deemed to be free from 
absorptions and by interpolating between these regions with a spline. An 
average signal-to-noise ratio per pixel of $\sim 25$, 45, and 40 was achieved 
in the final spectra at $\lambda \sim 3700$, 7000, and 9000 \AA, respectively.

%

\section{Data analysis and ionic column densities}
\label{data-analysis}

By combining our UVES/VLT spectra with the existing HIRES/Keck spectra of 
Q0841+129, Q1210+17, Q2230+02, and Q2348$-$1444, we covered the total spectral 
range from 3060 to 10\,000 \AA\ for the six selected QSOs. This gave us access 
to several metal-line transitions of 22 elements for each of their intervening 
DLAs. In this section we present the derived ionic column density measurements. 

The ionic column densities were obtained using the Voigt profile fitting 
technique. This technique consists of fitting theoretical Voigt profiles to 
the observed DLA absorption profiles. These profiles are described well as a 
complex of components, each defined by a redshift $z$, a Doppler parameter $b$, 
a column density $N$ and the corresponding errors. The fits were performed 
using an $\chi^2$ minimization routine {\tt FITLYMAN} in {\tt MIDAS} 
\citep{fontana95}. We assumed that metal species with similar ionization 
potentials can be fitted using identical component fitting parameters, i.e. the 
same $b$ (which means that macroturbulent motions dominate over thermal 
broadening) and the same $z$ in the same component, and allowing for variations 
from metal species to metal species in $N$ only. We distinguish three categories 
of metal species with similar ionization potentials: the low-ion transitions 
(i.e. the neutral and singly ionized species), the intermediate-ion transitions 
(e.g. \ion{Fe}{iii}, \ion{Al}{iii}), and the high-ion transitions (e.g. 
\ion{C}{iv}, \ion{Si}{iv}). By using relatively strong (but not saturated) lines  
to fix the component fitting parameters (the $b$ and $z$ values for each 
component), we then obtained excellent fitting results even for weak metal lines 
and for metal lines in the Ly$\alpha$ forest, where the probability of blending 
is high, by allowing only the column density to vary. We had a sufficient 
number of relatively strong metal lines to accurately constrain the fitting 
parameters in the seven DLAs studied that exhibited multicomponent velocity 
structures.

In Tables~\ref{Q0450-Ntable}--\ref{Q2348-Ntable} we present the results of the
component per component ionic column density measurements for the fitting model
solutions of the low- and intermediate-ion transitions for the seven DLA
systems analyzed. The reported errors are the 1\,$\sigma$ errors on the fits 
computed by {\tt FITLYMAN}. These errors do not take the uncertainties on the 
continuum level determination into account, unless it is mentioned in
Sects.~\ref{Q0450}--\ref{Q2348}. For the saturated components, the column 
densities are listed as lower limits. The values reported as upper limits are 
either cases with significant line blendings by \ion{H}{i} lines in the 
Ly$\alpha$ forest or telluric lines or cases of non-detection corresponding to 
4\,$\sigma$ limits. By adopting a conservative 4\,$\sigma$ upper limit based
on the statistical error, we account -- in part -- the continuum error. In 
Figs.~\ref{Q0450-metals}--\ref{Q2348-metals} (even numbers), we show the best
fitting solutions of all the low- and intermediate-ion transitions detected 
in the seven DLAs studied. In these velocity plots, $v=0$ corresponds to an 
arbitrary component, and all the identified components are marked by small 
vertical bars. The thin solid line represents the best fit. The telluric lines 
have been identified thanks to the spectrum of a hot, fast rotating star taken 
on the same night as the science exposures.

\begin{figure}[t]
\centering
\includegraphics[width=9cm]{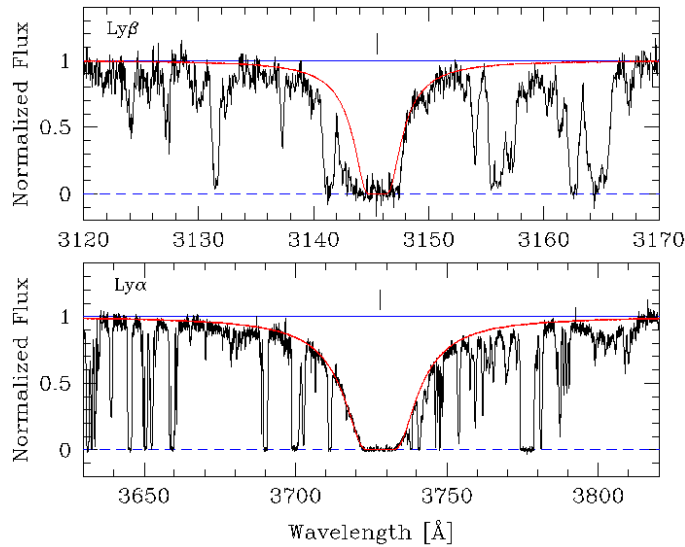}
\caption{Normalized UVES spectrum of Q0450$-$13 showing the DLA Ly$\alpha$ and
Ly$\beta$ line profiles with the Voigt profile fits. The vertical bar 
corresponds to the wavelength centroid of the component used for the best fit, 
$z=2.06666$. The measured \ion{H}{i} column density is $\log N$(\ion{H}{i}) 
$= 20.53\pm 0.08$.}
\label{Q0450-HI}
\end{figure}

The neutral hydrogen column densities were measured from the fit of the
Ly$\alpha$ damping line profile. The $b$-value was fixed at 20 km~s$^{-1}$,
and the redshift $z$ was left as a free parameter or fixed at the redshift of
the strongest component of the metal-line profile depending on the DLA system
(see the comments in the following sub-sections). When other lines of higher 
members of the Lyman series were accessible in our spectra, they were used in 
parallel to the Ly$\alpha$ line to derive the \ion{H}{i} column density.
In Figs.~\ref{Q0450-HI}--\ref{Q2348-HI} (odd numbers), we show the best 
\ion{H}{i} fitting solutions in the seven DLAs studied. The small vertical bar
corresponds to the redshift used in the best-fitting solution, and the thin
solid line represents the best fit.



Throughout the analysis we have adopted the list of atomic data, laboratory
wavelengths and oscillator strengths, compiled by J.\,X. Prochaska and
collaborators (version 0.5) and presented on the web site ``the HIRES Damped 
Ly$\alpha$ Abundance Database''\footnote{http://kingpin.ucsd.edu/$\sim$hiresdla/}.
The most recent measurements of $\lambda_{\rm rest}$ and $f$-values of the
metal-line transitions that impact the abundances of DLAs and their references
are reported there. In Appendix~\ref{appendix} we present an updated value of 
the \ion{Ni}{ii}\,$\lambda$\,1317 oscillator strength derived from different
\ion{Ni}{ii} transitions detected in the DLAs studied. We adopt the solar 
meteoritic abundances from \citet{grevesse98}.


%

%

\subsection{Q0450$-$13, z\mathversion{bold}$_{\rm abs}$\mathversion{normal} = 2.067}
\label{Q0450}

This quasar was discovered by C. Hazard and was first investigated by 
\citet{jaunsen95}. The presence of the DLA system at $z_{\rm abs} = 2.067$ on 
its line of sight was communicated to us by J.\,X. Prochaska (private 
communication). This is the first detailed analysis of its chemical composition. 
The high quality UVES/VLT spectra cover the total spectral range from 3060 
to 10\,000 \AA. These data permit analysis of some 39 metal-line transitions 
and help to obtain accurate column density measurements of \ion{Fe}{ii}, 
\ion{Si}{ii}, \ion{Al}{ii}, \ion{S}{ii}, \ion{N}{i}, \ion{P}{ii}, \ion{Mg}{ii}, 
\ion{C}{ii}$^*$, \ion{Al}{iii}, \ion{Fe}{iii}, and \ion{N}{ii}. In addition, we 
were able to put very reliable upper limits to the column densities of 
\ion{Ni}{ii}, \ion{Cr}{ii}, and \ion{Ar}{i} and to provide a lower limit to the 
column density of \ion{O}{i}.

%

\begin{figure*}[!]
\centering
\includegraphics[width=17.5cm]{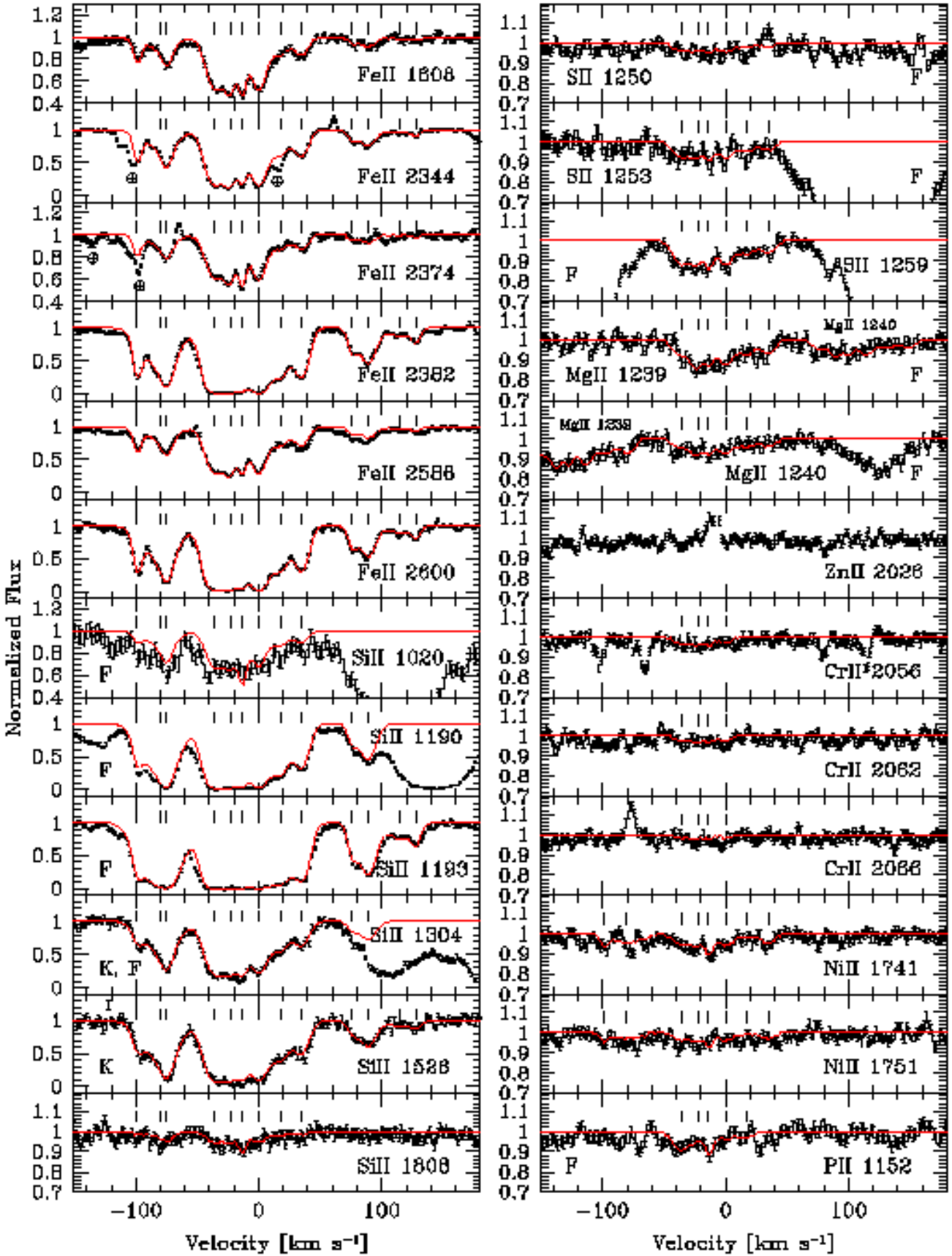}
\caption{Velocity plots of the metal line transitions (normalized intensities
shown by dots with 1\,$\sigma$ error bars) for the DLA toward Q0450$-$13. The
zero velocity is fixed at $z=2.06680$. For this and all the following figures 
with velocity plots, the vertical bars mark the positions of the fitted
velocity components, and the symbol $\oplus$ corresponds to telluric lines. The 
letter K refers to Keck spectra and the letter F to the Ly$\alpha$ forest.}
\label{Q0450-metals}
\end{figure*}

\addtocounter{figure}{-1}
\begin{figure*}[!]
\centering
\includegraphics[width=17.5cm]{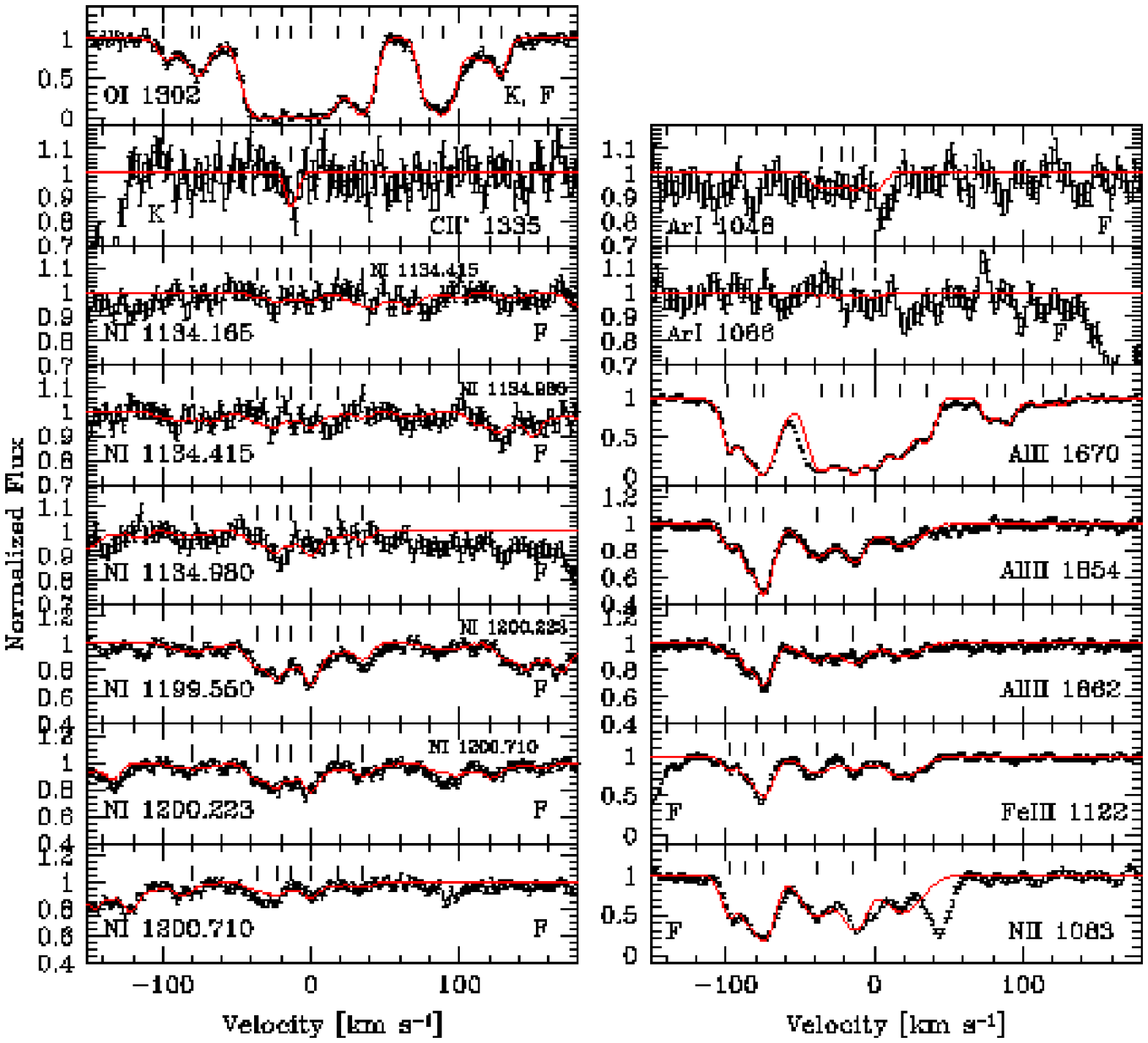}
\caption{{\em Continued}.}
\end{figure*}
%

The low-ion absorption line profiles of this DLA are characterized by 13
components  spread over 230 km~s$^{-1}$ in velocity space (see 
Fig.~\ref{Q0450-metals}). Their redshifts, $b$-values, and column densities are 
presented in Table~\ref{Q0450-Ntable}. Only the components 4--9 are detected in 
weak metal-line transitions, they contain approximately 80\,\% of the total 
column density obtained by summing the contribution of the 13 components. The 
accessible \ion{Ni}{ii} lines at $\lambda_{\rm rest} = 1741$, 1751 \AA, 
\ion{Cr}{ii} lines at $\lambda_{\rm rest} = 2056$, 2062, 2066 \AA, and 
\ion{Ar}{i} lines at $\lambda_{\rm rest} = 1048$, 1066 \AA\ are all so weak 
that we consider their derived column densities as upper limits. In the 
context of upper limits, the column density of P$^+$ is a borderline case. 
However, the \ion{P}{ii}\,$\lambda$\,1152 line has a slightly higher optical 
depth than the \ion{Ni}{ii} and \ion{Cr}{ii} lines, we thus assumed it is a 
detection. In contrast, the \ion{O}{i} line at $\lambda_{\rm rest} = 1302$ \AA\ 
is saturated in the components 4--7, so we only got a lower limit to 
its total column density. The Mg$^+$ column density was obtained from the 
unsaturated \ion{Mg}{ii}\,$\lambda$\,1239,\,1240 lines. Because these 
\ion{Mg}{ii} transitions are located in the red wing of the DLA Ly$\alpha$ 
damping profile, it is difficult to obtain an accurate column density 
measurement of Mg$^+$. To estimate the column density, we first normalized the 
spectra within the damped Ly$\alpha$ profile near the \ion{Mg}{ii} lines 
according to the fit of the Ly$\alpha$ damping wing profile. We applied the 
same procedure to the \ion{N}{i} triplet at $\lambda_{\rm rest} \simeq 1200$ 
\AA, which is located in the blue wing of the DLA Ly$\alpha$ damping profile. 
The 1\,$\sigma$ errors on the measured $N$(Mg$^+$) and $N$(N$^0$) column 
densities were estimated by varying the continuum level by 5\,\%. No 
\ion{Zn}{ii} line was detected in this DLA system. We provide a 4\,$\sigma$ 
upper limit to the Zn$^+$ column density of $\log N$(\ion{Zn}{ii}) $< 11.60$. 

The intermediate-ion absorption line profiles show a different velocity 
structure than the low-ion line profiles (see Fig.~\ref{Q0450-metals}). Their 
fitting solution is presented in the second part of Table~\ref{Q0450-Ntable}. 
However, in velocity space the positions of the 6 components characterizing 
the intermediate-ion line profiles correspond very closely to the positions of 
components 1, 3, 4, 6, and 8 of low-ion lines. When considering the column 
density ratios of different ionization species of the same element, we find the
following very interesting results: $\log N$(Al$^{++}$)/$N$(Al$^+$) 
$= -0.77\pm 0.08$, $\log N$(Fe$^{++}$)/$N$(Fe$^+$) $= +0.04\pm 0.04$ larger 
than $-1$, and $\log N$(N$^+$)/$N$(N$^0$) $= +0.88\pm 0.04$ larger than $-0.2$. 
Those ratios give a qualitative ``first-look'' analysis of the ionization 
state in a DLA.  According to the photoionization diagnostics described by
\citet{prochaska02a}, we have clear evidence in this DLA of a very high 
ionization level. This will be further discussed in Paper~IV of this series. 
In Sect.~\ref{cloud-to-cloud} we present additional indications toward the 
presence of strong ionization in this system.

%

\begin{table*}[!]
\begin{center}
\caption{Component structure of the $z_{\rm abs} = 2.067$ DLA system toward Q0450$-$13}
\label{Q0450-Ntable}
\begin{tabular}{l c c c l c | l c c c l c}
\hline\hline
Comp. & $z_{\rm abs}$ & $v_{\rm rel}^*$ & $b (\sigma_b)$ & Ion & $\log N (\sigma_{\log N})$ & Comp. & $z_{\rm abs}$ & 
$v_{\rm rel}^*$ & $b (\sigma_b)$ & Ion & $\log N (\sigma_{\log N})$ \\
      &               & [km s$^{-1}$]   & [km s$^{-1}$]  &     &                            &       &               & 
[km s$^{-1}$]   & [km s$^{-1}$]  &     &        
\smallskip
\\ 
\hline
\multicolumn{6}{l}{\hspace{0.3cm} Low-ion transitions} & & & & & & \\
\hline
\\[-0.28cm]
1 & 2.06579 &  $-$99 & \phantom{0}2.2{\scriptsize (0.1)} & \ion{Fe}{ii} & 12.98{\scriptsize (0.02)} & 7 & 2.06680 &	0  & \phantom{0}7.2{\scriptsize (0.1)} & \ion{Fe}{ii} & 13.59{\scriptsize (0.01)} \\  
  &         &        &                                   & \ion{Si}{ii} & 12.94{\scriptsize (0.05)} &	&	  &	   &				       & \ion{Si}{ii} & 13.87{\scriptsize (0.02)} \\  
  &         &        &                                   & \ion{O}{i}   & 13.13{\scriptsize (0.08)} &	&	  &	   &				       & \ion{O}{i}   & $>15.73$ \\   
  &         &        &                                   & \ion{Al}{ii} & 11.90{\scriptsize (0.03)} &	&	  &	   &				       & \ion{Al}{ii} & 12.60{\scriptsize (0.03)} \\  
  &         &        &                                   & \ion{Ni}{ii} & $<12.35$                  &	&	  &	   &				       & \ion{S}{ii}  & 13.53{\scriptsize (0.05)} \\  
\kern-5pt\raisebox{0pt}[0pt][0pt]{\framebox{\parbox[t]{0.2cm}{2\\ \\ \\ \\ \\ \\ 3}}}
  & 2.06598 &  $-$80 &           16.9{\scriptsize (0.3)} & \ion{Fe}{ii} & 13.14{\scriptsize (0.01)} &	&	  &	   &				       & \ion{Ni}{ii} & $<12.66$ \\   
  &         &        &                                   & \ion{Si}{ii} & 13.75{\scriptsize (0.02)} &	&	  &	   &				       & \ion{N}{i}   & 13.10{\scriptsize (0.03)} \\  
  &         &        &                                   & \ion{O}{i}   & 13.78{\scriptsize (0.04)} &	&	  &	   &				       & \ion{P}{ii}  & 11.90{\scriptsize (0.15)} \\  
  &         &        &                                   & \ion{Al}{ii} & 12.74{\scriptsize (0.01)} &	&	  &	   &				       & \ion{Mg}{ii} & 14.97{\scriptsize (0.04)} \\  
  &         &        &                                   & \ion{Ni}{ii} & $<12.86$                  &	&	  &	   &				       & \ion{Ar}{i}  & $<12.21$ \\   
  &         &        &                                   & \ion{N}{i}   & 12.71{\scriptsize (0.07)} &	&	  &	   &				       & \ion{Cr}{ii} & $<11.92$ \\   		      
  & 2.06603 &  $-$75 & \phantom{0}4.9{\scriptsize (0.2)} & \ion{Fe}{ii} & 12.94{\scriptsize (0.01)} & 8 & 2.06698 &  $+$18 & \phantom{0}8.6{\scriptsize (0.3)} & \ion{Fe}{ii} & 13.07{\scriptsize (0.02)} \\  
  &         &        &                                   & \ion{Si}{ii} & 13.51{\scriptsize (0.03)} &	&	  &	   &				       & \ion{Si}{ii} & 13.51{\scriptsize (0.01)} \\  
  &         &        &                                   & \ion{O}{i}   & 13.37{\scriptsize (0.08)} &	&	  &	   &				       & \ion{O}{i}   & 14.11{\scriptsize (0.04)} \\  
  &         &        &                                   & \ion{Al}{ii} & 12.48{\scriptsize (0.02)} &	&	  &	   &				       & \ion{Al}{ii} & 12.47{\scriptsize (0.01)} \\  
4 & 2.06643 &  $-$36 & \phantom{0}8.6{\scriptsize (0.2)} & \ion{Fe}{ii} & 13.63{\scriptsize (0.01)} &	&	  &	   &				       & \ion{S}{ii}  & 13.32{\scriptsize (0.07)} \\
  &         &        &                                   & \ion{Si}{ii} & 13.95{\scriptsize (0.02)} &	&	  &	   &				       & \ion{Ni}{ii} & $<12.17$ \\   		      
  &         &        &                                   & \ion{O}{i}   & $>14.63$                  &	&	  &	   &				       & \ion{N}{i}   & 12.51{\scriptsize (0.07)} \\  
  &         &        &                                   & \ion{Al}{ii} & 12.70{\scriptsize (0.08)} &	&	  &	   &				       & \ion{P}{ii}  & 11.85{\scriptsize (0.18)} \\  
  &         &        &                                   & \ion{S}{ii}  & 13.58{\scriptsize (0.05)} &	&	  &	   &				       & \ion{Mg}{ii} & 14.77{\scriptsize (0.06)} \\  
  &         &        &                                   & \ion{Ni}{ii} & $<12.68$                  & 9 & 2.06716 &  $+$35 & \phantom{0}6.6{\scriptsize (0.2)} & \ion{Fe}{ii} & 12.98{\scriptsize (0.01)} \\  
  &         &        &                                   & \ion{N}{i}   & 12.90{\scriptsize (0.04)} &	&	  &	   &				       & \ion{Si}{ii} & 13.30{\scriptsize (0.02)} \\  
  &         &        &                                   & \ion{P}{ii}  & 12.35{\scriptsize (0.10)} &	&	  &	   &				       & \ion{O}{i}   & 14.52{\scriptsize (0.03)} \\  
  &         &        &                                   & \ion{Mg}{ii} & 14.83{\scriptsize (0.07)} &	&	  &	   &				       & \ion{Al}{ii} & 12.13{\scriptsize (0.02)} \\  
  &         &        &                                   & \ion{Ar}{i}  & $<12.15$                  &	&	  &	   &				       & \ion{S}{ii}  & 13.19{\scriptsize (0.08)} \\
  &         &        &                                   & \ion{Cr}{ii} & $<12.01$                  &	&	  &	   &				       & \ion{Ni}{ii} & $<12.52$ \\   		      
5 & 2.06657 &  $-$22 & \phantom{0}5.2{\scriptsize (0.4)} & \ion{Fe}{ii} & 13.47{\scriptsize (0.03)} &	&	  &	   &				       & \ion{N}{i}   & 12.61{\scriptsize (0.07)} \\  
  &         &        &                                   & \ion{Si}{ii} & 13.75{\scriptsize (0.03)} &	&	  &	   &				       & \ion{Mg}{ii} & 14.66{\scriptsize (0.08)} \\  
  &         &        &                                   & \ion{O}{i}   & $>14.60$                  & 10& 2.06758 &  $+$76 & \phantom{0}3.2{\scriptsize (0.3)} & \ion{Fe}{ii} & 12.33{\scriptsize (0.02)} \\  
  &         &        &                                   & \ion{Al}{ii} & 12.36{\scriptsize (0.03)} &	&	  &	   &				       & \ion{Si}{ii} & 12.63{\scriptsize (0.08)} \\  
  &         &        &                                   & \ion{S}{ii}  & 13.37{\scriptsize (0.06)} &	&	  &	   &				       & \ion{O}{i}   & 13.83{\scriptsize (0.04)} \\  
  &         &        &                                   & \ion{Ni}{ii} & $<12.53$                  &	&	  &	   &				       & \ion{Al}{ii} & 11.54{\scriptsize (0.02)} \\  
  &         &        &                                   & \ion{N}{i}   & 12.90{\scriptsize (0.04)} & 11& 2.06771 &  $+$89 & \phantom{0}7.3{\scriptsize (0.2)} & \ion{Fe}{ii} & 12.75{\scriptsize (0.01)} \\  
  &         &        &                                   & \ion{P}{ii}  & 11.76{\scriptsize (0.19)} &	&	  &	   &				       & \ion{Si}{ii} & 13.17{\scriptsize (0.03)} \\  
  &         &        &                                   & \ion{Mg}{ii} & 14.90{\scriptsize (0.04)} &	&	  &	   &				       & \ion{O}{i}   & 14.53{\scriptsize (0.03)} \\  
  &         &        &                                   & \ion{Ar}{i}  & $<11.96$                  &	&	  &	   &				       & \ion{Al}{ii} & 11.88{\scriptsize (0.01)} \\  
  &         &        &                                   & \ion{Cr}{ii} & $<11.91$                  & \kern-5pt\raisebox{0pt}[0pt][0pt]{\framebox{\parbox[t]{0.3cm}{12\\ \\ \\ \\ 13}}}
                                                                                                        & 2.06797 & $+$114 &	       11.7{\scriptsize (2.5)} & \ion{Fe}{ii} & 12.17{\scriptsize (0.08)} \\  
6 & 2.06666 &  $-$14 & \phantom{0}2.3{\scriptsize (0.1)} & \ion{Fe}{ii} & 13.37{\scriptsize (0.02)} &	&	  &	   &				       & \ion{Si}{ii} & 12.60{\scriptsize (0.02)} \\  
  &         &        &                                   & \ion{Si}{ii} & 13.90{\scriptsize (0.10)} &	&	  &	   &				       & \ion{O}{i}   & 13.61{\scriptsize (0.03)} \\  
  &         &        &                                   & \ion{O}{i}   & $>13.85$                  &	&	  &	   &				       & \ion{Al}{ii} & 11.45{\scriptsize (0.03)} \\  
  &         &        &                                   & \ion{Al}{ii} & 13.70{\scriptsize (0.11)} & 
                                                                                                        & 2.06811 & $+$128 & \phantom{0}3.1{\scriptsize (0.8)} & \ion{Fe}{ii} & 12.00{\scriptsize (0.09)} \\  
  &         &        &                                   & \ion{S}{ii}  & 13.30{\scriptsize (0.06)} &	&	  &	   &				       & \ion{Si}{ii} & 12.36{\scriptsize (0.03)} \\  
  &         &        &                                   & \ion{Ni}{ii} & $<12.63$                  &	&	  &	   &				       & \ion{O}{i}   & 13.59{\scriptsize (0.04)} \\  
  &         &        &                                   & \ion{N}{i}   & 12.51{\scriptsize (0.06)} &	&	  &	   &				       & \ion{Al}{ii} & 10.74{\scriptsize (0.10)} \\
  &         &        &                                   & \ion{P}{ii}  & 12.15{\scriptsize (0.11)} &	&	  &	   & & & \\
  &         &        &                                   & \ion{Mg}{ii} & 14.69{\scriptsize (0.07)} &	&	  &	   & & & \\
  &         &        &                                   & \ion{Ar}{i}  & $<11.88$                  &	&	  &	   & & & \\ 
  &         &        &                                   & \ion{Cr}{ii} & $<11.71$                  &	&	  &	   & & & \\
  &         &        &                                & \ion{C}{ii}$^*$ & 12.57{\scriptsize (0.11)} &	&	  &	   & & & \\
\hline
\multicolumn{6}{l}{\hspace{0.3cm} Intermediate-ion transitions} & & & & & & \\
\hline
\\[-0.28cm]
1 & 2.06581 &  $-$97 & \phantom{0}6.6{\scriptsize (1.0)} & \ion{Al}{iii} & 12.00{\scriptsize (0.04)} & 3 & 2.06603 &  $-$75 & \phantom{0}8.0{\scriptsize (0.3)} & \ion{Al}{iii} & 12.65{\scriptsize (0.01)} \\
  &         &        &                                   & \ion{Fe}{iii} & 13.20{\scriptsize (0.06)} &   &	   &	    &					& \ion{Fe}{iii} & 13.86{\scriptsize (0.02)} \\
  &         &        &                                   & \ion{N}{ii}   & 13.54{\scriptsize (0.02)} &   &	   &	    &					& \ion{N}{ii}	& 13.99{\scriptsize (0.01)} \\
2 & 2.06591 &  $-$87 & \phantom{0}2.0{\scriptsize (1.0)} & \ion{Al}{iii} & 11.87{\scriptsize (0.05)} & 4 & 2.06640 &  $-$39 &		13.0{\scriptsize (0.9)} & \ion{Al}{iii} & 12.42{\scriptsize (0.03)} \\  
  &         &        &                                   & \ion{Fe}{iii} & 12.91{\scriptsize (0.12)} &   &	   &	    &					& \ion{Fe}{iii} & 13.56{\scriptsize (0.04)} \\  
  &         &        &                                   & \ion{N}{ii}   & 13.20{\scriptsize (0.04)} &   &	   &	    &					& \ion{N}{ii}	& 13.78{\scriptsize (0.01)} \\ 
\end{tabular}
\end{center}
\end{table*}

\addtocounter{table}{-1}
\begin{table*}[!]
\begin{center}
\caption{{\em Continued}}
\begin{tabular}{l c c c l c | l c c c l c}
\hline\hline
Comp. & $z_{\rm abs}$ & $v_{\rm rel}^*$ & $b (\sigma_b)$ & Ion & $\log N (\sigma_{\log N})$ & Comp. & $z_{\rm abs}$ & 
$v_{\rm rel}^*$ & $b (\sigma_b)$ & Ion & $\log N (\sigma_{\log N})$ \\
      &               & [km s$^{-1}$]   & [km s$^{-1}$]  &     &                            &       &               & 
[km s$^{-1}$]   & [km s$^{-1}$]  &     &        
\smallskip
\\ 
\hline
\\[-0.28cm]
5 & 2.06666 &  $-$14 & \phantom{0}9.8{\scriptsize (0.9)} & \ion{Al}{iii} & 12.35{\scriptsize (0.04)} & 6 & 2.06700 &  $+$19 &		16.2{\scriptsize (1.7)} & \ion{Al}{iii} & 12.31{\scriptsize (0.04)} \\   
  &	    &	     &  				 & \ion{Fe}{iii} & 13.32{\scriptsize (0.06)} &   &	   &	    &					& \ion{Fe}{iii} & 13.75{\scriptsize (0.04)} \\   
  &	    &	     &  				 & \ion{N}{ii}   & 13.86{\scriptsize (0.01)} &   &	   &	    &					& \ion{N}{ii}	& 13.79{\scriptsize (0.02)} \\
\hline
\end{tabular}
\begin{minipage}{160mm}
\smallskip
$^*$ Velocity relative to $z=2.06680$ \\
{\it Note.} In this and all the following tables describing the component structure of the DLAs 
studied, the encompassed components correspond to components which, according to our criterion, cannot 
be considered as individual ISM clouds on the QSO line of sight (see Sect.~\ref{indiv-clouds} for the 
definition of our criterion). 
\end{minipage}
\end{center}
\end{table*}
%

Finally, Fig.~\ref{Q0450-HI} shows the fitting solution of two Lyman lines 
of this DLA, Ly$\alpha$ and Ly$\beta$. The fits were obtained by fixing the 
$b$-value at 20 km~s$^{-1}$ and the redshift at $z=2.06666$, i.e. at the 
redshift of one of the stronger metal-line components (the component 6 at 
$v = -14$ km~s$^{-1}$). The derived \ion{H}{i} column density is 
$\log N$(\ion{H}{i}) $= 20.53\pm 0.08$. This is a relatively high value, 
especially in light of the evidence mentioned above for the high ionization 
level in this DLA \citep{vladilo01}.

%

\begin{figure}[t]
\centering
\includegraphics[width=9cm]{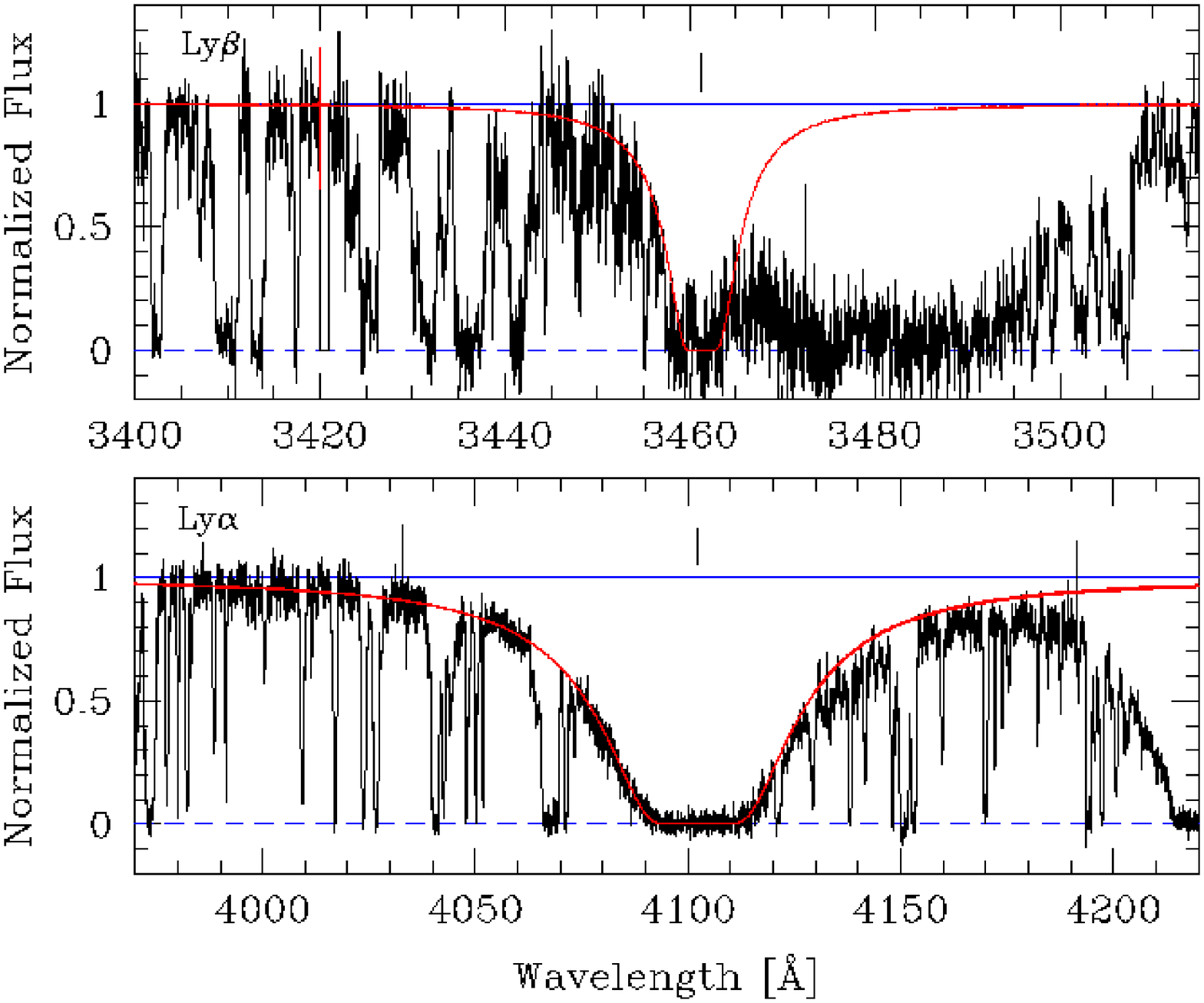}
\caption{Normalized UVES spectrum of Q0841+129 showing the $z_{\rm abs} = 2.375$
DLA Ly$\alpha$ and Ly$\beta$ line profiles with the Voigt profile fits. The 
vertical bar corresponds to the wavelength centroid of the component used for 
the best fit, $z=2.37452$. The measured \ion{H}{i} column density is $\log
N$(\ion{H}{i}) = $20.99\pm 0.08$.}
\label{Q0841-2p375-HI}
\end{figure}
%

\begin{table*}[!]
\begin{center}
\caption{Component structure of the $z_{\rm abs} = 2.375$ DLA system toward Q0841+129}
\label{Q0841-2p375-Ntable}
\begin{tabular}{l c c c l c | l c c c l c}
\hline\hline
\\[-0.3cm]
Comp. & $z_{\rm abs}$ & $v_{\rm rel}^*$ & $b (\sigma_b)$ & Ion & $\log N (\sigma_{\log N})$ & Comp. & $z_{\rm abs}$ & 
$v_{\rm rel}^*$ & $b (\sigma_b)$ & Ion & $\log N (\sigma_{\log N})$ \\
      &               & [km s$^{-1}$]   & [km s$^{-1}$]  &     &                            &       &               & 
[km s$^{-1}$]   & [km s$^{-1}$]  &     &        
\smallskip
\\ 
\hline
\multicolumn{6}{l}{\hspace{0.3cm} Low-ion transitions} & & & & & & \\
\hline
\\[-0.28cm]
\kern-5pt\raisebox{0pt}[0pt][0pt]{\framebox{\parbox[t]{0.2cm}{1\\ \\ \\ \\ 2}}}
  & 2.37407 & $-$40 & 2.0{\scriptsize (1.0)} & \ion{Fe}{ii} & 11.86{\scriptsize (0.05)} & 4 & 2.37452 &     0 & 8.8{\scriptsize (0.1)} & \ion{Fe}{ii} & 14.69{\scriptsize (0.01)} \\
  &         &       &                        & \ion{Si}{ii} & 12.67{\scriptsize (0.06)} &   &	      &       & 		       & \ion{Si}{ii} & 15.16{\scriptsize (0.03)} \\
  &         &       &                        & \ion{Al}{ii} & 11.37{\scriptsize (0.08)} &   &	      &       & 		       & \ion{Al}{ii} & $>13.67$ \\ 
  &         &       &                        & \ion{O}{i}   & 13.10{\scriptsize (0.05)} &   &	      &       & 		       & \ion{O}{i}   & $>15.76$ \\ 
  & 2.37421 & $-$28 & 5.7{\scriptsize (0.8)} & \ion{Fe}{ii} & 12.39{\scriptsize (0.05)} &   &	      &       & 		       & \ion{S}{ii}  & 14.65{\scriptsize (0.04)} \\
  &         &       &                        & \ion{Si}{ii} & 13.15{\scriptsize (0.04)} &   &	      &       & 		       & \ion{N}{i}   & 14.57{\scriptsize (0.01)} \\
  &         &       &                        & \ion{Al}{ii} & 11.81{\scriptsize (0.05)} &   &	      &       & 		       & \ion{Mn}{ii} & 12.42{\scriptsize (0.01)} \\ 
  &         &       &                        & \ion{O}{i}   & 13.85{\scriptsize (0.05)} &   &	      &       & 		       & \ion{Cr}{ii} & 13.05{\scriptsize (0.01)} \\
3 & 2.37435 & $-$15 & 6.6{\scriptsize (0.4)} & \ion{Fe}{ii} & 13.81{\scriptsize (0.03)} &   &	      &       & 		       & \ion{Zn}{ii} & 12.10{\scriptsize (0.02)} \\
  &         &       &                        & \ion{Si}{ii} & 14.04{\scriptsize (0.15)} &   &	      &       & 		       & \ion{Ni}{ii} & 13.45{\scriptsize (0.03)} \\
  &         &       &                        & \ion{Al}{ii} & $>12.54$                  &   &	      &       & 		       & \ion{Mg}{ii} & 15.13{\scriptsize (0.10)} \\
  &         &       &                        & \ion{O}{i}   & $>15.55$                  &   &	      &       & 		       & \ion{P}{ii}  & 12.82{\scriptsize (0.06)} \\
  &         &       &                        & \ion{S}{ii}  & 13.63{\scriptsize (0.15)} &   &	      &       & 		       & \ion{Ar}{i}  & 13.53{\scriptsize (0.08)} \\
  &         &       &                        & \ion{N}{i}   & 13.49{\scriptsize (0.02)} &   &	      &       & 		    & \ion{C}{ii}$^*$ & 12.96{\scriptsize (0.08)} \\
  &         &       &                        & \ion{Mn}{ii} & 11.70{\scriptsize (0.07)} &   &	      &       & 		       & \ion{C}{i}   & $<12.89$ \\
  &         &       &                        & \ion{Cr}{ii} & 11.63{\scriptsize (0.12)} & 5 & 2.37479 & $+$24 & 4.2{\scriptsize (0.1)} & \ion{Fe}{ii} & 13.27{\scriptsize (0.01)} \\
  &         &       &                        & \ion{Ni}{ii} & 12.74{\scriptsize (0.13)} &   &	      &       & 		       & \ion{Si}{ii} & 13.56{\scriptsize (0.01)} \\
  &         &       &                        &              &                           &   &	      &       & 		       & \ion{Al}{ii} & 11.98{\scriptsize (0.03)} \\
  &         &       &                        &              &                           &   &	      &       & 		       & \ion{O}{i}   & $>15.27$ \\
  &         &       &                        &              &                           & 6 & 2.37493 & $+$36 & 3.0{\scriptsize (1.0)} & \ion{Fe}{ii} & 12.22{\scriptsize (0.08)} \\
  &         &       &                        &              &                           &   &	      &       & 		       & \ion{Si}{ii} & 12.43{\scriptsize (0.07)} \\
  &         &       &                        &              &                           &   &	      &       & 		       & \ion{Al}{ii} & 11.31{\scriptsize (0.08)} \\
  &         &       &                        &              &                           &   &	      &       & 		       & \ion{O}{i}   & 13.47{\scriptsize (0.04)} \\
\hline
\multicolumn{6}{l}{\hspace{0.3cm} Intermediate-ion transitions} & & & & & & \\
\hline
\\[-0.28cm]
1 & 2.37435 & $-$15 & 6.6{\scriptsize (0.0)} & \ion{Al}{iii} & 11.82{\scriptsize (0.07)} & 2 & 2.37454 & $+$2  & 11.0{\scriptsize (1.1)} & \ion{Al}{iii} & 12.40{\scriptsize (0.04)} \\ 
  &         &       &                        & \ion{Fe}{iii} & 13.00{\scriptsize (0.12)} &   &	       &       &       		         & \ion{Fe}{iii} & 13.75{\scriptsize (0.03)} \\ 
  &         &       &                        & \ion{S}{iii}  & $<13.55$                  &   &	       &       &                         & \ion{S}{iii}  & $<14.04$ \\
\hline
\end{tabular}
\begin{minipage}{160mm}
\smallskip
$^*$ Velocity relative to $z=2.37452$ 
\end{minipage}
\end{center}
\end{table*}
%

\subsection{Q0841+129, z\mathversion{bold}$_{\rm abs}$\mathversion{normal} = 2.375}
\label{Q0841-2p375}

The two DLA systems at $z_{\rm abs} = 2.375$ and at $z_{\rm abs} = 2.476$ 
toward the quasar Q0841+129 were identified by C. Hazard and were first studied 
at low resolution by \citet{pettini97}. More recently, high resolution spectra 
obtained with HIRES/Keck and UVES/VLT were analyzed by \citet{prochaska99}, 
\citet{prochaska01}, \citet{centurion03}, \citet{ledoux03}, and 
\citet{vladilo03}. 

In the case of the first DLA at $z_{\rm abs} = 2.375$, we confirm the
\citet{prochaska99} and \citet{prochaska01} column density measurements of
\ion{Fe}{ii}, \ion{Si}{ii}, \ion{Ni}{ii}, \ion{Cr}{ii}, and \ion{Zn}{ii}.
However, our column density measurements of \ion{Fe}{ii}, \ion{Zn}{ii},
\ion{S}{ii}, and \ion{Ar}{i} differ by almost 0.1~dex from those of
\citet{centurion03} and \citet{vladilo03}. This discrepancy is likely to be the 
result of the lower signal-to-noise ratio in their UVES/VLT spectra and an 
underestimation of their uncertainties. In the cases of Fe$^+$, Zn$^+$, and 
Ar$^0$, we used at least two different metal-line transitions to derive the 
column densities; e.g. for $N$(Fe$^+$) 9 transitions are considered (see 
Fig.~\ref{Q0841-2p375-metals}). We, nevertheless, confirm the 
\citet{centurion03} $N$(N$^0$) measurement, obtained in our analysis using 
both the \ion{N}{i} triplet at $\lambda_{\rm rest} \sim 1134$ and $\sim 1200$ 
\AA, except the \ion{N}{i}\,$\lambda$\,1134.165 line which is blended with 
\ion{H}{i} lines in the Ly$\alpha$ forest.

From a total of 46 metal-line transitions detected and analyzed (see 
Fig.~\ref{Q0841-2p375-metals}), we obtained the column density measurements of 
\ion{Mg}{ii}, \ion{Mn}{ii}, \ion{P}{ii}, \ion{C}{ii}$^*$, \ion{Al}{iii}, and 
\ion{Fe}{iii}, in addition to the ions discussed above. We also derived 
lower limits to the column densities of \ion{O}{i} and \ion{Al}{ii} and upper 
limits to $N$(\ion{C}{i}) and $N$(\ion{S}{iii}).

The low-ion absorption line profiles of this DLA are characterized by a simple
velocity structure composed of 6 components spread over 80 km~s$^{-1}$. Their 
properties are described in Table~\ref{Q0841-2p375-Ntable}. Only the components 
3 and 4 are detected in the weaker metal-line transitions.  They contain, 
however, $\sim 95$\,\% of the total column density of the fully integrated
profile. Due to the high \ion{H}{i} column density of this DLA system and the 
proximity of the second DLA at $z_{\rm abs} = 2.476$ on the same QSO line of 
sight, the \ion{N}{i} triplet at $\lambda_{\rm rest} \sim 1200$ \AA\ is located 
in the blue wing of the DLA Ly$\alpha$ damping line profile, the 
\ion{Mg}{ii}\,$\lambda$\,1239,\,1240 lines in the red wing of the DLA Ly$\alpha$ 
line, and the \ion{S}{ii}\,$\lambda$\,1259 line, the only unblended \ion{S}{ii} 
line available, in the red wing of the Ly$\alpha$ damping line profile of the 
second DLA. Consequently, we renormalized the damped Ly$\alpha$ profiles of both 
DLAs according to their best values, in order to derive accurate column 
densities of N$^0$, Mg$^+$, and S$^+$. The 1\,$\sigma$ errors on the measured 
column densities are estimated by varying the continuum level by 5\,\%. The 
\ion{S}{ii}\,$\lambda$\,1259 line shows a slight asymmetry in its profile 
compared to other low-ion metal lines, which is not perfectly modeled in our 
solution. The \ion{Mg}{i}\,$\lambda$\,2026 line in Fig.~\ref{Q0841-2p375-metals} 
is plotted to illustrate that there is no contamination of 
\ion{Zn}{ii}\,$\lambda$\,2026 by Mg$^0$.

The intermediate-ion line profiles of this DLA are composed of only two 
components with very similar characteristics to the main components 3 and 4 of 
the low-ion lines (see the second part of Table~\ref{Q0841-2p375-Ntable}). The
derived S$^{++}$ column density measurement is a borderline case between a 
detection and an upper limit due to possible blends with \ion{H}{i} lines in the 
Ly$\alpha$ forest. To be conservative, we assumed it is an upper limit. 
Following the \citet{prochaska02a} photoionization diagnostics, we used the 
column density ratios of different ionization species of the same element 
measured in this DLA to infer the ionization state. Specifically the ratios
$-$~$\log N$(Al$^{++}$)/$N$(Al$^+$) $< -1.22$, $\log N$(Fe$^{++}$)/$N$(Fe$^+$) 
$= -0.94\pm 0.04$, and $\log N$(S$^{++}$)/$N$(S$^+$) $< -0.53$~$-$ indicate 
that this system has a low ionization level, with an ionization fraction, 
defined as the ratio of H$^+$ over (H$^0$ + H$^+$), lower than 10\,\%.

The determination of the \ion{H}{i} column densities from the Ly$\alpha$ damping
lines of the two DLAs toward Q0841+129 was particularly difficult in this case 
because of the proximity of the two systems (separated by 9040 km~s$^{-1}$ 
only). The detections of the Ly$\beta$ line in the DLA at $z_{\rm abs} = 2.375$ 
and the Ly$\beta$ and Ly$\gamma$ lines in the DLA at $z_{\rm abs} = 2.476$ in 
our UVES/VLT spectra provide an indirect check on the fits. In 
Figs.~\ref{Q0841-2p375-HI} and \ref{Q0841-2p476-HI}, we show the derived results
for the two DLAs. In both systems, we fitted all the lines of the Lyman series
simultaneously, and we fixed the $b$-value at 20 km~s$^{-1}$ and the redshift 
at $z = 2.37452$ and $z = 2.47621$, respectively, i.e. at the redshift of the 
strongest metal-line component (the component 4 and 2, respectively). The
derived \ion{H}{i} column densities, $\log N$(\ion{H}{i}) $= 20.99\pm 0.08$ for
the first DLA and $\log N$(\ion{H}{i}) $= 20.78\pm 0.08$ for the second DLA, 
are in a very good agreement with the values from \citet{centurion03} and 
\citet{vladilo03}.

%
\begin{figure*}[!]
\centering
\includegraphics[width=17.5cm]{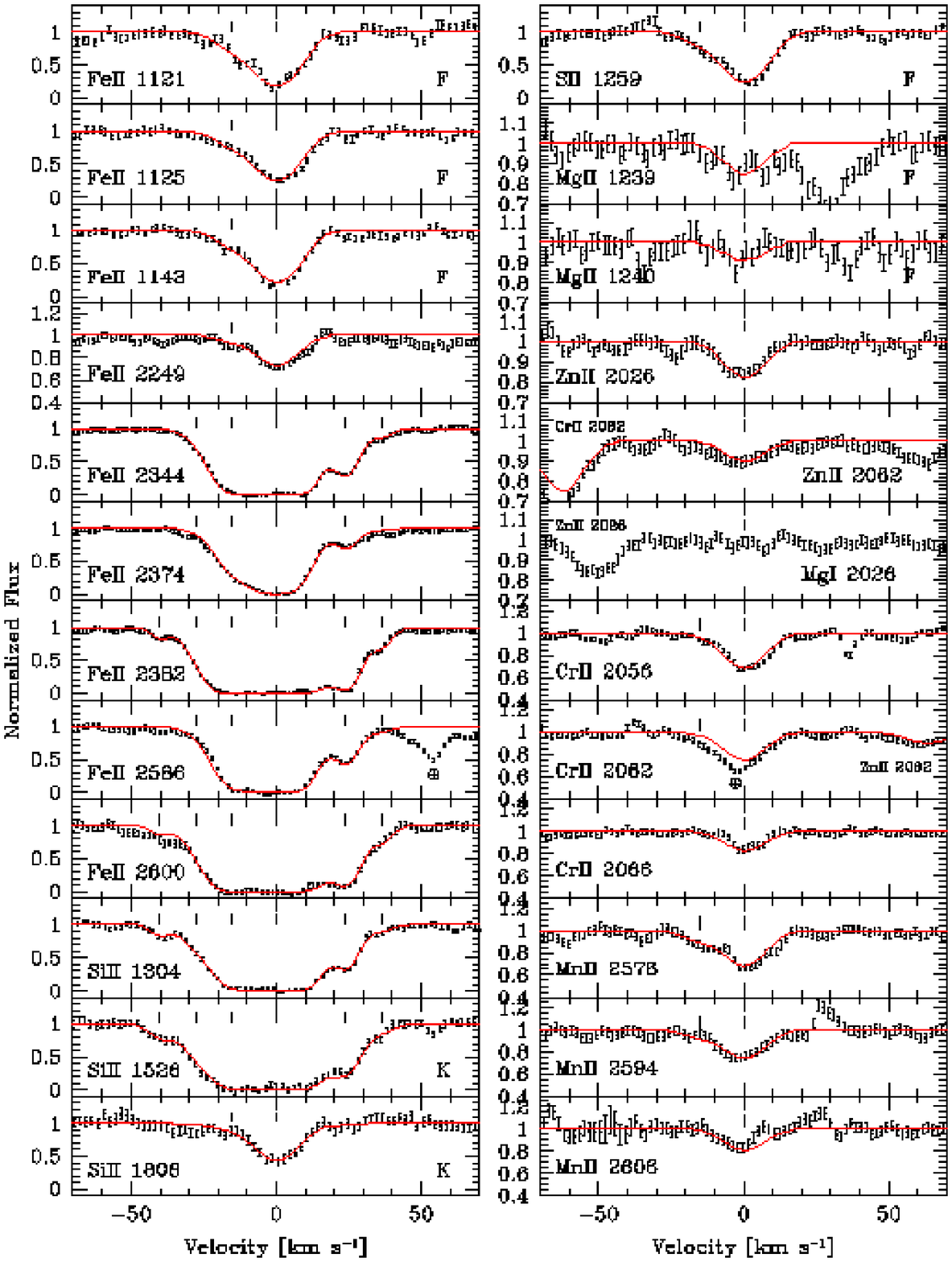}
\caption{Same as Fig.~\ref{Q0450-metals} for the first DLA toward Q0841+129 
at $z_{\rm abs} = 2.375$. The zero velocity is fixed at $z=2.37452$.}
\label{Q0841-2p375-metals}
\end{figure*}

\addtocounter{figure}{-1}
\begin{figure*}[!]
\centering
\includegraphics[width=17.5cm]{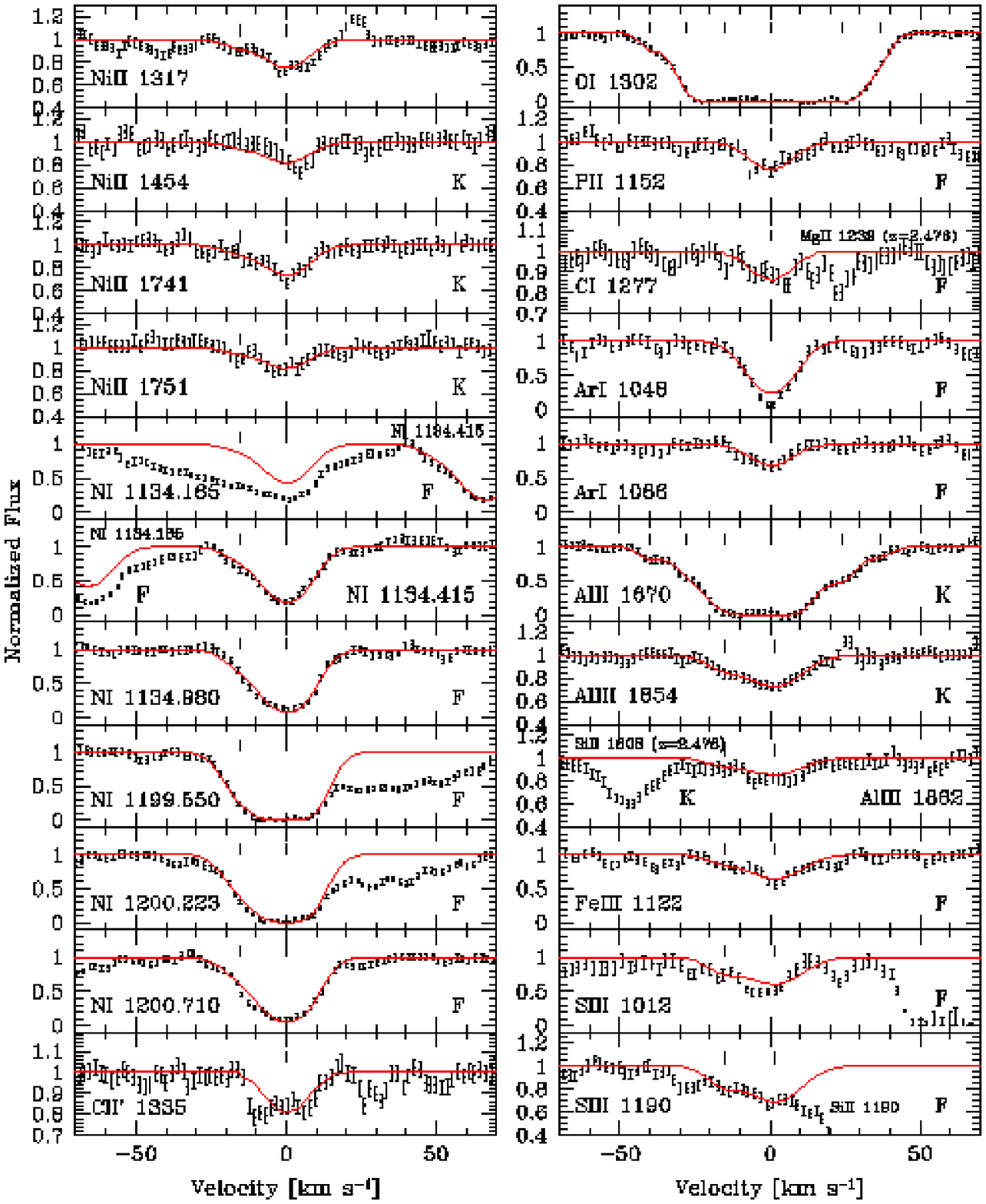}
\caption{{\em Continued}.}
\end{figure*}
%

\begin{figure}[t]
\centering
\includegraphics[width=9cm]{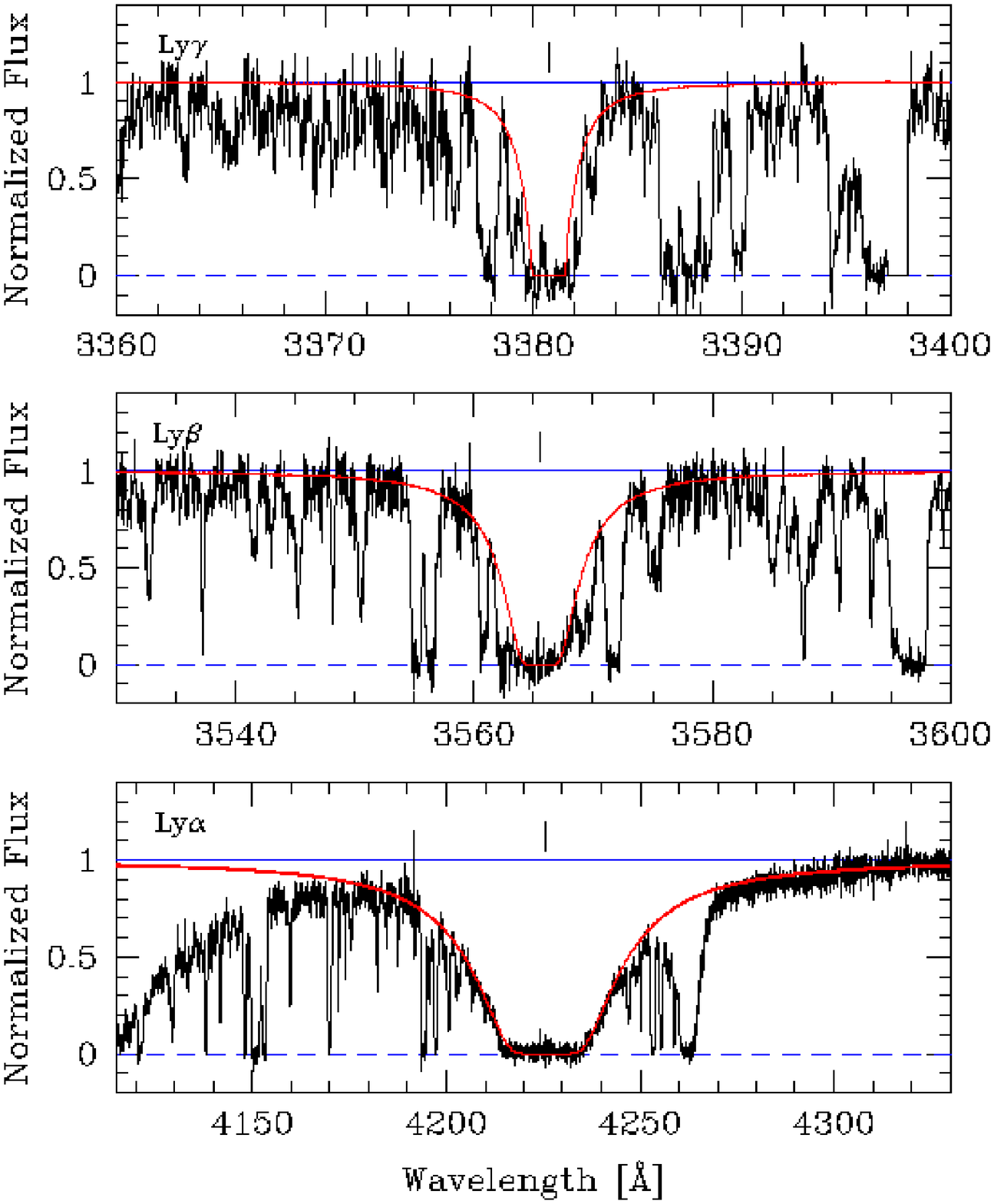}
\caption{Normalized UVES spectrum of Q0841+129 showing the $z_{\rm abs} = 2.476$
DLA Ly$\alpha$, Ly$\beta$, and Ly$\gamma$ line profiles with the Voigt profile 
fits. The vertical bar corresponds to the wavelength centroid of the component 
used for the best fit, $z=2.47621$. The measured \ion{H}{i} column density is 
$\log N$(\ion{H}{i}) = $20.78\pm 0.08$.}
\label{Q0841-2p476-HI}
\end{figure}
%

\begin{figure*}[!]
\centering
\includegraphics[width=17.5cm]{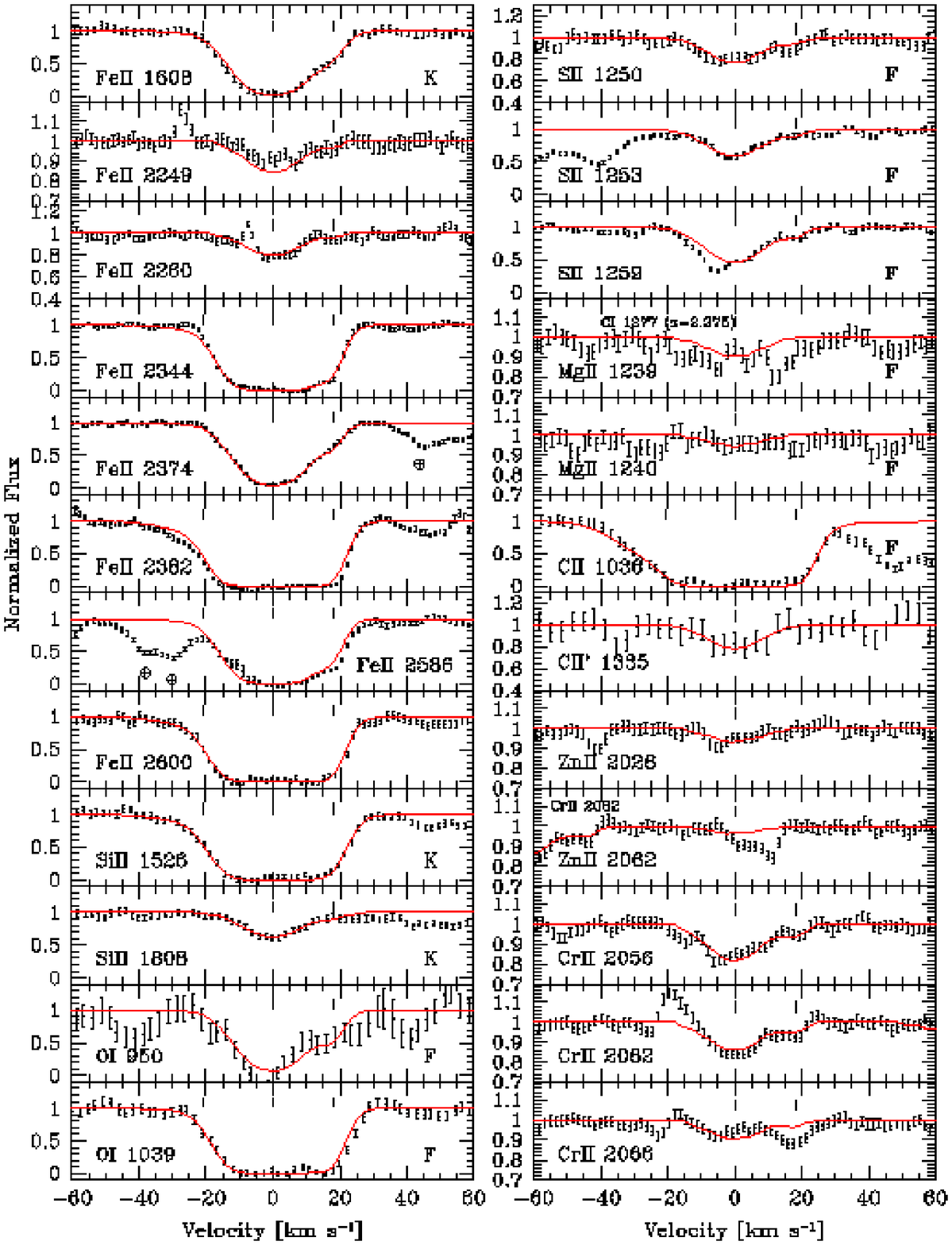}
\caption{Same as Fig.~\ref{Q0450-metals} for the second DLA toward Q0841+129 
at $z_{\rm abs} = 2.476$. The zero velocity is fixed at $z=2.47621$.}
\label{Q0841-2p476-metals}
\end{figure*}

\addtocounter{figure}{-1}
\begin{figure*}[!]
\centering
\includegraphics[width=17.5cm]{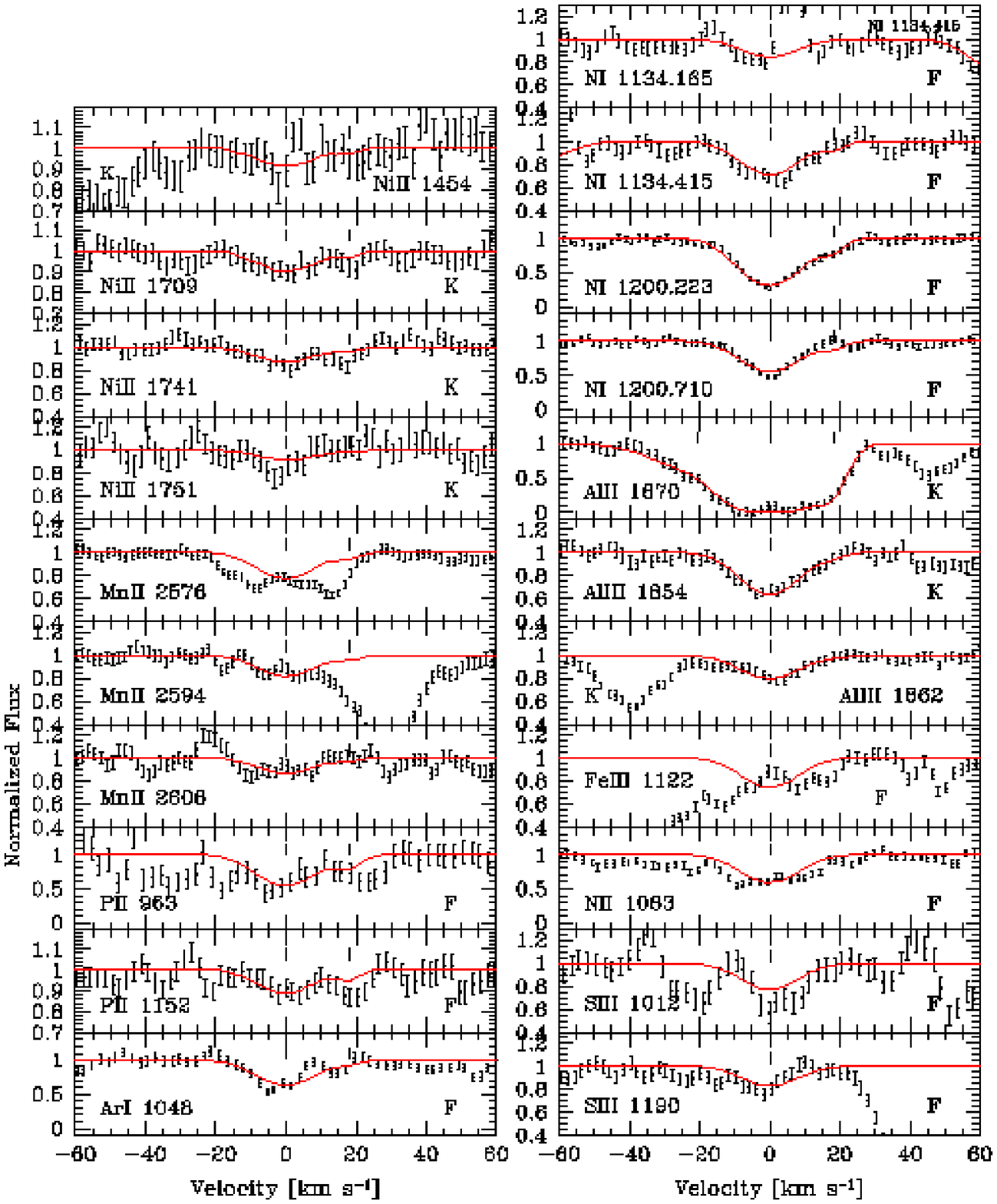}
\caption{{\em Continued}.}
\end{figure*}
%

\begin{table}[t]
\begin{center}
\caption{Component structure of the $z_{\rm abs} = 2.476$ DLA system toward Q0841+129}
\label{Q0841-2p476-Ntable}
\vspace{-0.4cm}
\begin{tabular}{l c c c l c}
\hline\hline
\\[-0.3cm]
Comp. & $z_{\rm abs}$ & $v_{\rm rel}^*$ & $b (\sigma_b)$ & Ion & $\log N (\sigma_{\log N})$ \\
      &               & [km s$^{-1}$]   & [km s$^{-1}$]  &     &                            
\smallskip
\\ 
\hline
\multicolumn{6}{l}{\hspace{0.3cm} Low- and intermediate-ion transitions} \\
\hline
\\[-0.28cm]
\kern-5pt\raisebox{0pt}[0pt][0pt]{\framebox{\parbox[t]{0.2cm}{1\\ \\ \\ \\ \\ 2}}}
  & 2.47597 & $-$21 &           15.0{\scriptsize (2.0)} & \ion{Fe}{ii}  & 12.43{\scriptsize (0.12)} \\
  &         &       &   		                & \ion{Si}{ii}  & 12.83{\scriptsize (0.12)} \\
  &         &       &   		                & \ion{O}{i}    & 13.74{\scriptsize (0.20)} \\
  &         &       &   		                & \ion{Al}{ii}  & 12.22{\scriptsize (0.03)} \\
  &         &       &   		                & \ion{C}{ii}   & $>13.88$ \\
  & 2.47621 &     0 &           10.2{\scriptsize (0.1)} & \ion{Fe}{ii}  & 14.48{\scriptsize (0.03)} \\
  &         &       &   		                & \ion{Si}{ii}  & 14.96{\scriptsize (0.02)} \\
  &         &       &   		                & \ion{N}{i}    & 13.91{\scriptsize (0.08)} \\
  &         &       &   		                & \ion{S}{ii}   & 14.45{\scriptsize (0.10)} \\
  &         &       &   		                & \ion{O}{i}    & 16.11{\scriptsize (0.09)} \\
  &         &       &   		                & \ion{Ar}{i}   & 13.09{\scriptsize (0.08)} \\
  &         &       &   		                & \ion{Al}{ii}  & $>13.26$ \\ 
  &         &       &   		                & \ion{Cr}{ii}  & 12.84{\scriptsize (0.06)} \\
  &         &       &   		                & \ion{Zn}{ii}  & 11.69{\scriptsize (0.10)} \\
  &         &       &   		                & \ion{Mn}{ii}  & 12.30{\scriptsize (0.15)} \\
  &         &       &   		                & \ion{P}{ii}   & 12.50{\scriptsize (0.07)} \\
  &         &       &   		                & \ion{Ni}{ii}  & 13.15{\scriptsize (0.07)} \\
  &         &       &   		                & \ion{Mg}{ii}  & 14.99{\scriptsize (0.15)} \\
  &         &       &   		                & \ion{C}{ii}   & $>15.52$ \\
  &         &       &   		             & \ion{C}{ii}$^*$  & $<13.08$ \\
  &         &       &                                   & \ion{Al}{iii} & 12.54{\scriptsize (0.02)} \\
  &         &       &   		                & \ion{Fe}{iii} & $<13.55$ \\
  &         &       &   		                & \ion{N}{ii}	& $<13.54$ \\
  &         &       &   		                & \ion{S}{iii}  & $<13.72$ \\
3 & 2.47642 & $+$18 & \phantom{0}2.3{\scriptsize (0.3)} & \ion{Fe}{ii}  & 13.15{\scriptsize (0.05)} \\
  &	    &	    &			                & \ion{Si}{ii}  & 13.83{\scriptsize (0.16)} \\
  &	    &	    &			                & \ion{N}{i}    & 12.80{\scriptsize (0.10)} \\
  &	    &	    &			                & \ion{S}{ii}   & 13.31{\scriptsize (0.08)} \\
  &	    &	    &			                & \ion{O}{i}    & 15.11{\scriptsize (0.19)} \\
  &	    &	    &			                & \ion{Ar}{i}   & 11.79{\scriptsize (0.15)} \\
  &	    &	    &			                & \ion{Al}{ii}  & $>12.54$ \\ 
  &	    &	    &			                & \ion{Cr}{ii}  & 11.91{\scriptsize (0.09)} \\
  &	    &	    &			                & \ion{Mn}{ii}  & 11.17{\scriptsize (0.20)} \\
  &	    &	    &			                & \ion{P}{ii}   & 11.70{\scriptsize (0.20)} \\
  &	    &	    &			                & \ion{Ni}{ii}  & 12.15{\scriptsize (0.20)} \\ 
  &	    &	    &			                & \ion{C}{ii}   & $>15.74$ \\		  
\hline
\end{tabular}
\begin{minipage}{160mm}
\smallskip
$^*$ Velocity relative to $z=2.47621$ 
\end{minipage}
\end{center}
\end{table}
%

\subsection{Q0841+129, z\mathversion{bold}$_{\rm abs}$\mathversion{normal} = 2.476}
\label{Q0841-2p476}

As mentioned in Sect.~\ref{Q0841-2p375}, this DLA system toward Q0841+129 has
already been studied by several authors at high resolution. Using our UVES/VLT
spectra combined with the existing HIRES/Keck spectra, we obtained a high
signal-to-noise ratio and a total optical wavelength coverage, which allowed us
to analyze 45 different metal-line transitions (see
Fig.~\ref{Q0841-2p476-metals}). We obtained the column density measurements of
\ion{Fe}{ii}, \ion{Si}{ii}, \ion{N}{i}, \ion{S}{ii}, \ion{O}{i}, \ion{Ar}{i},
\ion{Mg}{ii}, \ion{P}{ii}, \ion{Mn}{ii}, \ion{Ni}{ii}, \ion{Cr}{ii}, 
\ion{Zn}{ii}, and \ion{Al}{iii}. We also derived lower limits to the column 
densities of \ion{C}{ii} and \ion{Al}{ii} from saturated lines, and upper limits 
to the column densities of \ion{C}{ii}$^*$, \ion{Fe}{iii}, \ion{N}{ii}, and 
\ion{S}{iii}. Discrepancies between 0.05 and 0.1~dex are observed between 
our column density measurements and those obtained by \citet{prochaska99}, 
\citet{prochaska01}, \citet{centurion03}, and \citet{vladilo03}. In the majority 
of cases, we used a larger number of metal-line transitions to measure the 
column density of a given ion, which appreciably increases the accuracy of the 
measurement, especially in those cases where the transitions lie outside the 
Ly$\alpha$ forest.  In addition, in the case of the \ion{N}{i}, \ion{S}{ii}, and 
\ion{Ar}{i} lines, we benefited from a signal-to-noise ratio that is twice 
as high as the spectra used by \citet{centurion03} and \citet{vladilo03}. 

The low-ion absorption line profiles of this DLA are characterized by only 3
components spread over 20 km~s$^{-1}$ (see Table~\ref{Q0841-2p476-Ntable}). The
component~1 is only observed in the stronger metal-line transitions and it
contributes negligibly to the total column density. In weak metal-line 
transitions, the component 3 is only marginally detected, like for instance in 
\ion{Ar}{i}, \ion{P}{ii}, \ion{Mn}{ii}, and \ion{Ni}{ii} (see 
Fig.~\ref{Q0841-2p476-metals}). Thus, for several ions we adopted large errors 
on their component~3 column density measurements. The column density 
measurement of O$^0$ is a borderline case between a detection and a limit. It 
was derived from the saturated \ion{O}{i}\,$\lambda$\,1039 line, which 
gives a constraint on the lower limit to $N$(O$^0$) and the 
\ion{O}{i}\,$\lambda$\,950 line far in the blue, exposed to several possible 
line blendings, which provides a constraint on the upper limit to $N$(O$^0$). 
By fitting these two lines simultaneously, we assumed that we get a reliable 
column density measurement of O$^0$. Granted the large uncertainty, we adopted 
a conservative error estimate. The determination of $N$(P$^+$) was also 
challenging. However, the value derived from the \ion{P}{ii}\,$\lambda$\,963 
line, far in the blue (perhaps blended with \ion{H}{i} lines in the Ly$\alpha$ 
forest), agrees perfectly with the weak \ion{P}{ii}\,$\lambda$\,1152 
line. Therefore, we assumed the measured $N$(P$^+$) value is a good detection. 
Finally, the measurements of the column densities of Mg$^+$ and Zn$^+$ are 
borderline cases between detections and upper limits. Indeed, the observed 
\ion{Zn}{ii}\,$\lambda$\,2026,\,2062 and \ion{Mg}{ii}\,$\lambda$\,1239,\,1240 
lines are very weak, detected at 3\,$\sigma$ only. Therefore, we report 
conservatively large errors for their column density measurements.

In the intermediate-ion line profiles of this DLA, we detected a single 
component which has the same characteristics as the strongest component of 
low-ion lines, component 2 (see Table~\ref{Q0841-2p476-Ntable}). We obtained 
the column density measurements of four intermediate-ions, Al$^{++}$, Fe$^{++}$, 
N$^+$, and S$^{++}$. To be conservative, we considered three of them, 
$-$~$N$(Fe$^{++}$), $N$(N$^+$), and $N$(S$^{++}$)~$-$, as upper limits due to 
possible blends with \ion{H}{i} lines in the Ly$\alpha$ forest (see 
Fig.~\ref{Q0841-2p476-metals}). These measurements lead to the following column 
density ratios for different ionization species of the same element: 
$\log N$(Al$^{++}$)/$N$(Al$^+$) $< -0.78$, $\log N$(Fe$^{++}$)/$N$(Fe$^+$) 
$<-0.95$, $\log N$(N$^{+}$)/$N$(N$^0$) $<-0.40$, and 
$\log N$(S$^{++}$)/$N$(S$^+$) $< -0.76$. According to the photoionization 
diagnostics of \citet{prochaska02a}, these ratios show the gas is predominantly 
neutral.

%

\begin{figure}[t]
\centering
\includegraphics[width=9cm]{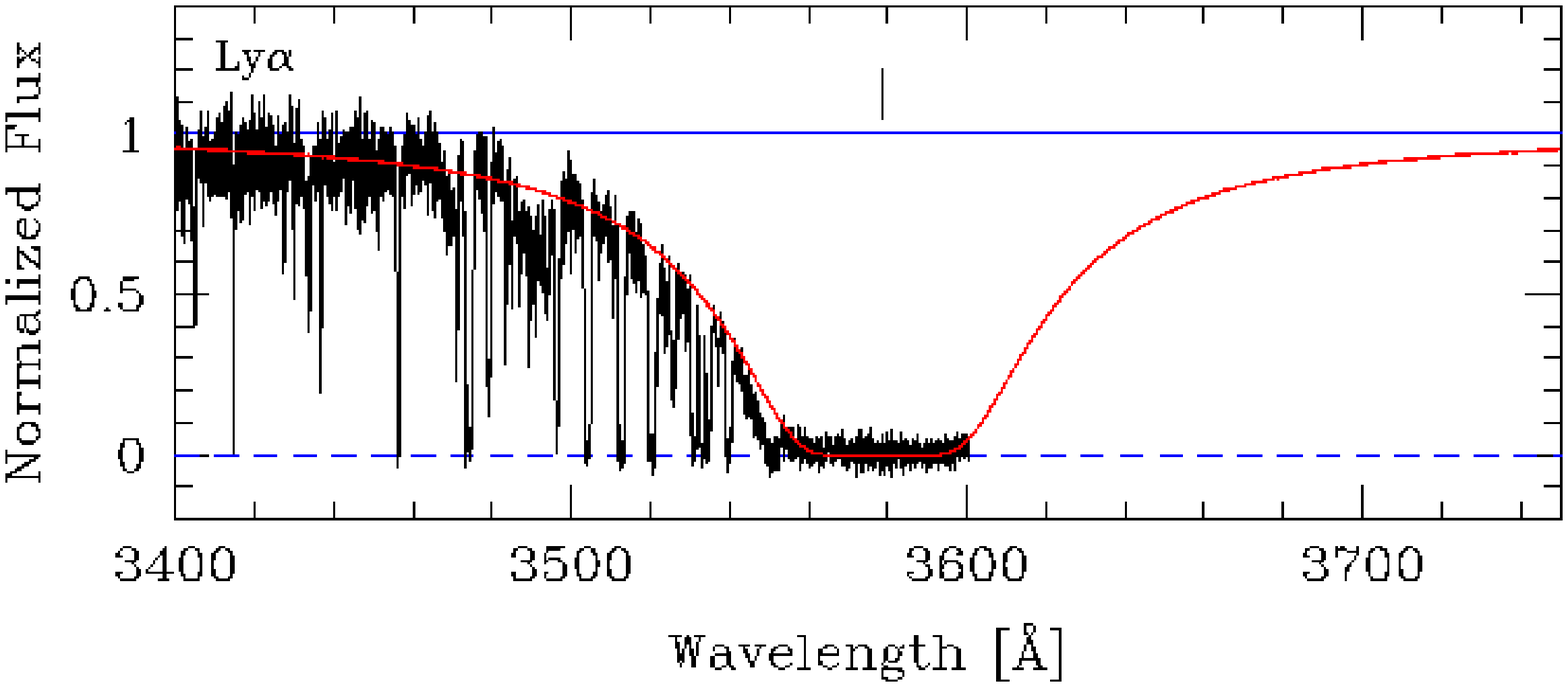}
\caption{Normalized UVES spectrum of Q1157+014 showing the DLA blue wing of 
the Ly$\alpha$ line profile with the Voigt profile fit. The vertical bar 
corresponds to the wavelength centroid of the component used for the best fit, 
$z=1.94366$. The measured \ion{H}{i} column density is $\log N$(\ion{H}{i}) = 
$21.60\pm 0.10$.}
\label{Q1157-HI}
\end{figure}
%

\begin{figure*}[!]
\centering
\includegraphics[width=17.5cm]{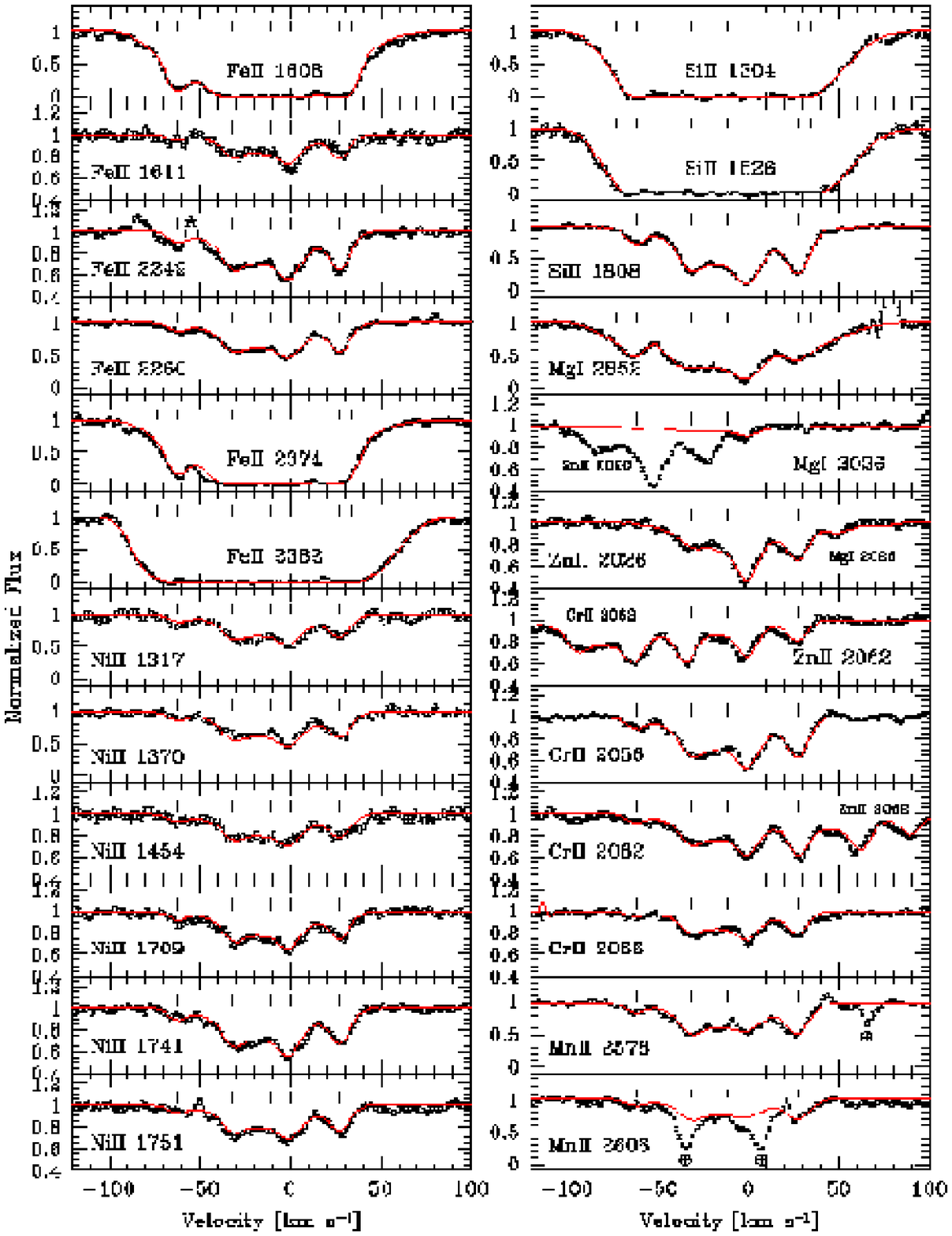}
\caption{Same as Fig.~\ref{Q0450-metals} for the DLA toward Q1157+014. The 
zero velocity is fixed at $z=1.94377$.}
\label{Q1157-metals}
\end{figure*}

\addtocounter{figure}{-1}
\begin{figure*}[!]
\centering
\includegraphics[width=17.5cm]{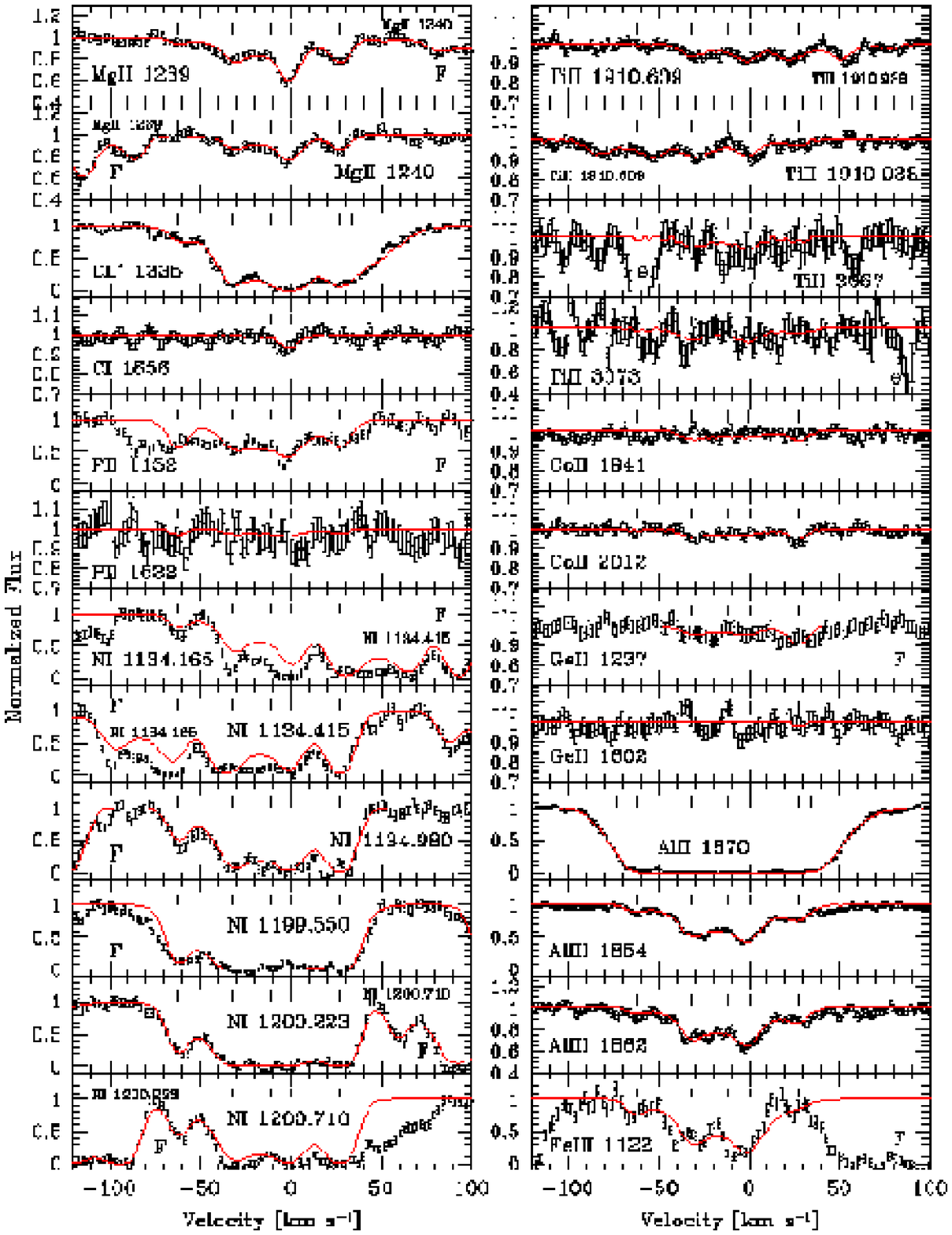}
\caption{{\em Continued}.}
\end{figure*}
%

\begin{table*}[!]
\begin{center}
\caption{Component structure of the $z_{\rm abs} = 1.944$ DLA system toward Q1157+014}
\label{Q1157-Ntable}
\begin{tabular}{l c c c l c | l c c c l c}
\hline\hline
\\[-0.3cm]
Comp. & $z_{\rm abs}$ & $v_{\rm rel}^*$ & $b (\sigma_b)$ & Ion & $\log N (\sigma_{\log N})$ & Comp. & $z_{\rm abs}$ & 
$v_{\rm rel}^*$ & $b (\sigma_b)$ & Ion & $\log N (\sigma_{\log N})$ \\
      &               & [km s$^{-1}$]   & [km s$^{-1}$]  &     &                            &       &               & 
[km s$^{-1}$]   & [km s$^{-1}$]  &     &        
\smallskip
\\ 
\hline
\multicolumn{8}{l}{\hspace{0.3cm} Low- and intermediate-ion transitions} & & & & \\
\hline
\\[-0.28cm]
\kern-5pt\raisebox{0pt}[0pt][0pt]{\framebox{\parbox[t]{0.2cm}{1\\ \\ \\ \\ 2}}}
  & 1.94305 & $-$73 &           12.2{\scriptsize (0.7)} & \ion{Fe}{ii}  & 13.39{\scriptsize (0.05)} & \kern-5pt\raisebox{0pt}[0pt][0pt]{\framebox{\parbox[b]{0.2cm}{\mbox{} \\ 5}}}
                                                                                                        & 1.94377 &	0 & \phantom{0}5.7{\scriptsize (0.2)} & \ion{Fe}{ii}  & 14.47{\scriptsize (0.03)} \\
  &         &       &                                   & \ion{Si}{ii}  & 13.60{\scriptsize (0.01)} &   &	  &	  &				      & \ion{Si}{ii}  & 15.27{\scriptsize (0.02)} \\
  &         &       &                                   & \ion{Al}{ii}  & $>12.05$                  &   &	  &	  &				      & \ion{Al}{ii}  & $>14.81$ \\
  &         &       &                                   & \ion{Mg}{i}   & 11.69{\scriptsize (0.02)} &   &	  &	  &				      & \ion{P}{ii}   & 12.76{\scriptsize (0.10)} \\
  & 1.94316 & $-$62 & \phantom{0}5.9{\scriptsize (0.4)} & \ion{Fe}{ii}  & 14.00{\scriptsize (0.10)} &   &	  &	  &				      & \ion{Mn}{ii}  & 11.98{\scriptsize (0.06)} \\
  &         &       &                                   & \ion{Si}{ii}  & 14.52{\scriptsize (0.03)} &   &	  &	  &				      & \ion{Cr}{ii}  & 12.82{\scriptsize (0.03)} \\
  &         &       &                                   & \ion{Al}{ii}  & $>12.48$                  &   &	  &	  &				      & \ion{N}{i}    & $>14.39$ \\
  &         &       &                                   & \ion{P}{ii}   & 13.02{\scriptsize (0.14)} &   &	  &	  &				      & \ion{Zn}{ii}  & 12.38{\scriptsize (0.06)} \\
  &         &       &                                   & \ion{Mn}{ii}  & 11.93{\scriptsize (0.04)} &   &	  &	  &				      & \ion{Ni}{ii}  & 13.22{\scriptsize (0.04)} \\
  &         &       &                                   & \ion{Cr}{ii}  & 12.32{\scriptsize (0.03)} &   &	  &	  &				      & \ion{Mg}{ii}  & 15.34{\scriptsize (0.03)} \\
  &         &       &                                   & \ion{N}{i}    & 13.84{\scriptsize (0.03)} &   &	  &	  &				      & \ion{Ti}{ii}  & 12.11{\scriptsize (0.07)} \\
  &         &       &                                   & \ion{Ni}{ii}  & 12.80{\scriptsize (0.03)} &   &	  &	  &				      & \ion{Co}{ii}  & 11.43{\scriptsize (0.25)} \\
  &         &       &                                   & \ion{Mg}{ii}  & 14.52{\scriptsize (0.12)} &   &	  &	  &				      & \ion{Mg}{i}   & 12.03{\scriptsize (0.04)} \\
  &         &       &                                   & \ion{Ti}{ii}  & 11.30{\scriptsize (0.23)} &   &	  &	  &				    & \ion{C}{ii}$^*$ & 14.95{\scriptsize (0.12)} \\
  &         &       &                                   & \ion{Co}{ii}  & 11.85{\scriptsize (0.24)} &   &	  &	  &				      & \ion{C}{i}    & $<12.07$ \\ 
  &         &       &                                   & \ion{Mg}{i}   & 11.76{\scriptsize (0.02)} &   &	  &	  &				      & \ion{Ge}{ii}  & 10.73{\scriptsize (0.27)} \\
  &         &       &                                 & \ion{C}{ii}$^*$ & 12.78{\scriptsize (0.05)} &   &	  &	  &				      & \ion{Al}{iii} & 12.28{\scriptsize (0.03)} \\
  &         &       &                                   & \ion{Al}{iii} & 11.63{\scriptsize (0.05)} &   &	  &	  &				      & \ion{Fe}{iii} & $<13.97$ \\
  &         &       &                                   & \ion{Fe}{iii} & $<13.29$                  & \kern-5pt\raisebox{0pt}[0pt][0pt]{\framebox{\parbox[t]{0.2cm}{6\\ \\ \\ \\ \\ \\ \\ \\ \\ \\ \\ \\ \\ \\ \\ \\ \\ 7}}}
                                                                                                        & 1.94404 & $+$27 & \phantom{0}6.3{\scriptsize (0.2)} & \ion{Fe}{ii}  & 14.66{\scriptsize (0.01)} \\
3 & 1.94346 & $-$32 & \phantom{0}6.3{\scriptsize (0.4)} & \ion{Fe}{ii}  & 14.40{\scriptsize (0.04)} &   &	  &	  &				      & \ion{Si}{ii}  & 15.19{\scriptsize (0.01)} \\
  &         &       &                                   & \ion{Si}{ii}  & 14.90{\scriptsize (0.02)} &   &	  &	  &				      & \ion{Al}{ii}  & $>9.00$ \\		    
  &         &       &                                   & \ion{Al}{ii}  & $>9.01$                   &   &	  &	  &				      & \ion{P}{ii}   & 12.95{\scriptsize (0.09)} \\
  &         &       &                                   & \ion{P}{ii}   & 12.65{\scriptsize (0.11)} &   &	  &	  &				      & \ion{Mn}{ii}  & 12.55{\scriptsize (0.01)} \\
  &         &       &                                   & \ion{Mn}{ii}  & 12.36{\scriptsize (0.03)} &   &	  &	  &				      & \ion{Cr}{ii}  & 12.98{\scriptsize (0.02)} \\
  &         &       &                                   & \ion{Cr}{ii}  & 12.71{\scriptsize (0.04)} &   &	  &	  &				      & \ion{N}{i}    & $>14.87$ \\		    
  &         &       &                                   & \ion{N}{i}    & $>14.27$                  &   &	  &	  &				      & \ion{Zn}{ii}  & 12.22{\scriptsize (0.05)} \\
  &         &       &                                   & \ion{Zn}{ii}  & 11.63{\scriptsize (0.06)} &   &	  &	  &				      & \ion{Ni}{ii}  & 13.36{\scriptsize (0.02)} \\
  &         &       &                                   & \ion{Ni}{ii}  & 13.19{\scriptsize (0.02)} &   &	  &	  &				      & \ion{Mg}{ii}  & 15.20{\scriptsize (0.03)} \\
  &         &       &                                   & \ion{Mg}{ii}  & 14.95{\scriptsize (0.06)} &   &	  &	  &				      & \ion{Ti}{ii}  & 11.76{\scriptsize (0.20)} \\
  &         &       &                                   & \ion{Ti}{ii}  & 12.07{\scriptsize (0.07)} &   &	  &	  &				      & \ion{Co}{ii}  & 12.46{\scriptsize (0.16)} \\
  &         &       &                                   & \ion{Co}{ii}  & 12.24{\scriptsize (0.18)} &   &	  &	  &				      & \ion{Mg}{i}   & 11.50{\scriptsize (0.04)} \\
  &         &       &                                   & \ion{Mg}{i}   & 11.92{\scriptsize (0.03)} &   &	  &	  &				    & \ion{C}{ii}$^*$ & 13.80{\scriptsize (0.04)} \\
  &         &       &                                 & \ion{C}{ii}$^*$ & 13.70{\scriptsize (0.04)} &   &	  &	  &				      & \ion{Ge}{ii}  & 11.51{\scriptsize (0.12)} \\
  &         &       &                                   & \ion{Ge}{ii}  & 10.88{\scriptsize (0.18)} &   &	  &	  &				      & \ion{Al}{iii} & 12.00{\scriptsize (0.03)} \\
  &         &       &                                   & \ion{Al}{iii} & 12.24{\scriptsize (0.03)} &   &	  &	  &				      & \ion{Fe}{iii} & $<12.31$ \\
  &         &       &                                   & \ion{Fe}{iii} & $<13.76$                  & 
                                                                                                        & 1.94411 & $+$35 &	      24.0{\scriptsize (1.3)} & \ion{Fe}{ii}  & 13.54{\scriptsize (0.06)} \\
\kern-5pt\raisebox{0pt}[0pt][0pt]{\framebox{\parbox[t]{0.2cm}{4\\ \\ \\ \\ \\ \\ \\ \\ \\ \\ \\ \\ \\ \\ \\ \\ \\ \\}}}
  & 1.94366 & $-$11 &           28.8{\scriptsize (0.4)} & \ion{Fe}{ii}  & 15.24{\scriptsize (0.01)} &   &	  &	  &				      & \ion{Si}{ii}  & 14.16{\scriptsize (0.06)} \\
  &         &       &                                   & \ion{Si}{ii}  & 15.67{\scriptsize (0.01)} &   &	  &	  &				      & \ion{Al}{ii}  & $>12.54$ \\		    
  &         &       &                                   & \ion{Al}{ii}  & $>14.56$                  &   &	  &	  &				      & \ion{Mg}{i}   & 12.10{\scriptsize (0.02)} \\
  &         &       &                                   & \ion{P}{ii}   & 13.64{\scriptsize (0.08)} &   &	  &	  &				    & \ion{C}{ii}$^*$ & 14.05{\scriptsize (0.01)} \\
  &         &       &                                   & \ion{Mn}{ii}  & 13.01{\scriptsize (0.01)} &   &   &    &    &    & \\
  &         &       &                                   & \ion{Cr}{ii}  & 13.53{\scriptsize (0.01)} &   &   &	 &    &    & \\
  &         &       &                                   & \ion{N}{i}    & $>14.86$                  &   &   &	 &    &    & \\
  &         &       &                                   & \ion{Zn}{ii}  & 12.73{\scriptsize (0.04)} &   &   &	 &    &    & \\
  &         &       &                                   & \ion{Ni}{ii}  & 14.00{\scriptsize (0.01)} &   &   &	 &    &    & \\
  &         &       &                                   & \ion{Mg}{ii}  & 15.67{\scriptsize (0.03)} &   &   &	 &    &    & \\
  &         &       &                                   & \ion{Ti}{ii}  & 12.55{\scriptsize (0.16)} &   &   &	 &    &    & \\
  &         &       &                                   & \ion{Co}{ii}  & 12.89{\scriptsize (0.10)} &   &   &	 &    &    & \\
  &         &       &                                   & \ion{Mg}{i}   & 12.65{\scriptsize (0.01)} &   &   &	 &    &    & \\
  &         &       &                                 & \ion{C}{ii}$^*$ & 14.34{\scriptsize (0.02)} &   &   &	 &    &    & \\
  &         &       &                                   & \ion{C}{i}    & $<12.10$                  &   &   &	 &    &    & \\
  &         &       &                                   & \ion{Ge}{ii}  & 11.71{\scriptsize (0.14)} &   &   &	 &    &    & \\
  &         &       &                                   & \ion{Al}{iii} & 13.02{\scriptsize (0.01)} &   &   &    &    &    & \\
  &         &       &                                   & \ion{Fe}{iii} & $<14.45$                  &   &   &    &    &    & \\
\hline
\end{tabular}
\begin{minipage}{160mm}
\smallskip
$^*$ Velocity relative to $z=1.94377$ 
\end{minipage}
\end{center}
\end{table*}
%

\subsection{Q1157+014, z\mathversion{bold}$_{\rm abs}$\mathversion{normal} = 1.944}
\label{Q1157}

This DLA system was first identified by \citet{wolfe81} at 21~cm. They obtained
an estimation of its H$^0$ column density of $\log N$(\ion{H}{i}) $=21.80\pm 
0.10$. This value is an important reference to check our measurement of
$N$(\ion{H}{i}) derived from UVES/VLT high resolution spectra. In 
Fig.~\ref{Q1157-HI} we show the Ly$\alpha$ line of the DLA system. Only the 
blue wing is plotted, because of the very uncertain normalization of the red
Ly$\alpha$ damping wing, given its proximity to the QSO Ly$\alpha$ emission peak 
($\lambda \sim 3635$ \AA). By fixing the $b$-value at 20 km~s$^{-1}$ and the 
redshift at $z = 1.94366$, i.e. at the redshift of the strongest metal-line 
component (the component 4), we obtained a good fit of the blue Ly$\alpha$ 
damping wing profile. The measured \ion{H}{i} column density, 
$\log N$(\ion{H}{i}) $= 21.60\pm 0.10$, is consistent with the value obtained 
by \citet{wolfe81}. In addition, our measurement is in a very good agreement 
with the H$^0$ column density, $\log N$(\ion{H}{i}) $=21.70\pm 0.10$, also
measured from high resolution spectra by \citet{ledoux03}.

This DLA, with its high \ion{H}{i} column density and high metallicity 
($>10^{21}$ cm$^{-2}$, [Zn/H] $= -1.27\pm 0.12$), belongs to the category of 
``metal-strong'' DLA systems defined by \citet{prochaska03b}. These systems 
show strong metal absorption lines and, as a consequence, allow the ionic 
column density measurements of a very large number of ions and elements. Indeed, 
in this DLA system, for instance, 54 metal-line transitions were detected 
(see Fig.~\ref{Q1157-metals}). We obtained accurate column density measurements 
of \ion{Fe}{ii}, \ion{Si}{ii}, \ion{Mg}{ii}, \ion{P}{ii}, \ion{C}{ii}$^*$, 
\ion{Mn}{ii}, \ion{Ni}{ii}, \ion{Cr}{ii}, \ion{Zn}{ii}, \ion{Ti}{ii}, 
\ion{Co}{ii}, \ion{Ge}{ii}, \ion{Mg}{i}, and \ion{Al}{iii}. We also derived 
lower limits to the column densities of \ion{Al}{ii} and \ion{N}{i} and upper 
limits to the column densities of \ion{C}{i}, \ion{B}{ii}, \ion{Cl}{i}, 
\ion{Cu}{ii}, \ion{Kr}{i}, and \ion{As}{ii}.

The low-ion absorption line profiles of this DLA are characterized by 7
components spread over 145 km~s$^{-1}$. Their redshifts, $b$-values, and column
densities are presented in Table~\ref{Q1157-Ntable}. The components 1 and 7 are
only observed in the stronger metal lines and make a negligible contribution to
the total ionic column densities. The problem of contamination of the 
\ion{Zn}{ii}\,$\lambda$\,2026 line by the \ion{Mg}{i}\,$\lambda$\,2026 line was 
first raised by \citet{prochaska01}. To control the impact of this blend on the 
Zn$^+$ column density measurement, we needed to know the level of the 
contribution of Mg$^0$ in the \ion{Zn}{ii} line. For this purpose, we used the 
\ion{Mg}{i}\,$\lambda$\,2852 line, which gives an accurate estimate of the 
contamination at 2026 \AA\ (see Fig.~\ref{Q1157-metals}). One drawback 
of the high \ion{H}{i} column density of this DLA is that the blue Ly$\alpha$
damping line wing overlaps the \ion{N}{i} triplet at $\lambda_{\rm rest} \sim 
1200$ \AA. Therefore, we locally renormalized the spectrum with the fit of the 
Ly$\alpha$ damping wing profile to derive a column density of N$^0$. An 
additional difficulty related to the $N$(\ion{N}{i}) measurement is that
the profile is probably saturated. Indeed, the components 5 and 6 look almost 
saturated even in the weaker \ion{N}{i} transitions, e.g. in the lines at 
$\lambda_{\rm rest} = 1134.415$, 1134.980, and 1200.710 \AA, and the 
weakest \ion{N}{i} line at $\lambda_{\rm rest} = 1134.165$ \AA\ is strongly 
blended with \ion{H}{i} lines in the Ly$\alpha$ forest to be relevant. We are 
thus inclined to assume that the derived N$^0$ column density is a lower limit. 
In contrast, the \ion{P}{ii} column density measurement is a borderline case 
between a detection and an upper limit. Indeed, the \ion{P}{ii}\,$\lambda$\,1152 
line shows signs of possible blends with \ion{H}{i} lines in the Ly$\alpha$ 
forest mainly toward the blue edge. However, its fit agrees perfectly with the 
weaker \ion{P}{ii}\,$\lambda$\,1532 line (see Fig.~\ref{Q1157-metals}). We thus 
assume that the measured $N$(P$^+$) is a detection. 

The high \ion{H}{i} column density and the relatively high metallicity of this 
DLA allowed us to obtain column density measurements and reliable upper 
limits of some very weak metal-line transitions. We obtained a very accurate 
column density measurement of \ion{Ti}{ii} from the detection and analysis of 
several \ion{Ti}{ii} lines at $\lambda_{\rm rest} = 1910.609$, 1910.938, 3067, 
and 3073 \AA. More exceptionally, at 3\,$\sigma$ we detected the 
\ion{Co}{ii}\,$\lambda$\,2012 and \ion{Ge}{ii}\,$\lambda$\,1237 lines (see 
Fig.~\ref{Q1157-metals}). Only two detections of Co$^+$ in DLAs have been 
reported by \citet{ellison01} and \citet{rao05} and one previous 
measurement of Ge$^+$ by \citet{prochaska03b}. We are concerned that the 
observed Co and Ge profiles do not show significant optical depth at $v=0$, 
but we suppose that this is within the statistical error adopted on their 
column density measurement. We also obtained very reliable 4\,$\sigma$ upper 
limits to the column densities of B$^+$, Cl$^0$, Cu$^+$, Kr$^0$, and As$^+$ 
from the undetected but clean, i.e. free from blending and with a high 
signal-to-noise ratio, \ion{B}{ii}\,$\lambda$\,1362, 
\ion{Cl}{i}\,$\lambda$\,1347, \ion{Cu}{ii}\,$\lambda$\,1358, 
\ion{Kr}{i}\,$\lambda$\,1164, and \ion{As}{ii}\,$\lambda$\,1355 lines. 
Previously, only \citet{prochaska03b} have put meaningful upper limits to the 
column densities of these ions in another metal-strong DLA system.

The intermediate-ion line profiles have the same shape and characteristics as 
the low-ion line profiles in this DLA (see Fig.~\ref{Q1157-metals} and 
Table~\ref{Q1157-Ntable}). We obtained the column density measurement of 
\ion{Al}{iii} and an upper limit to $N$(\ion{Fe}{iii}) from the 
\ion{Fe}{iii}\,$\lambda$\,1122 line located in the Ly$\alpha$ forest and 
probably blended with \ion{H}{i} lines, by using the same fitting parameters as 
those defined for the low-ion lines. The measured very low column density ratio, 
$\log N$(Al$^{++}$)/$N$(Al$^+$) $< -1.82$, clearly shows that the ionization is 
negligible in this DLA, strongly self-shielded by the high neutral hydrogen 
column density gas \citep{prochaska02a}.

%

\begin{figure}[t]
\centering
\includegraphics[width=9cm]{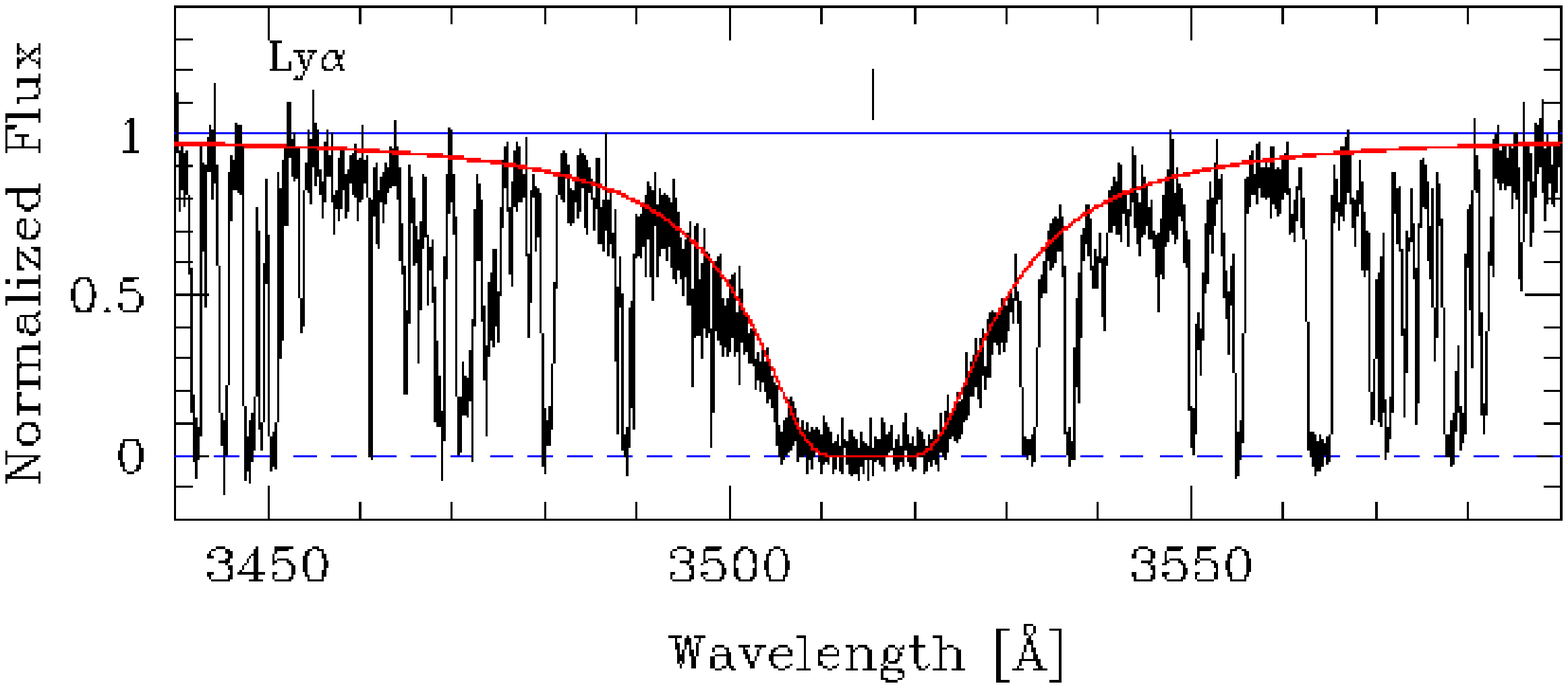}
\caption{Normalized UVES spectrum of Q1210+17 showing the DLA Ly$\alpha$ line 
profile with the Voigt profile fit. The vertical bar corresponds to the 
wavelength centroid of the component used for the best fit, $z=1.89177$. The 
measured \ion{H}{i} column density is $\log N$(\ion{H}{i}) = $20.63\pm 0.08$.}
\label{Q1210-HI}
\end{figure}
%

\begin{figure*}[!]
\centering
\includegraphics[width=17.5cm]{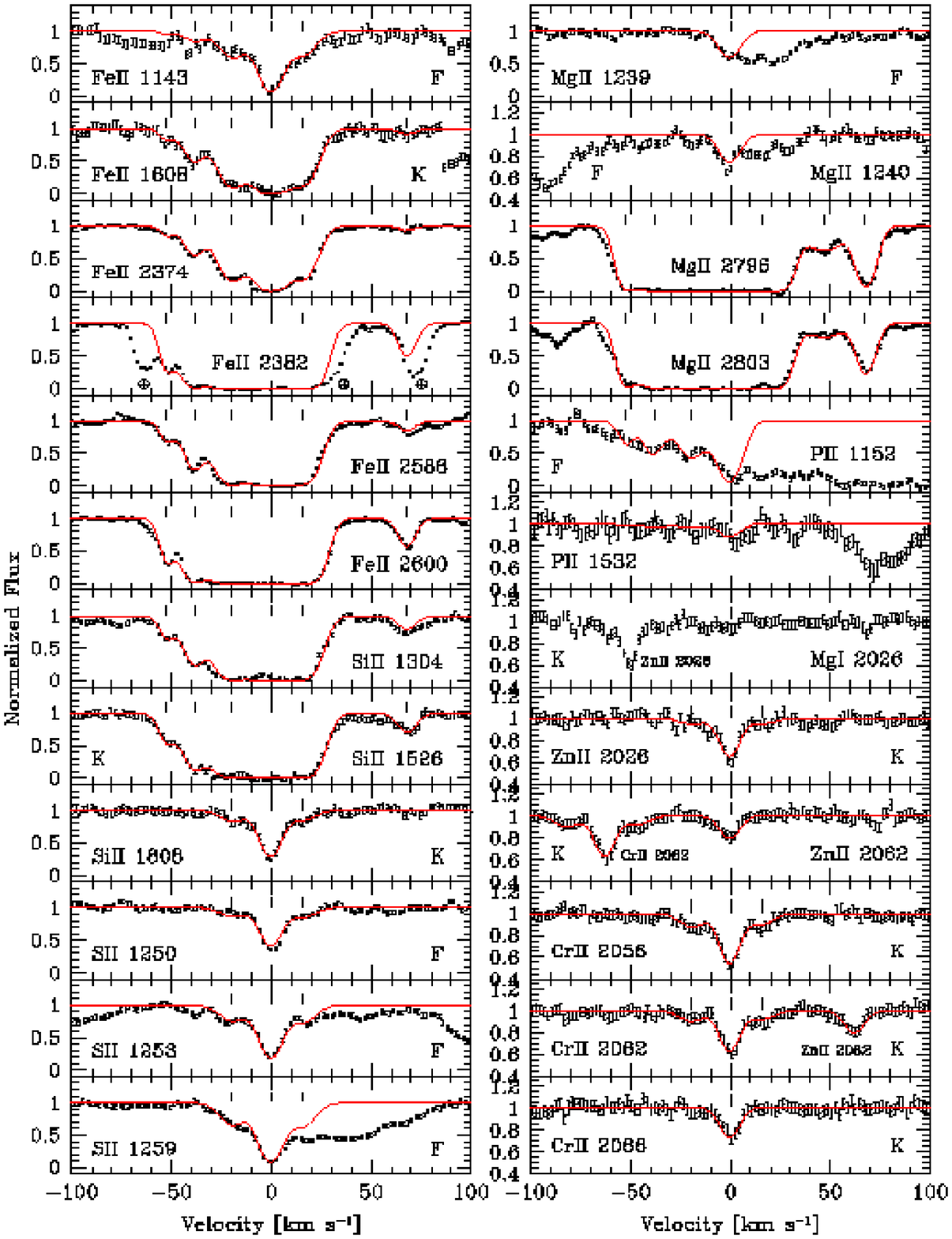}
\caption{Same as Fig.~\ref{Q0450-metals} for the DLA toward Q1210+17. The 
zero velocity is fixed at $z=1.89177$.}
\label{Q1210-metals}
\end{figure*}

\addtocounter{figure}{-1}
\begin{figure*}[!]
\centering
\includegraphics[width=17.5cm]{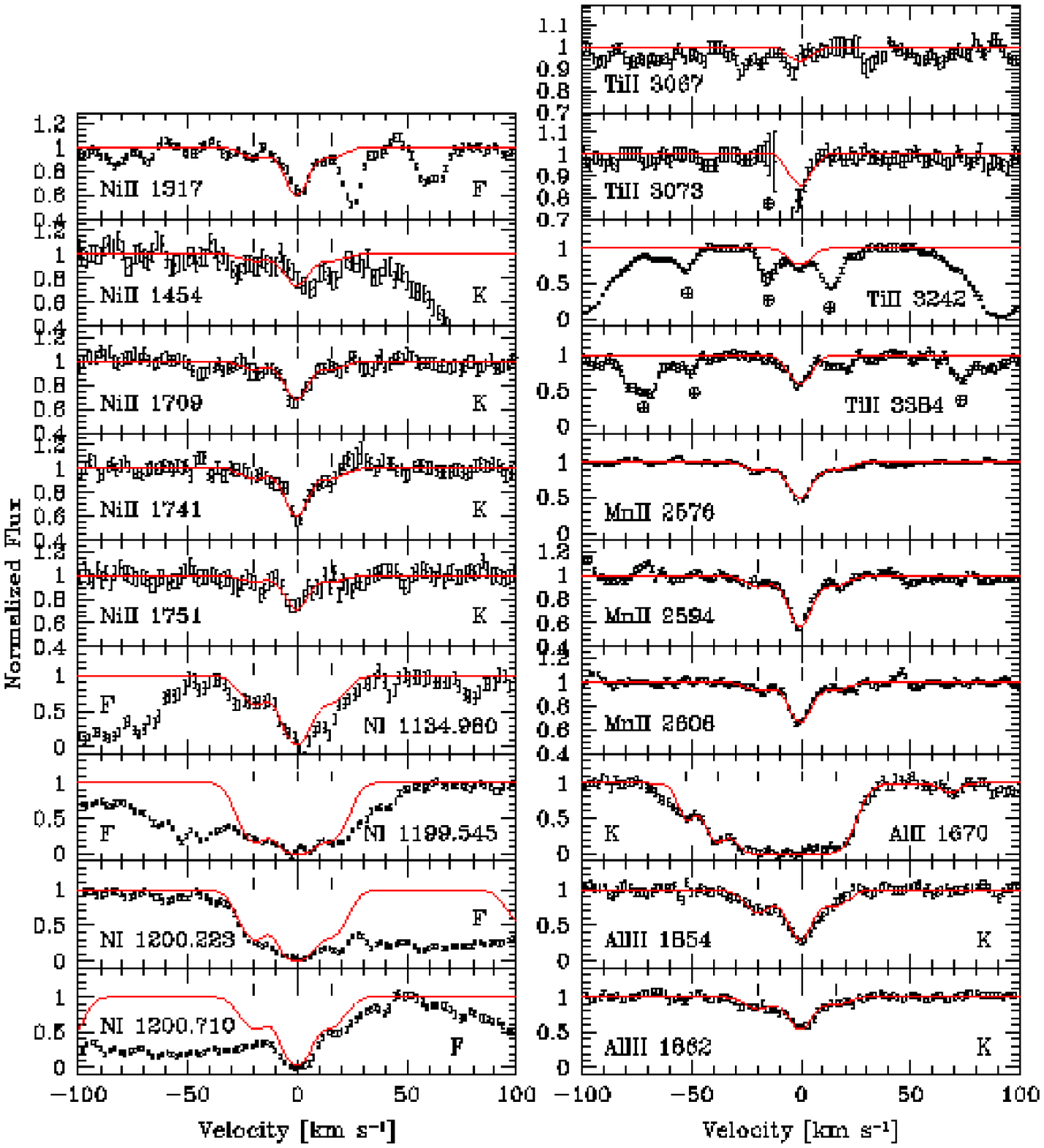}
\caption{{\em Continued}.}
\end{figure*}
%

\begin{table}[!t]
\begin{center}
\caption{Component structure of the $z_{\rm abs} = 1.892$ DLA system toward Q1210+17}
\label{Q1210-Ntable}
\vspace{-0.4cm}
\begin{tabular}{l c c c l c}
\hline\hline
\\[-0.3cm]
Comp. & $z_{\rm abs}$ & $v_{\rm rel}^*$ & $b (\sigma_b)$ & Ion & $\log N (\sigma_{\log N})$ \\
      &               & [km s$^{-1}$]   & [km s$^{-1}$]  &     &                       
\smallskip
\\ 
\hline
\multicolumn{6}{l}{\hspace{0.3cm} Low- and intermediate-ion transitions} \\
\hline
\\[-0.28cm]
1 & 1.89126 & $-$53 & \phantom{0}3.0{\scriptsize (0.4)} & \ion{Fe}{ii}  & 12.85{\scriptsize (0.02)} \\	
  &         &       &                                   & \ion{Si}{ii}  & 13.10{\scriptsize (0.03)} \\
  &         &       &                                   & \ion{Al}{ii}  & 11.93{\scriptsize (0.04)} \\	
  &         &       &                                   & \ion{P}{ii}   & $<12.77$ \\
  &         &       &                                   & \ion{Mg}{ii}  & $>13.74$ \\
2 & 1.89140 & $-$38 & \phantom{0}5.4{\scriptsize (0.2)} & \ion{Fe}{ii}  & 13.55{\scriptsize (0.01)} \\	
  &         &       &                                   & \ion{Si}{ii}  & 13.75{\scriptsize (0.02)} \\	
  &         &       &                                   & \ion{Al}{ii}  & 12.41{\scriptsize (0.03)} \\	
  &         &       &                                   & \ion{P}{ii}   & $<13.09$ \\
  &         &       &                                   & \ion{Mg}{ii}  & $>13.58$ \\
3 & 1.89158 & $-$20 & \phantom{0}7.7{\scriptsize (0.8)} & \ion{Fe}{ii}  & 14.19{\scriptsize (0.02)} \\	
  &         &       &                                   & \ion{Si}{ii}  & 14.43{\scriptsize (0.06)} \\	
  &         &       &                                   & \ion{Al}{ii}  & $>13.20$ \\	
  &         &       &                                   & \ion{S}{ii}   & 14.09{\scriptsize (0.02)} \\	
  &         &       &                                   & \ion{Mn}{ii}  & 11.91{\scriptsize (0.02)} \\	
  &         &       &                                   & \ion{Cr}{ii}  & 12.55{\scriptsize (0.05)} \\	
  &         &       &                                   & \ion{N}{i}    & 13.83{\scriptsize (0.06)} \\	
  &         &       &                                   & \ion{Zn}{ii}  & 11.45{\scriptsize (0.17)} \\	
  &         &       &                                   & \ion{Ni}{ii}  & 12.86{\scriptsize (0.08)} \\	
  &         &       &                                   & \ion{P}{ii}   & $<13.27$ \\
  &         &       &                                   & \ion{Mg}{ii}  & $>14.22$ \\
  &         &       &                                   & \ion{Al}{iii} & 12.36{\scriptsize (0.03)} \\
4 & 1.89177 &	  0 & \phantom{0}5.8{\scriptsize (0.1)} & \ion{Fe}{ii}  & 14.84{\scriptsize (0.03)} \\  
  &         &       &                                   & \ion{Si}{ii}  & 15.19{\scriptsize (0.02)} \\	
  &         &       &                                   & \ion{Al}{ii}  & $>14.93$ \\	
  &         &       &                                   & \ion{S}{ii}   & 14.81{\scriptsize (0.01)} \\	
  &         &       &                                   & \ion{Mn}{ii}  & 12.57{\scriptsize (0.01)} \\	
  &         &       &                                   & \ion{Cr}{ii}  & 13.11{\scriptsize (0.02)} \\	
  &         &       &                                   & \ion{N}{i}    & 14.59{\scriptsize (0.10)} \\	
  &         &       &                                   & \ion{Zn}{ii}  & 12.29{\scriptsize (0.02)} \\	
  &         &       &                                   & \ion{Ni}{ii}  & 13.52{\scriptsize (0.05)} \\	
  &         &       &                                   & \ion{Ti}{ii}  & 12.34{\scriptsize (0.08)} \\
  &         &       &                                   & \ion{P}{ii}   & $<13.72$ \\
  &         &       &                                   & \ion{Mg}{ii}  & $<15.51$ \\	
  &         &       &                                   & \ion{Al}{iii} & 12.77{\scriptsize (0.02)} \\
5 & 1.89192 & $+$16 & \phantom{0}7.6{\scriptsize (0.1)} & \ion{Fe}{ii}  & 14.13{\scriptsize (0.01)} \\	
  &         &       &                                   & \ion{Si}{ii}  & 14.38{\scriptsize (0.06)} \\	
  &         &       &                                   & \ion{Al}{ii}  & $>12.76$ \\	
  &         &       &                                   & \ion{S}{ii}   & 14.13{\scriptsize (0.05)} \\	
  &         &       &                                   & \ion{Mn}{ii}  & 11.88{\scriptsize (0.02)} \\	
  &         &       &                                   & \ion{Cr}{ii}  & 12.45{\scriptsize (0.07)} \\	
  &         &       &                                   & \ion{N}{i}    & 13.78{\scriptsize (0.10)} \\	
  &         &       &                                   & \ion{Zn}{ii}  & 11.44{\scriptsize (0.17)} \\	
  &         &       &                                   & \ion{Ni}{ii}  & 12.82{\scriptsize (0.08)} \\	
  &         &       &                                   & \ion{Mg}{ii}  & $>14.71$ \\
  &         &       &                                   & \ion{Al}{iii} & 12.17{\scriptsize (0.05)} \\
6 & 1.89222 & $+$47 &           11.7{\scriptsize (0.5)} & \ion{Mg}{ii}  & 12.34{\scriptsize (0.04)} \\
7 & 1.89242 & $+$67 & \phantom{0}4.4{\scriptsize (0.2)} & \ion{Fe}{ii}  & 12.57{\scriptsize (0.01)} \\
  &         &       &                                   & \ion{Si}{ii}  & 12.90{\scriptsize (0.04)} \\	
  &         &       &                                   & \ion{Al}{ii}  & 11.26{\scriptsize (0.10)} \\	
  &         &       &                                   & \ion{Mg}{ii}  & 12.89{\scriptsize (0.01)} \\
\hline
\end{tabular}
\begin{minipage}{160mm}
\smallskip
$^*$ Velocity relative to $z=1.89177$ 
\end{minipage}
\end{center}
\end{table} 
%

\subsection{Q1210+17, z\mathversion{bold}$_{\rm abs}$\mathversion{normal} = 1.892}
\label{Q1210}

This DLA system toward Q1210+17 was discovered by \citet{wolfe95} and is a 
member of the Large Bright QSO Survey. It has already been studied at high 
resolution by \citet{prochaska01}. Using our UVES/VLT spectra combined with 
their HIRES/Keck spectra, we confirm their column density measurements of 
\ion{Fe}{ii}, \ion{Si}{ii}, \ion{Ni}{ii}, \ion{Cr}{ii}, and \ion{Zn}{ii} at 
0.05~dex. In addition, we obtained the column density measurements of 
\ion{S}{ii}, \ion{N}{i}, \ion{Mn}{ii}, \ion{Ti}{ii}, and \ion{Al}{iii}, lower 
limits to the column density of \ion{Al}{ii} and \ion{Mg}{ii}, and an upper 
limit to the column density of \ion{P}{ii} from a total of 45 metal-line
transitions analyzed (see Fig.~\ref{Q1210-metals}).

The low-ion absorption line profiles of this DLA are characterized by a simple
velocity structure, composed of 7 components spread over 130 km~s$^{-1}$. Their
properties are described in Table~\ref{Q1210-Ntable}. Only the components 3, 4, 
and 5 were detected in the weak metal-line transitions. They contain nearly
all of the column density of the profile ($\sim 95$\,\%). The derived column 
density of N$^0$ is a borderline case between a detection and an upper limit 
due to possible blends with \ion{H}{i} lines in the Ly$\alpha$ forest. All the 
available \ion{N}{i} lines seem to be blended, although luckily not over the 
whole profile. In this way, we succeeded in constraining the column densities 
of components 3, 4, and 5 by fitting them to the various \ion{N}{i} lines. We 
believe that we obtained a reliable $N$(N$^0$) but report a conservatively large 
error estimate. The measurement of the column density of Mg$^+$ is also tricky. 
The saturated \ion{Mg}{ii}\,$\lambda$\,2796,\,2803 lines provide a lower limit 
to $N$(Mg$^+$), while the \ion{Mg}{ii}\,$\lambda$\,1239,\,1240 lines in the 
Ly$\alpha$ forest give an upper limit to the column density of the component 4. 
The upper limit to the column density of P$^+$ was obtained from the
\ion{P}{ii}\,$\lambda$\,1152,\,1532 lines. The \ion{Mg}{i}\,$\lambda$\,2026 
line in Fig.~\ref{Q1210-metals} illustrates that the 
\ion{Zn}{ii}\,$\lambda$\,2026 line is not contaminated by Mg$^0$.

The intermediate-ion lines of this DLA show exactly the same profiles as the
low-ion lines (see Fig.~\ref{Q1210-metals} and Table~\ref{Q1210-Ntable}). We
obtained the column density measurement of Al$^{++}$ only. This DLA has a very
low column density ratio $\log N$(Al$^{++}$)/$N$(Al$^+$) $< -1.96$, which 
clearly indicates that the ionization is very low in this system 
\citep{prochaska02a}.

Figure~\ref{Q1210-HI} shows the fitting solution of the Ly$\alpha$ line of this
DLA. The fit was obtained by fixing the $b$-value at 20 km~s$^{-1}$ and the
redshift at $z=1.89177$, i.e. at the redshift of the strongest metal-line
component (the component 4). The derived \ion{H}{i} column density is 
$\log N$(\ion{H}{i}) $= 20.63\pm 0.08$, which agrees well with
the value obtained by \citet{wolfe95} from low resolution spectra. This DLA 
system exhibits a high metallicity of [Zn/H] $= -0.88\pm 0.10$, and hence 
almost belongs to the category of metal-strong DLAs \citep{prochaska03b}. 
Therefore, we were able to derive a very accurate Ti$^+$ column density 
measurement from the detection of several \ion{Ti}{ii} lines at $\lambda_{\rm 
rest} = 3067$, 3073, 3242, and 3384 \AA. In addition, we obtained reliable 
4\,$\sigma$ upper limits to the column densities of Kr$^0$ and As$^+$ from the 
undetected \ion{Kr}{i}\,$\lambda$\,1164 and \ion{As}{ii}\,$\lambda$\,1263 
lines, respectively. The \ion{Ge}{ii}\,$\lambda$\,1237 line is unfortunately 
blended, and the \ion{B}{ii}\,$\lambda$\,1362, \ion{Cu}{ii}\,$\lambda$\,1358, 
\ion{O}{i}\,$\lambda$\,1355, and \ion{Co}{ii} lines are not covered in our high
signal-to-noise ratio UVES/VLT spectra, with which we could probably have 
provided reliable upper limits or detections of these ions in this DLA (see 
Sect.~\ref{Q1157}). 

%

\begin{figure}[t]
\centering
\includegraphics[width=9cm]{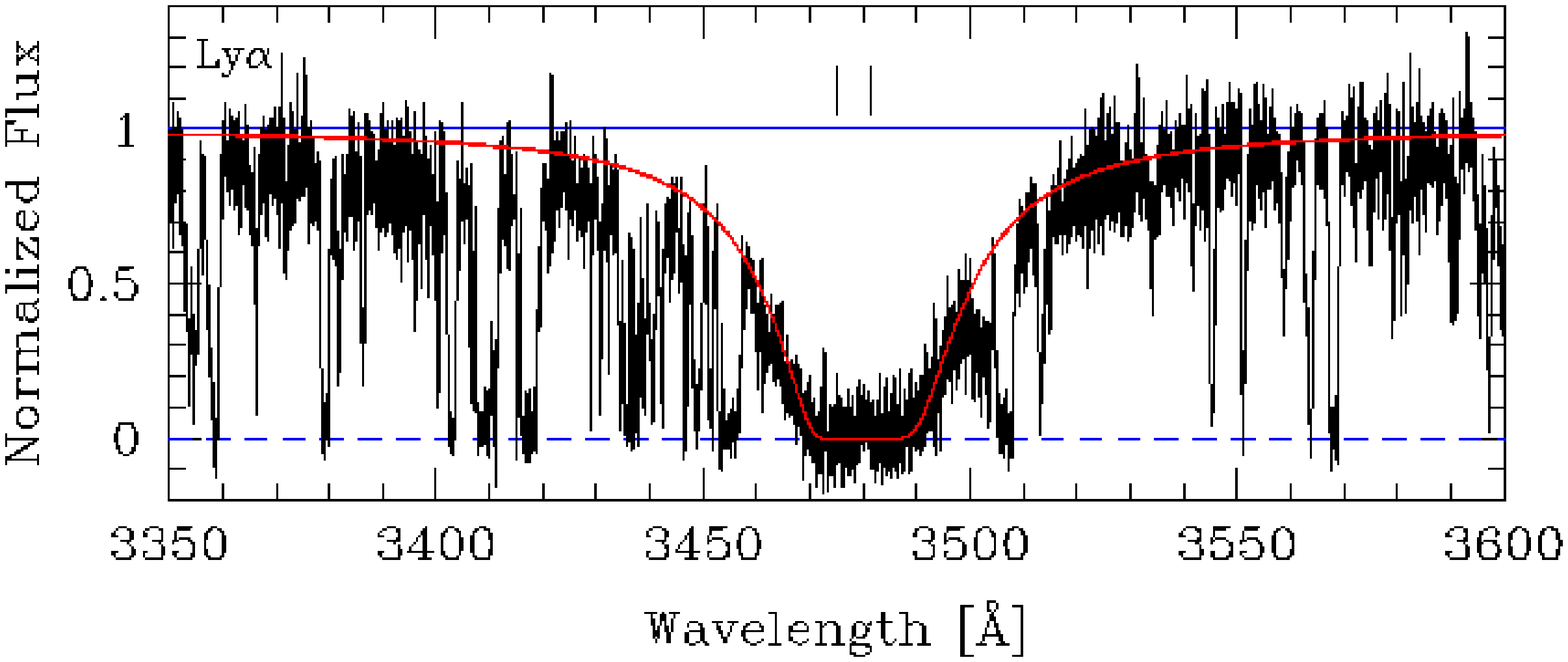}
\caption{Normalized UVES spectrum of Q2230+02 showing the DLA Ly$\alpha$ line 
profile with the Voigt profile fit. The vertical bars correspond to the 
wavelength centroids of the components used for the best fit, $z=1.86375$ and
$z=1.85851$, referring to the DLA system and an additional absorber, 
respectively. The measured \ion{H}{i} column densities are $\log N$(\ion{H}{i}) 
= $20.83\pm 0.05$ and $20.00\pm 0.10$, respectively.}
\label{Q2230-HI}
\end{figure}
%

\begin{table*}[!]
\begin{center}
\caption{Component structure of the $z_{\rm abs} = 1.864$ DLA system toward Q2230+02}
\label{Q2230-Ntable}
\begin{tabular}{l c c c l c | l c c c l c}
\hline\hline
comp. & $z_{\rm abs}$ & $v_{\rm rel}^*$ & $b (\sigma_b)$ & Ion & $\log N (\sigma_{\log N})$ & Comp. & $z_{\rm abs}$ & 
$v_{\rm rel}^*$ & $b (\sigma_b)$ & Ion & $\log N (\sigma_{\log N})$ \\				      
      &               & [km s$^{-1}$]   & [km s$^{-1}$]  &     &                            &       &               & 
[km s$^{-1}$]   & [km s$^{-1}$]  &     &        						      
\smallskip											      
\\ 												      
\hline												      
\multicolumn{6}{l}{\hspace{0.3cm} Low-ion transitions} & & & & & & \\ 			      
\hline	
\\[-0.28cm]											      
\kern-5pt\raisebox{0pt}[0pt][0pt]{\framebox{\parbox[t]{0.2cm}{1\\ \\ \\ 2}}}
  & 1.86183 & $-$184 & \phantom{0}4.3{\scriptsize (0.3)} & \ion{Fe}{ii} & 12.94{\scriptsize (0.04)} & 12& 1.86375 &  $+$17 & \phantom{0}6.0{\scriptsize (0.4)} & \ion{Fe}{ii} & 14.40{\scriptsize (0.04)} \\ 
  &         &        &                                   & \ion{Si}{ii} & 13.60{\scriptsize (0.08)} &   &	  &	   &				       & \ion{Si}{ii} & 14.96{\scriptsize (0.02)} \\
  &         &        &                                   & \ion{Al}{ii} & 12.16{\scriptsize (0.06)} &   &	  &	   &				       & \ion{S}{ii}  & 14.52{\scriptsize (0.04)} \\ 
  & 1.86191 & $-$176 &           10.0{\scriptsize (0.5)} & \ion{Fe}{ii} & 12.93{\scriptsize (0.04)} &   &	  &	   &				       & \ion{Mn}{ii} & 12.35{\scriptsize (0.02)} \\ 
  &         &        &                                   & \ion{Si}{ii} & 13.16{\scriptsize (0.08)} &   &	  &	   &				       & \ion{Mg}{i}  & 12.11{\scriptsize (0.01)} \\	
  &         &        &                                   & \ion{Al}{ii} & 12.06{\scriptsize (0.06)} &   &	  &	   &				    & \ion{C}{ii}$^*$ & 13.23{\scriptsize (0.01)} \\
\kern-5pt\raisebox{0pt}[0pt][0pt]{\framebox{\parbox[t]{0.2cm}{3\\ \\ \\ \\ 4}}} 
  & 1.86214 & $-$152 & \phantom{0}5.5{\scriptsize (0.3)} & \ion{Fe}{ii} & 12.91{\scriptsize (0.04)} &   &	  &	   &				       & \ion{N}{i}   & 14.44{\scriptsize (0.14)} \\
  &         &        &                                   & \ion{Si}{ii} & 13.52{\scriptsize (0.06)} &   &	  &	   &				       & \ion{Cr}{ii} & 12.80{\scriptsize (0.04)} \\	
  &         &        &                                   & \ion{O}{i}   & 13.70{\scriptsize (0.03)} &   &	  &	   &				       & \ion{Zn}{ii} & 12.09{\scriptsize (0.04)} \\ 
  &         &        &                                   & \ion{Al}{ii} & 12.28{\scriptsize (0.04)} &   &	  &	   &				       & \ion{Ni}{ii} & 13.33{\scriptsize (0.02)} \\
  & 1.86223 & $-$142 & \phantom{0}9.0{\scriptsize (1.3)} & \ion{Fe}{ii} & 12.52{\scriptsize (0.10)} &   &	  &	   &				       & \ion{P}{ii}  & 13.07{\scriptsize (0.08)} \\ 
  &         &        &                                   & \ion{Si}{ii} & 12.84{\scriptsize (0.17)} &   &	  &	   &				       & \ion{Ti}{ii} & 12.19{\scriptsize (0.08)} \\   
  &         &        &                                   & \ion{O}{i}   & 13.33{\scriptsize (0.06)} & 13& 1.86399 &  $+$42 &	       14.5{\scriptsize (0.7)} & \ion{Fe}{ii} & 14.34{\scriptsize (0.03)} \\ 
  &         &        &                                   & \ion{Al}{ii} & 11.57{\scriptsize (0.17)} &   &	  &	   &				       & \ion{Si}{ii} & 14.81{\scriptsize (0.03)} \\
5 & 1.86262 & $-$102 & \phantom{0}6.7{\scriptsize (0.3)} & \ion{Fe}{ii} & 13.52{\scriptsize (0.01)} &   &	  &	   &				       & \ion{S}{ii}  & 14.61{\scriptsize (0.08)} \\
  &         &        &                                   & \ion{Si}{ii} & 13.91{\scriptsize (0.02)} &   &	  &	   &				       & \ion{Mn}{ii} & 12.18{\scriptsize (0.04)} \\	
  &         &        &                                   & \ion{Mg}{i}  & 11.47{\scriptsize (0.03)} &   &	  &	   &				       & \ion{Mg}{i}  & 12.22{\scriptsize (0.01)} \\
6 & 1.86274 &  $-$89 & \phantom{0}2.5{\scriptsize (0.4)} & \ion{Fe}{ii} & 13.23{\scriptsize (0.04)} &   &	  &	   &				    & \ion{C}{ii}$^*$ & 13.18{\scriptsize (0.02)} \\ 
  &         &        &                                   & \ion{Si}{ii} & 13.58{\scriptsize (0.07)} &   &	  &	   &				       & \ion{N}{i}   & 14.12{\scriptsize (0.03)} \\
  &         &        &                                   & \ion{Mg}{i}  & 10.93{\scriptsize (0.08)} &   &	  &	   &				       & \ion{Cr}{ii} & 12.46{\scriptsize (0.11)} \\
7 & 1.86284 &  $-$79 & \phantom{0}7.4{\scriptsize (0.4)} & \ion{Fe}{ii} & 13.74{\scriptsize (0.02)} &   &	  &	   &				       & \ion{Zn}{ii} & 11.88{\scriptsize (0.09)} \\ 
  &         &        &                                   & \ion{Si}{ii} & 14.34{\scriptsize (0.02)} &   &	  &	   &				       & \ion{Ni}{ii} & 13.30{\scriptsize (0.03)} \\
  &         &        &                                   & \ion{Mg}{i}  & 11.54{\scriptsize (0.03)} &   &	  &	   &				       & \ion{P}{ii}  & 13.08{\scriptsize (0.08)} \\
8 & 1.86305 &  $-$57 & \phantom{0}9.4{\scriptsize (0.6)} & \ion{Fe}{ii} & 13.91{\scriptsize (0.03)} & \kern-5pt\raisebox{0pt}[0pt][0pt]{\framebox{\parbox[t]{0.3cm}{14\\ \\ \\ \\ \\ \\ \\ \\ \\ \\ \\ \\ 15}}} 
                                                                                                        & 1.86430 &  $+$74 &	       17.3{\scriptsize (0.7)} & \ion{Fe}{ii} & 14.70{\scriptsize (0.03)} \\ 
  &         &        &                                   & \ion{Si}{ii} & 14.29{\scriptsize (0.07)} &   &	  &	   &				       & \ion{Si}{ii} & 15.11{\scriptsize (0.02)} \\	
  &         &        &                                   & \ion{Mn}{ii} & 11.92{\scriptsize (0.05)} &   &	  &	   &				       & \ion{S}{ii}  & 14.80{\scriptsize (0.06)} \\ 
  &         &        &                                   & \ion{Mg}{i}  & 11.42{\scriptsize (0.03)} &   &	  &	   &				       & \ion{Mn}{ii} & 12.48{\scriptsize (0.02)} \\ 
  &         &        &                                   & \ion{N}{i}   & 13.68{\scriptsize (0.03)} &   &	  &	   &				       & \ion{Mg}{i}  & 12.30{\scriptsize (0.10)} \\ 
  &         &        &                                   & \ion{Ni}{ii} & 12.90{\scriptsize (0.17)} &   &	  &	   &				    & \ion{C}{ii}$^*$ & 13.22{\scriptsize (0.02)} \\
9 & 1.86320 &  $-$41 & \phantom{0}8.3{\scriptsize (0.4)} & \ion{Fe}{ii} & 14.58{\scriptsize (0.02)} &   &	  &	   &				       & \ion{N}{i}   & 14.46{\scriptsize (0.03)} \\ 
  &         &        &                                   & \ion{Si}{ii} & 15.00{\scriptsize (0.02)} &   &	  &	   &				       & \ion{Cr}{ii} & 12.98{\scriptsize (0.04)} \\	 
  &         &        &                                   & \ion{S}{ii}  & 14.61{\scriptsize (0.04)} &   &	  &	   &				       & \ion{Zn}{ii} & 12.30{\scriptsize (0.04)} \\ 
  &         &        &                                   & \ion{Mn}{ii} & 12.45{\scriptsize (0.02)} &   &	  &	   &				       & \ion{Ni}{ii} & 13.58{\scriptsize (0.02)} \\  
  &         &        &                                   & \ion{Mg}{i}  & 11.93{\scriptsize (0.02)} &   &	  &	   &				       & \ion{P}{ii}  & 13.17{\scriptsize (0.07)} \\ 
  &         &        &                                & \ion{C}{ii}$^*$ & $<13.32$                  &   &	  &	   &				       & \ion{Ti}{ii} & 12.25{\scriptsize (0.12)} \\ 
  &         &        &                                   & \ion{N}{i}   & 14.39{\scriptsize (0.03)} & 
                                                                                                        & 1.86445 &  $+$90 & \phantom{0}5.1{\scriptsize (1.5)} & \ion{Fe}{ii} & 13.42{\scriptsize (0.04)} \\  
  &         &        &                                   & \ion{Cr}{ii} & 12.94{\scriptsize (0.03)} &   &	  &	   &				       & \ion{Si}{ii} & 13.83{\scriptsize (0.10)} \\   
  &         &        &                                   & \ion{Zn}{ii} & 12.09{\scriptsize (0.06)} &   &	  &	   &				       & \ion{Al}{ii} & 14.08{\scriptsize (0.11)} \\  
  &         &        &                                   & \ion{Ni}{ii} & 13.54{\scriptsize (0.02)} &   &	  &	   &				       & \ion{Mg}{i}  & 11.31{\scriptsize (0.04)} \\   
  &         &        &                                   & \ion{P}{ii}  & 13.01{\scriptsize (0.09)} & \kern-5pt\raisebox{0pt}[0pt][0pt]{\framebox{\parbox[t]{0.3cm}{16\\ \\ \\ 17}}}
                                                                                                        & 1.86468 & $+$114 & \phantom{0}5.6{\scriptsize (0.3)} & \ion{Fe}{ii} & 13.11{\scriptsize (0.04)} \\  
  &         &        &                                   & \ion{Ti}{ii} & 12.17{\scriptsize (0.08)} &   &	  &	   &				       & \ion{Si}{ii} & 13.61{\scriptsize (0.13)} \\
10& 1.86338 &  $-$22 &           12.9{\scriptsize (0.9)} & \ion{Fe}{ii} & 13.81{\scriptsize (0.03)} &   &	  &	   &				       & \ion{Al}{ii} & 12.38{\scriptsize (0.04)} \\  
  &         &        &                                   & \ion{Si}{ii} & 14.26{\scriptsize (0.02)} & 
                                                                                                        & 1.86478 & $+$125 &	       10.2{\scriptsize (0.4)} & \ion{Fe}{ii} & 13.26{\scriptsize (0.03)} \\	 
  &         &        &                                   & \ion{S}{ii}  & 13.99{\scriptsize (0.10)} &   &	  &	   &				       & \ion{Si}{ii} & 13.64{\scriptsize (0.07)} \\  
  &         &        &                                   & \ion{Mg}{i}  & 11.54{\scriptsize (0.03)} &   &	  &	   &				       & \ion{Al}{ii} & 12.23{\scriptsize (0.03)} \\  
  &         &        &                                & \ion{C}{ii}$^*$ & 13.15{\scriptsize (0.02)} &   &	  &	   &				       & \ion{Mg}{i}  & 11.62{\scriptsize (0.03)} \\ 
  &         &        &                                   & \ion{N}{i}   & 13.67{\scriptsize (0.04)} & 18& 1.86502 & $+$150 & \phantom{0}6.9{\scriptsize (0.2)} & \ion{Fe}{ii} & 13.19{\scriptsize (0.01)} \\
11& 1.86359 &      0 & \phantom{0}6.5{\scriptsize (0.2)} & \ion{Fe}{ii} & 13.65{\scriptsize (0.01)} &   &	  &	   &				       & \ion{Si}{ii} & 13.57{\scriptsize (0.10)} \\ 
  &         &        &                                   & \ion{Si}{ii} & 14.11{\scriptsize (0.09)} &   &	  &	   &				       & \ion{O}{i}   & 14.47{\scriptsize (0.02)} \\	
  &         &        &                                   & \ion{S}{ii}  & 13.83{\scriptsize (0.21)} &   &	  &	   &				       & \ion{Al}{ii} & 12.30{\scriptsize (0.02)} \\ 
  &         &        &                                   & \ion{Mn}{ii} & 11.54{\scriptsize (0.10)} &   &	  &	   &				       & \ion{Mg}{i}  & 11.58{\scriptsize (0.03)} \\ 
  &         &        &                                   & \ion{Mg}{i}  & 11.68{\scriptsize (0.02)} & 19& 1.86519 & $+$168 & \phantom{0}9.0{\scriptsize (0.6)} & \ion{Fe}{ii} & 12.53{\scriptsize (0.02)} \\ 
  &         &        &                                & \ion{C}{ii}$^*$ & 12.69{\scriptsize (0.03)} &   &	  &	   &				       & \ion{Si}{ii} & 13.12{\scriptsize (0.09)} \\
  &         &        &                                   & \ion{N}{i}   & 13.21{\scriptsize (0.06)} &   &	  &	   &				       & \ion{O}{i}   & 13.82{\scriptsize (0.02)} \\ 
  &         &        &                                   & \ion{Ni}{ii} & 12.41{\scriptsize (0.20)} &   &	  &	   &				       & \ion{Al}{ii} & 11.87{\scriptsize (0.04)} \\
\end{tabular}
\end{center}
\end{table*}

\addtocounter{table}{-1}
\begin{table*}[!]
\begin{center}
\caption{{\em Continued}}
\begin{tabular}{l c c c l c | l c c c l c}
\hline\hline
No & $z_{\rm abs}$ & $v_{\rm rel}^*$ & $b (\sigma_b)$ & Ion & $\log N (\sigma_{\log N})$ & No & $z_{\rm abs}$ & 
$v_{\rm rel}^*$ & $b (\sigma_b)$ & Ion & $\log N (\sigma_{\log N})$ \\				      
   &               & km s$^{-1}$     & km s$^{-1}$    &     &                            &    &               & 
km s$^{-1}$     & km s$^{-1}$    &     &        						      
\smallskip											      
\\ 	
\hline												      
\\[-0.28cm]
20& 1.86548 & $+$198 &  	 14.6{\scriptsize (0.8)} & \ion{Fe}{ii} & 12.73{\scriptsize (0.02)} & 21& 1.86576 & $+$227 & \phantom{0}6.0{\scriptsize (1.7)} & \ion{Fe}{ii} & 11.85{\scriptsize (0.07)} \\
  &	    &	     &  				 & \ion{Si}{ii} & 13.23{\scriptsize (0.08)} &   &   &   &   &   & \\
  &	    &	     &  				 & \ion{O}{i}	& 14.07{\scriptsize (0.01)} &   &   &   &   &   & \\
  &	    &	     &  				 & \ion{Al}{ii} & 11.81{\scriptsize (0.05)} &   &   &   &   &   & \\
\hline												      
\multicolumn{6}{l}{\hspace{0.3cm} Intermediate-ion transitions} & & & & & & \\ 			      
\hline	
\\[-0.28cm] 
1 & 1.86262 & $-$102 & \phantom{0}6.7{\scriptsize (0.3)} & \ion{Al}{iii} & 12.25{\scriptsize (0.03)} & 8 & 1.86375 &  $+$17 & \phantom{0}6.0{\scriptsize (0.4)} & \ion{Al}{iii} & 12.75{\scriptsize (0.01)} \\
2 & 1.86274 &  $-$89 & \phantom{0}2.5{\scriptsize (0.4)} & \ion{Al}{iii} & 12.28{\scriptsize (0.04)} &   &	   &	    &					& \ion{Fe}{iii} & $<13.74$ \\
3 & 1.86284 &  $-$79 & \phantom{0}7.4{\scriptsize (0.4)} & \ion{Al}{iii} & 12.53{\scriptsize (0.02)} & 9 & 1.86399 &  $+$42 &		14.5{\scriptsize (0.7)} & \ion{Al}{iii} & 12.71{\scriptsize (0.02)} \\
4 & 1.86305 &  $-$57 & \phantom{0}9.4{\scriptsize (0.6)} & \ion{Al}{iii} & 12.31{\scriptsize (0.03)} &   &	   &	    &					& \ion{Fe}{iii} & $<13.72$ \\
5 & 1.86320 &  $-$41 & \phantom{0}8.3{\scriptsize (0.4)} & \ion{Al}{iii} & 12.91{\scriptsize (0.01)} & 10& 1.86432 &  $+$76 &		11.2{\scriptsize (0.6)} & \ion{Al}{iii} & 12.91{\scriptsize (0.02)} \\
  &         &        &                                   & \ion{Fe}{iii} & $<13.90$                  &   &	   &	    &					& \ion{Fe}{iii} & $<13.92$ \\
6 & 1.86338 &  $-$22 &           12.9{\scriptsize (0.9)} & \ion{Al}{iii} & 12.18{\scriptsize (0.04)} & 11& 1.86450 &  $+$95 & \phantom{0}4.8{\scriptsize (1.5)} & \ion{Al}{iii} & 11.99{\scriptsize (0.09)} \\ 
  &         &        &                                   & \ion{Fe}{iii} & $<13.73$                  &   &	   &	    &					& \ion{Fe}{iii} & $<13.05$ \\
7 & 1.86359 &      0 & \phantom{0}6.5{\scriptsize (0.2)} & \ion{Al}{iii} & 11.82{\scriptsize (0.07)} &   &   &   &   &   & \\
  &         &        &                                   & \ion{Fe}{iii} & $<13.46$                  &   &   &   &   &   & \\
\hline
\end{tabular}
\begin{minipage}{160mm}
\smallskip
$^*$ Velocity relative to $z=1.86359$ 
\end{minipage}
\end{center}
\end{table*} 
%

\subsection{Q2230+02, z\mathversion{bold}$_{\rm abs}$\mathversion{normal} = 1.864}
\label{Q2230}

This DLA system is an object from the Large Bright QSO Survey \citep{wolfe95}.
It has been studied at low resolution by \citet{pettini94}, and more recently 
at high resolution by \citet{prochaska99} and \citet{prochaska01}. 
\citet{prochaska99} have analyzed this DLA extensively thanks to the detection 
of 28 metal-line transitions in their HIRES/Keck spectra. We confirm their
column density measurements of \ion{Fe}{ii}, \ion{Si}{ii}, \ion{Ni}{ii},
\ion{Cr}{ii}, and \ion{Al}{iii} at 0.05~dex. However, our column density
measurements of \ion{Zn}{ii} and \ion{Ti}{ii} differ by more than 0.05~dex, as
described below. From a total of 46 metal-line transitions detected and 
analyzed (see Fig.~\ref{Q2230-metals}), we obtained, in addition, the column
density measurements of \ion{S}{ii}, \ion{N}{i}, \ion{P}{ii}, \ion{Mn}{ii}, 
\ion{C}{ii}$^*$, and \ion{Mg}{i}, and an upper limit to the column density of  
\ion{Fe}{iii}.

The low-ion absorption line profiles of this DLA are characterized by a very
complex velocity structure extended over 400 km~s$^{-1}$ in velocity space and 
composed of 21 components presented in Table~\ref{Q2230-Ntable}. Three important
clumps at $-41$, +17, and +74 km~s$^{-1}$ composed of mainly 4 components, the
components 9, 12, 13, and 14, dominate the profile. They contain about 75\,\% 
of the total column density obtained by summing the contribution of the 21 
components. This complex velocity structure of low-ion lines made the 
determination of the N$^0$ column density very difficult. Indeed, due to the
large velocity spread of the profiles, the three lines of the \ion{N}{i}
triplets at $\lambda_{\rm rest} \sim 1134$ and at $\sim 1200$ \AA\ are 
blended together, but luckily not too heavily contaminated by \ion{H}{i} lines 
in the Ly$\alpha$ forest (see Fig.~\ref{Q2230-metals}). By careful work on 
the 6 \ion{N}{i} lines available in our UVES/VLT spectra and by using the 
fitting parameters constrained with unblended and uncontaminated low-ion lines, 
we finally derived an accurate N$^0$ column density. 
The Zn$^+$ column density measurement also needed some attention, due to a 
possible blend of the \ion{Zn}{ii}\,$\lambda$\,2026 line with the 
\ion{Mg}{i}\,$\lambda$\,2026 line. The \ion{Mg}{i}\,$\lambda$\,2852 line 
allowed us to very accurately determine the contamination of the \ion{Zn}{ii} 
column density by Mg$^0$ by obtaining a fit of the \ion{Mg}{i}\,$\lambda$\,2026 
line. This contamination is lower than 0.1~dex, but it explains the discrepancy 
between \citet{prochaska99} Zn$^+$ column density measurement, 
$\log N$(\ion{Zn}{ii}) $= 12.80\pm 0.03$, and our weaker value, 
$\log N$(\ion{Zn}{ii}) $= 12.72\pm 0.05$. 

In the saturated \ion{O}{i}\,$\lambda$\,1302 and \ion{Al}{ii}\,$\lambda$\,1670 
lines, we measured the column densities of the unsaturated components only (see 
Table~\ref{Q2230-Ntable}). The column density of the component 9 of 
\ion{C}{ii}$^*$ is referred to as an upper limit because the blue edge of the 
\ion{C}{ii}$^*$\,$\lambda$\,1335 line is blended with the 
\ion{C}{ii}\,$\lambda$\,1334 line. The \ion{Mg}{ii}\,$\lambda$\,1239,\,1240
lines were undetected in this DLA, so we provide only a 4\,$\sigma$ upper 
limit to the column density of Mg$^+$. Despite the low signal-to-noise ratio 
per pixel of 8 in the wavelength regions covering the \ion{S}{ii} and 
\ion{P}{ii} lines, we obtained accurate S$^+$ and P$^+$ column densities.
Finally, we derived a very accurate column density measurement of Ti$^+$, 
thanks to the access to several \ion{Ti}{ii} lines at $\lambda_{\rm rest} = 
1910.609$, 1910.938, 3073, 3242, and 3384 \AA. Our $N$(Ti$^+$) differs from the 
one of \citet{prochaska99} obtained from the two \ion{Ti}{ii} lines at 
$\lambda_{\rm rest} \sim 1910$ \AA, which have a 2--3 times weaker oscillator 
strength than the \ion{Ti}{ii}\,$\lambda$\,3242,\,3384 lines used in our
measurement.

Only 11 components were detected in the intermediate-ion line profiles of this 
DLA. They show very similar characteristics as the components 5 to 15 of the 
low-ion lines (see the second part of Table~\ref{Q2230-Ntable} and 
Fig.~\ref{Q2230-metals}). We obtained a column density measurement of Al$^{++}$
and an upper limit to $N$(Fe$^{++}$) from the \ion{Fe}{iii}\,$\lambda$\,1122
line probably blended with \ion{H}{i} lines in the Ly$\alpha$ forest. The 
measured low column density ratios, $\log N$(Al$^{++}$)/$N$(Al$^+$) $< -0.43$ 
and $\log N$(Fe$^{++}$)/$N$(Fe$^+$) $< -0.61$, clearly show that this DLA 
system has a low ionization level \citep{prochaska02a}.

%

\begin{figure*}[!]
\centering
\includegraphics[width=17.5cm]{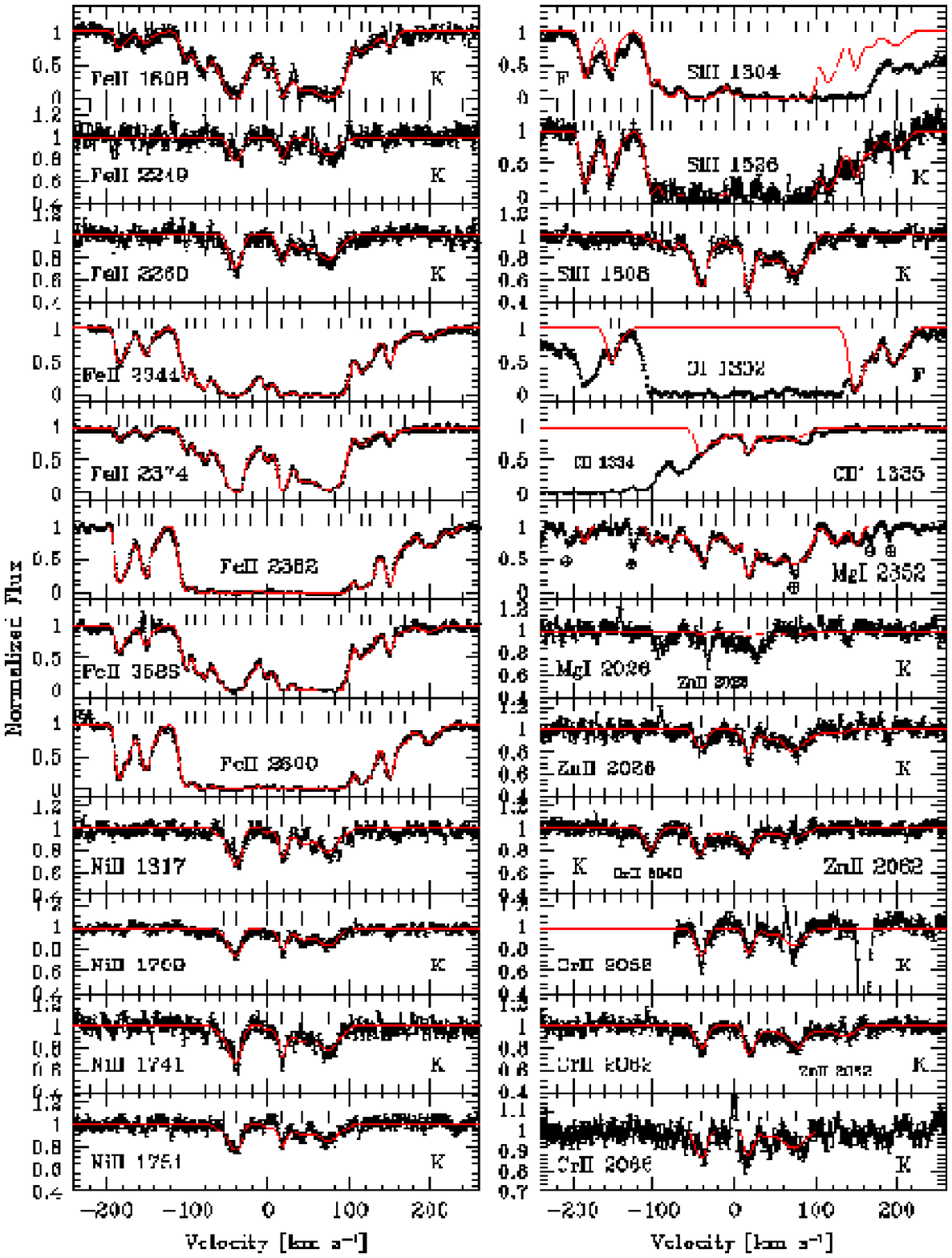}
\caption{Same as Fig.~\ref{Q0450-metals} for the DLA toward Q2230+02. The 
zero velocity is fixed at $z=1.86359$.}
\label{Q2230-metals}
\end{figure*}

\addtocounter{figure}{-1}
\begin{figure*}[!]
\centering
\includegraphics[width=17.5cm]{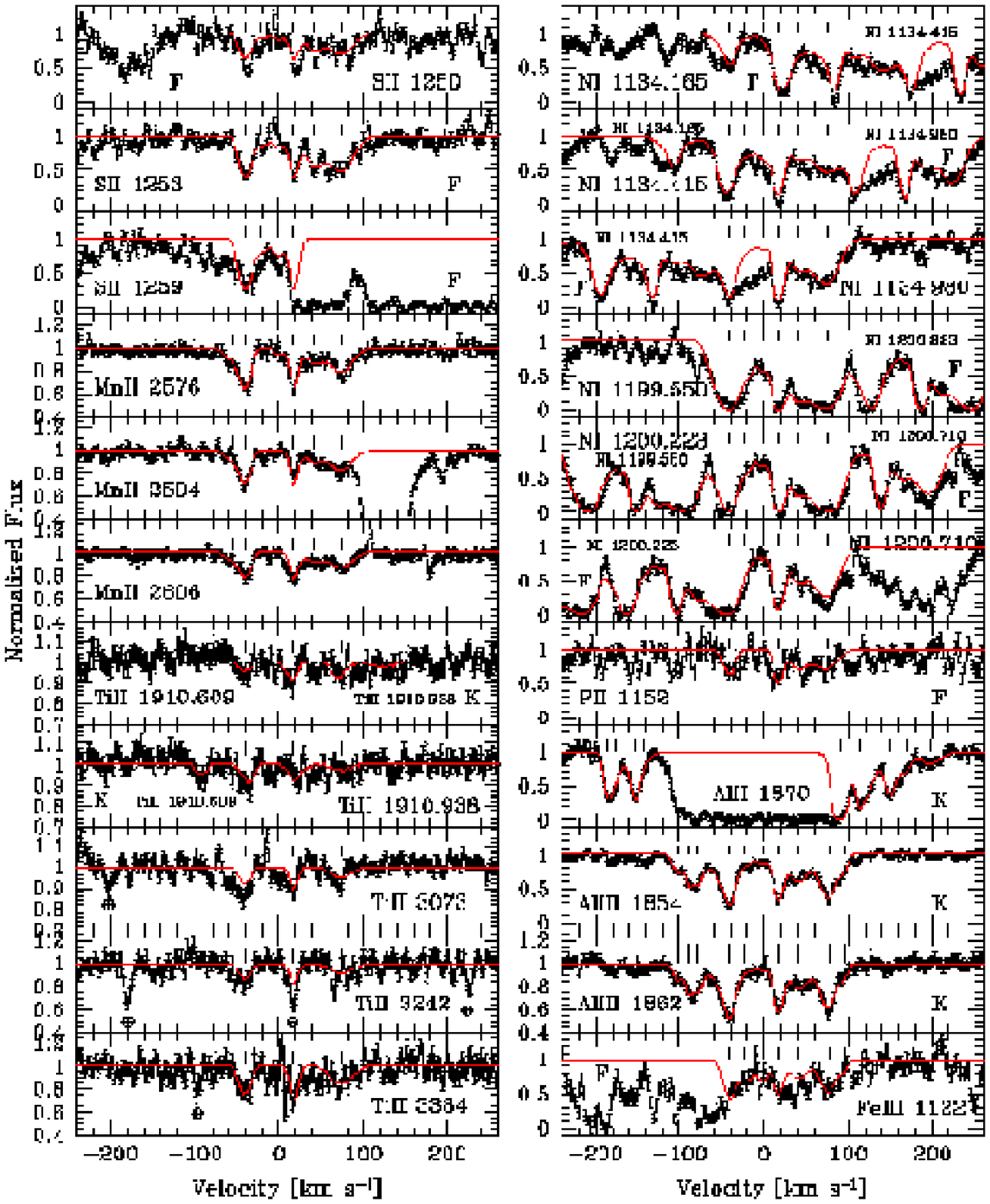}
\caption{{\em Continued}.}
\end{figure*}
%

While fitting the Ly$\alpha$ damping line profile at $z=1.86375$ (redshift of 
one of the stronger metal-line components, the component 12), we found it 
necessary to include the contribution of a second absorber shifted by about 550 
km~s$^{-1}$ bluewards the DLA system (see Fig.~\ref{Q2230-HI}). The redshift of 
this second absorber, $z=1.85851$, was accurately determined from several 
associated metal lines. \citet{pettini94} already discovered the presence of 
these two closeby absorption line systems in their low resolution spectra by 
observing that the metal lines consist of two main components separated by 
$\sim 550$ km~s$^{-1}$. The derived \ion{H}{i} column densities are 
$\log N$(\ion{H}{i}) $= 20.83\pm 0.05$ for the DLA and $\log N$(\ion{H}{i}) 
$= 20.00\pm 0.10$ for the second absorber. They agree well with the values 
obtained by \citet{pettini94}.

%

\begin{table}[!t]
\begin{center}
\caption{Component structure of the $z_{\rm abs} = 2.279$ DLA system toward Q2348$-$1444}
\label{Q2348-Ntable}
\vspace{-0.4cm}
\begin{tabular}{l c c c l c}
\hline\hline
\\[-0.3cm]
Comp. & $z_{\rm abs}$ & $v_{\rm rel}^*$ & $b (\sigma_b)$ & Ion & $\log N (\sigma_{\log N})$ \\
      &               & [km s$^{-1}$]   & [km s$^{-1}$]  &     &                          
\smallskip
\\ 
\hline
\multicolumn{6}{l}{\hspace{0.3cm} Low- and intermediate-ion transitions} \\
\hline
\\[-0.28cm]
\kern-5pt\raisebox{0pt}[0pt][0pt]{\framebox{\parbox[t]{0.2cm}{1\\ \\ \\ \\ \\ \\ \\ \\ \\ \\ \\ \\ \\ 2\\ \\ \\ \\ \\ \\ \\ \\ \\ \\ \\ \\ \\ \\ 3}}}
  & 2.27923 & $-$15 & 4.0{\scriptsize (0.7)} & \ion{Fe}{ii}  & 13.05{\scriptsize (0.07)} \\
  &         &       &                        & \ion{Si}{ii}  & 13.37{\scriptsize (0.07)} \\  
  &         &       &                        & \ion{S}{ii}   & 13.01{\scriptsize (0.11)} \\  
  &         &       &                        & \ion{N}{i}    & 12.45{\scriptsize (0.08)} \\  
  &         &       &                        & \ion{Al}{ii}  & $<11.82$ \\  
  &         &       &                        & \ion{O}{i}    & $>14.71$ \\   
  &         &       &                        & \ion{Mn}{ii}  & 10.93{\scriptsize (0.20)} \\  
  &         &       &                    & \ion{Mg}{ii}$^1$  & $<14.12$ \\  
  &         &       &                    & \ion{Mg}{ii}$^2$  & $>13.21$ \\  
  &         &       &                        & \ion{Cr}{ii}  & 11.56{\scriptsize (0.11)} \\
  &         &       &                        & \ion{Al}{iii} & 11.31{\scriptsize (0.11)} \\  
  &         &       &                        & \ion{Fe}{iii} & 12.35{\scriptsize (0.09)} \\
  &         &       &                        & \ion{S}{iii}  & $<13.33$ \\
  & 2.27930 &  $-$8 & 2.0{\scriptsize (1.0)} & \ion{Fe}{ii}  & 13.22{\scriptsize (0.06)} \\
  &         &       &                        & \ion{Si}{ii}  & 13.58{\scriptsize (0.11)} \\ 
  &         &       &                        & \ion{S}{ii}   & 13.00{\scriptsize (0.11)} \\
  &         &       &                        & \ion{N}{i}    & 12.78{\scriptsize (0.02)} \\ 
  &         &       &                        & \ion{Al}{ii}  & $<11.84$ \\ 
  &         &       &                        & \ion{O}{i}    & $>13.02$ \\ 
  &         &       &                        & \ion{Mn}{ii}  & 11.20{\scriptsize (0.15)} \\ 
  &         &       &                    & \ion{Mg}{ii}$^1$  & $<14.36$ \\ 
  &         &       &                    & \ion{Mg}{ii}$^2$  & $>12.76$ \\
  &         &       &                        & \ion{Cr}{ii}  & 11.60{\scriptsize (0.09)} \\
  &         &       &                        & \ion{Zn}{ii}  & $<10.53$ \\
  &         &       &                        & \ion{Al}{iii} & 11.45{\scriptsize (0.09)} \\
  &         &       &                        & \ion{Fe}{iii} & 12.93{\scriptsize (0.10)} \\
  &         &       &                        & \ion{S}{iii}  & $<13.18$ \\ 
  & 2.27939 &     0 & 4.8{\scriptsize (0.2)} & \ion{Fe}{ii}  & 13.60{\scriptsize (0.04)} \\
  &         &       &                        & \ion{Si}{ii}  & 13.92{\scriptsize (0.03)} \\ 
  &         &       &                        & \ion{S}{ii}   & 13.56{\scriptsize (0.03)} \\
  &         &       &                        & \ion{N}{i}    & 13.13{\scriptsize (0.05)} \\ 
  &         &       &                        & \ion{Al}{ii}  & $<12.47$ \\ 
  &         &       &                        & \ion{O}{i}    & $>14.75$ \\ 
  &         &       &                        & \ion{Ni}{ii}  & $<12.30$ \\ 
  &         &       &                     & \ion{C}{ii}$^*$  & $<12.25$ \\ 
  &         &       &                        & \ion{Mn}{ii}  & 11.33{\scriptsize (0.12)} \\
  &         &       &                    & \ion{Mg}{ii}$^1$  & $<14.44$ \\
  &         &       &                    & \ion{Mg}{ii}$^2$  & $>13.29$ \\ 
  &         &       &                        & \ion{Cr}{ii}  & 12.09{\scriptsize (0.08)} \\ 
  &         &       &                        & \ion{Zn}{ii}  & $<11.20$ \\
  &         &       &                        & \ion{Al}{iii} & 11.80{\scriptsize (0.04)} \\
  &         &       &                        & \ion{Fe}{iii} & 13.12{\scriptsize (0.15)} \\
  &         &       &                        & \ion{S}{iii}  & $<13.39$ \\
4 & 2.27952 & $+$12 & 8.0{\scriptsize (1.4)} & \ion{Fe}{ii}  & 12.05{\scriptsize (0.08)} \\ 
  &         &       &                        & \ion{Si}{ii}  & 12.80{\scriptsize (0.04)} \\ 
  &         &       &                        & \ion{Al}{ii}  & $<11.91$ \\ 
  &         &       &                        & \ion{O}{i}    & 13.12{\scriptsize (0.05)} \\ 
  &         &       &                    & \ion{Mg}{ii}$^2$  & 12.78{\scriptsize (0.15)} \\
\hline
\end{tabular}
\begin{minipage}{160mm}
\smallskip
$^*$ Velocity relative to $z=2.27939$ \\
$^1$ \ion{Mg}{ii}\,$\lambda$1239,1240 \\
$^2$ \ion{Mg}{ii}\,$\lambda$2793,2803
\end{minipage}
\end{center}
\end{table}
%

\begin{figure}[t]
\centering
\includegraphics[width=9cm]{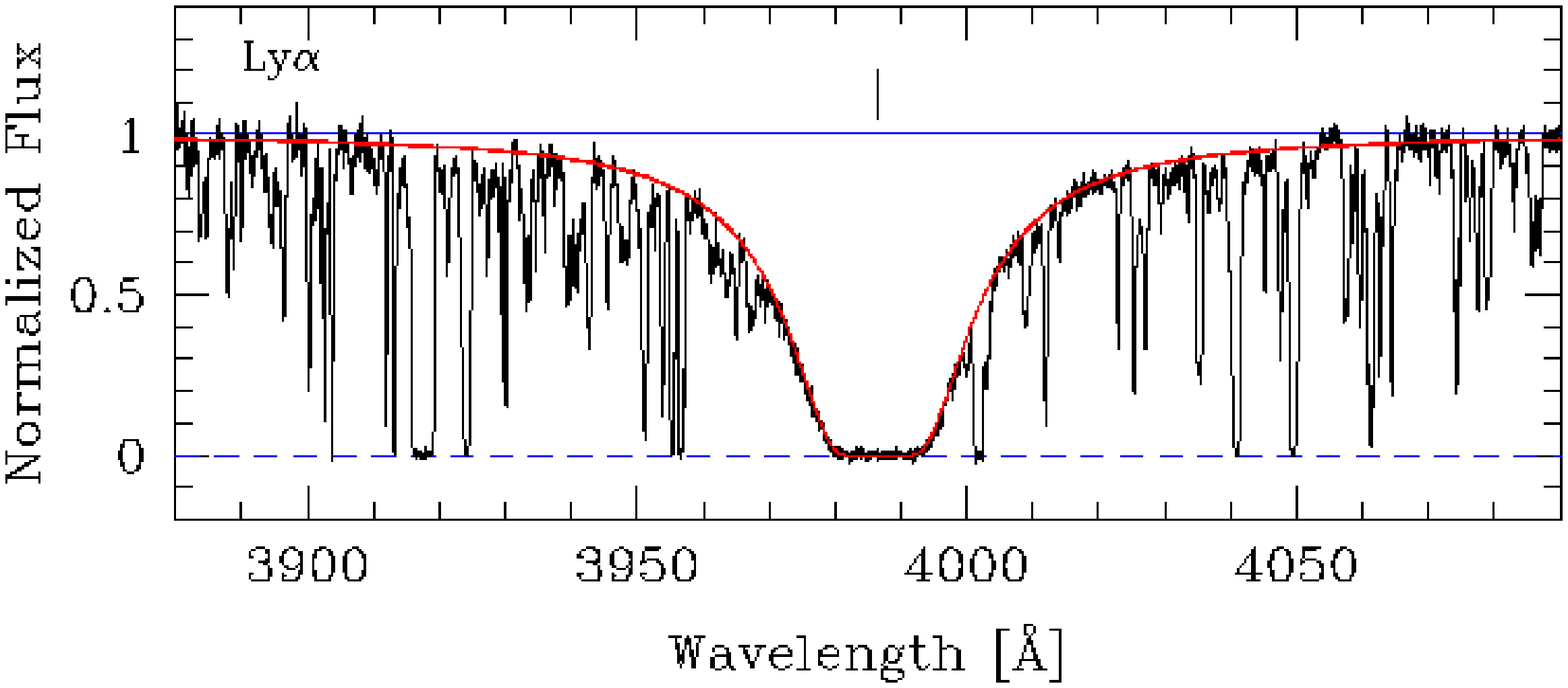}
\caption{Normalized UVES spectrum of Q2348$-$1444 showing the DLA Ly$\alpha$ 
line profile with the Voigt profile fit. The vertical bar corresponds to the 
wavelength centroid of the component used for the best fit, $z=2.27939$. The 
measured \ion{H}{i} column density is $\log N$(\ion{H}{i}) = $20.59\pm 0.08$.}
\label{Q2348-HI}
\end{figure}
%

\subsection{Q2348$-$1444, z\mathversion{bold}$_{\rm abs}$\mathversion{normal} = 2.279}
\label{Q2348}

This DLA system was first discussed by \citet{pettini94} and first studied at
high resolution by \citet{pettini95}. Subsequently it was carefully
analyzed by \citet{prochaska99} and \citet{prochaska02b} using HIRES/Keck
spectra. We confirm their column density measurements of \ion{Fe}{ii},
\ion{Si}{ii}, \ion{S}{ii}, \ion{Al}{ii}, and \ion{C}{ii}$^*$ at 0.05~dex, and
their lower limit to the column density of \ion{O}{i} derived from the saturated
\ion{O}{i}\,$\lambda$\,1302 line. From a total of 36 metal-line transitions 
detected and analyzed (see Fig.~\ref{Q2348-metals}), we obtained the column 
density measurements of \ion{N}{i}, \ion{Mn}{ii}, \ion{Cr}{ii}, \ion{Al}{iii}, 
and \ion{Fe}{iii}, in addition, and upper limits to the column densities of 
\ion{Mg}{ii}, \ion{Ni}{ii}, \ion{Zn}{ii}, and \ion{S}{iii}.

The low-ion absorption line profiles of this DLA are characterized by a very
simple velocity structure composed of 4 components spread over 40 km~s$^{-1}$.
Their properties are described in Table~\ref{Q2348-Ntable}. Component 4 was
only detected in the stronger metal-line transitions and its contribution to 
the total column density is negligible. Many metal lines observed in this DLA
are very weak, and their column density measurements are borderline cases
between detections and upper limits. This is the case for the \ion{Mg}{ii},
\ion{Mn}{ii}, \ion{Ni}{ii}, \ion{Cr}{ii}, and \ion{Zn}{ii} lines. Only the
\ion{Mn}{ii} and \ion{Cr}{ii} lines were detected at 3\,$\sigma$; hence, by
adopting large errors we obtained reliable Mn$^+$ and Cr$^+$ column densities. 
For the other lines detected at less than 3\,$\sigma$, we provided valuable
upper limits. While the very weak \ion{Mg}{ii}\,$\lambda$\,1239,\,1240 lines 
gave an upper limit to $N$(Mg$^+$), the saturated 
\ion{Mg}{ii}\,$\lambda$\,2796,\,2803 lines led to a lower limit to $N$(Mg$^+$)
(see Table~\ref{Q2348-Ntable}). We consider the derived \ion{C}{ii}$^*$ column 
density in agreement with the value obtained by \citet{prochaska99} as an upper 
limit due to possible blends of the \ion{C}{ii}$^*$\,$\lambda$\,1335 line with 
\ion{H}{i} lines in the Ly$\alpha$ forest. Similarly we prefer to assume that 
the measured \ion{Al}{ii} column density is an upper limit due to a possible 
blend of the \ion{Al}{ii}\,$\lambda$\,1670 line with a metal line (see 
Fig.~\ref{Q2348-metals}). Thanks to our access to several \ion{N}{i} lines, we 
could derive a value for the first time and not only an upper limit to the 
N$^0$ column density \citep{pettini95,prochaska02b}. However, before fitting 
the \ion{N}{i} lines, we first had to locally renormalize the spectrum around 
the \ion{N}{i} triplet at $\lambda_{\rm rest} \sim 1200$ \AA\ with the fit of 
the blue DLA Ly$\alpha$ damping wing profile. 
%

\begin{figure*}[!]
\centering
\includegraphics[width=17.5cm]{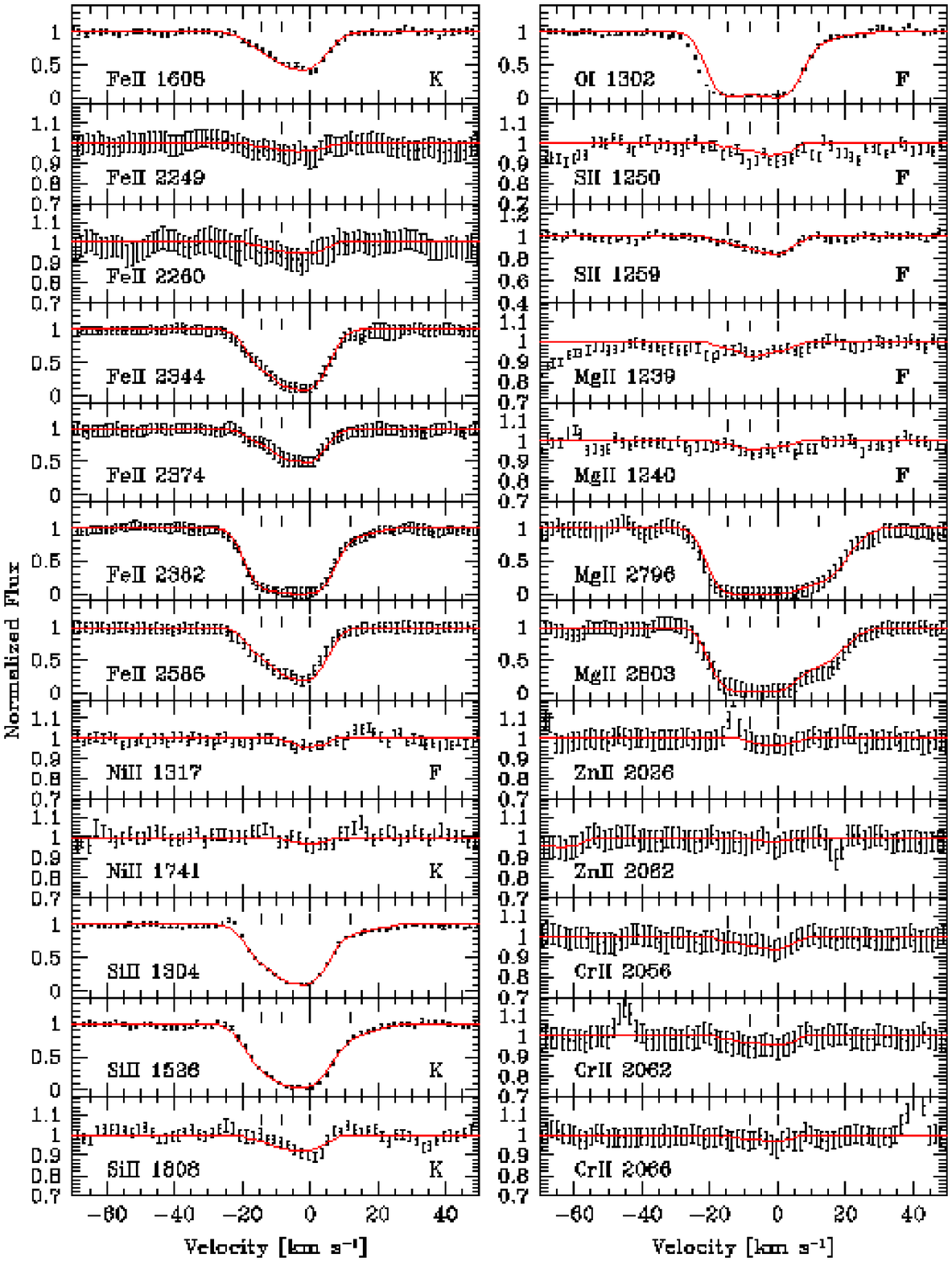}
\caption{Same as Fig.~\ref{Q0450-metals} for the DLA toward Q2348$-$1444. The 
zero velocity is fixed at $z=2.27939$.}
\label{Q2348-metals}
\end{figure*}

\addtocounter{figure}{-1}
\begin{figure*}[!]
\centering
\includegraphics[width=17.5cm]{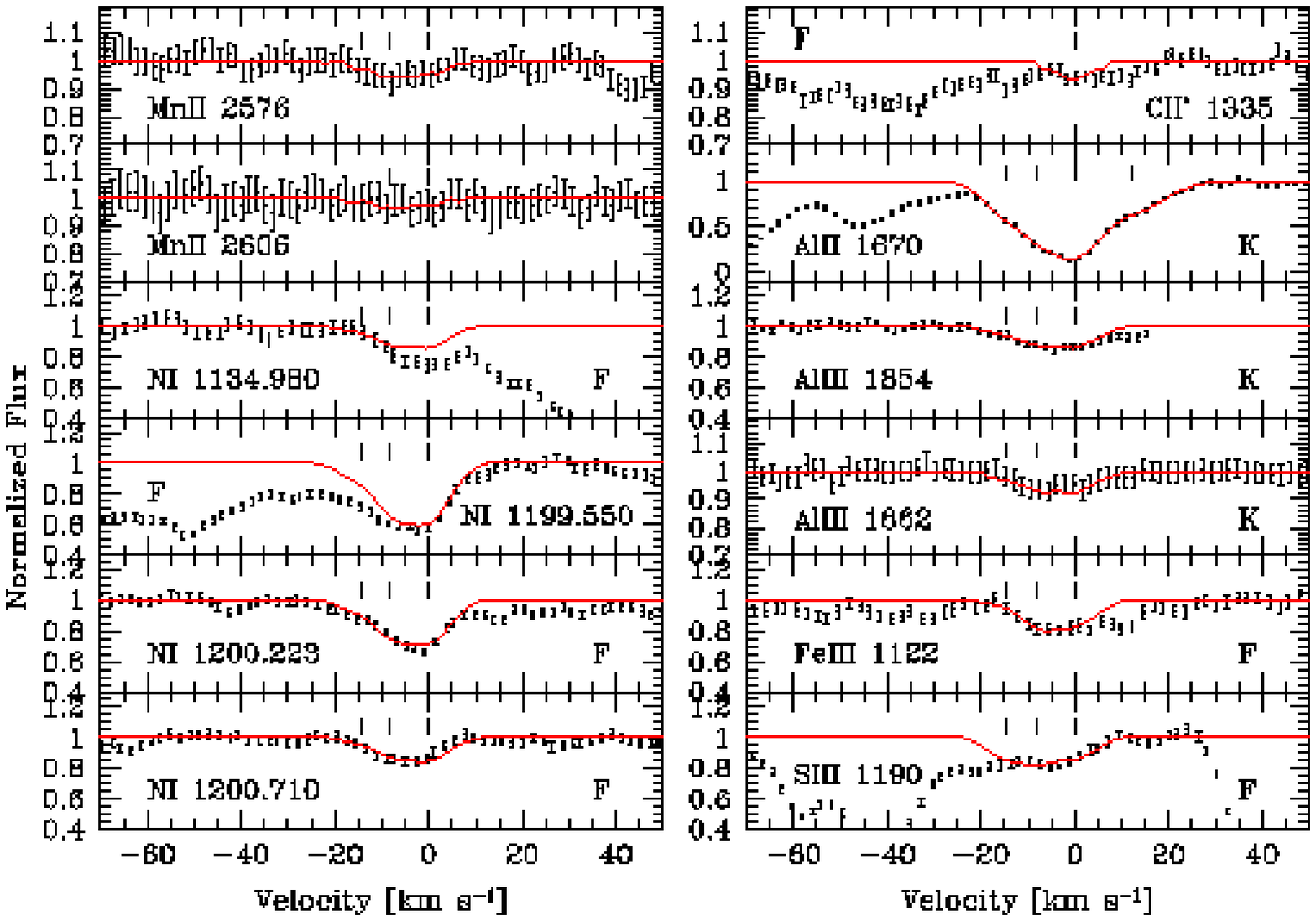}
\caption{{\em Continued}.}
\end{figure*}
%

The intermediate-ion lines of this DLA show exactly the same profiles as the
low-ion lines (see Fig.~\ref{Q2348-metals} and Table~\ref{Q2348-Ntable}). We
obtained the column density measurements of Al$^{++}$ and Fe$^{++}$ from the 
\ion{Fe}{iii}\,$\lambda$\,1122 line located in the Ly$\alpha$ forest. The
\ion{Fe}{iii} line might be blended with \ion{H}{i} lines, so we adopted a 
large error on $N$(Fe$^{++}$). We also derived an upper limit to the column 
density of \ion{S}{iii}. These measurements lead to the following column 
density ratios of different ionization species of the same element: 
$\log N$(Al$^{++}$)/$N$(Al$^+$) $> -0.66$, $\log N$(Fe$^{++}$)/$N$(Fe$^+$) 
$= -0.45\pm 0.14$, and $\log N$(S$^{++}$)/$N$(S$^+$) $< +0.04$. According to the 
photoionization diagnostics of \citet{prochaska02a} , these ratios show that 
the DLA system is likely to be partially ionized. The ionization fraction is, 
however, lower than 50\,\% and the expected ionization corrections on the 
measured ionic column densities are on the order of only 0.1~dex. The most 
important corrections to the observed gas-phase abundances are that the 
intrinsic [Si/Fe] ratio is slightly lower and the intrinsic [N/Si,S] ratios are 
slightly higher.

Figure~\ref{Q2348-HI} shows the fitting solution of the Ly$\alpha$ line of this
DLA system. The fit was obtained by fixing the $b$-value at 20 km~s$^{-1}$ and
the redshift at $z=2.27939$, i.e. at the redshift of the strongest metal-line
component (component 3). The derived \ion{H}{i} column density is $\log
N$(\ion{H}{i}) $= 20.59\pm 0.08$, which agrees well with the value obtained by 
\citet{pettini94} from low resolution spectra.

%

\section{Global gas-phase abundance patterns}
\label{DLA-trends}

In the next three sections we will use the results obtained for the seven DLAs 
studied here (see Sect.~\ref{data-analysis}) in combination with the results 
derived in Paper~I for four DLAs. These eleven DLAs constitute the first sample 
of high redshift objects for which we have the following at our disposal: 
accurate column density measurements of individual interstellar medium 
``clouds'' within the DLAs detected along the velocity profiles; the column 
density measurements of up to 30 ions; and the abundance measurements of up to 
22 elements.

%

\begin{figure*}[t]
\centering
\includegraphics[width=16.8cm]{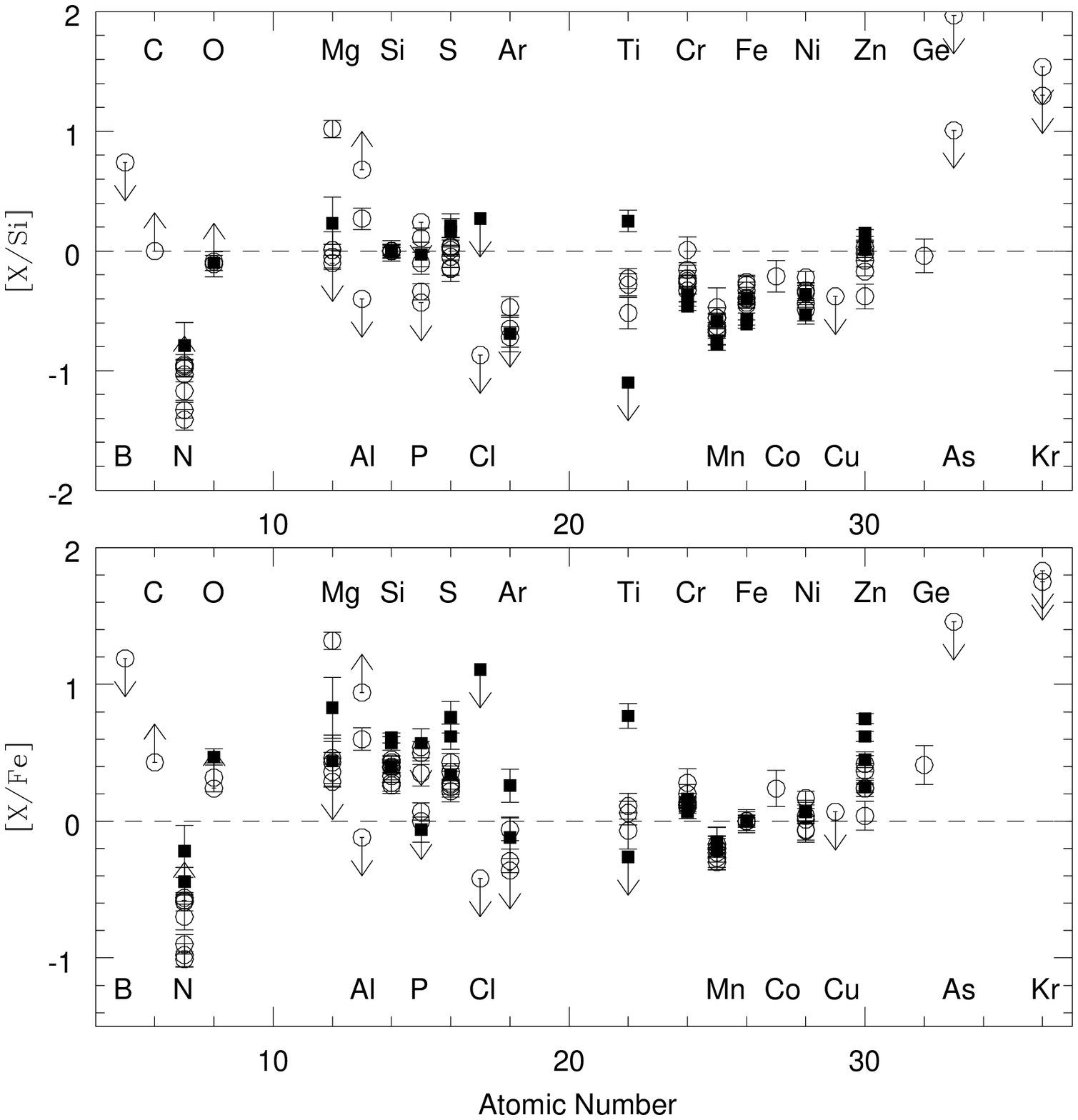}
\caption{Nucleosynthetic abundance patterns [X/Si] (upper panel) and [X/Fe] 
(lower panel) for our sample of 11 DLAs. We consider the entire set of detected 
elements X covering a large range of atomic numbers from 5 to 36. The plotted 
abundance ratios are the raw abundance ratio measurements, i.e. the directly 
observed gas-phase abundance ratios, free from any correction (dust or 
ionization). Our sample of data is composed of the 7 DLAs studied in this paper 
(open circles) and of 4 DLAs analyzed in Paper~I (filled squares). In 
Table~\ref{general-trends} we present some statistical results relative to the 
observed abundance pattern trends.}
\label{general-trends-Fig}
\end{figure*}
%

The absolute abundances, [X/H], are obtained by summing the contributions of 
all the components detected in the element X profile and reported in 
Tables~\ref{Q0450-Ntable}--\ref{Q2348-Ntable}. In what follows we will discuss 
the abundance ratios, [X/Y]. These are computed by considering only the column 
densities of components detected in both the element X and Y profiles. This is 
equivalent to comparing X and Y over the exact same velocity intervals. In this 
way we prevent an overestimation of abundances derived from strong metal-line 
profiles relative to the abundances derived from weaker metal-line profiles in 
which only the stronger components are usually detected. In the case of very 
weak metal lines, like the \ion{Ti}{ii} lines for instance, one can indeed 
underestimate the [X/Fe] ratios by up to $0.3-0.4$~dex by considering the total 
Fe abundance generally derived by summing a much larger number of components 
than for the Ti abundance. This effect is particularly important in DLAs with 
complex metal-line profiles extended over a large velocity range, in which a 
large difference is observed between the number of components detected in strong 
and weak metal lines. In our sample this is mainly the case of the DLAs 
toward Q0450$-$13 and toward Q2230+02. The same approach was already used 
in Paper~I, where the DLAs toward Q2231$-$00 and toward Q2343+12 were subject 
to such uncertainties.

%

\begin{table}[t]
\begin{center}
\caption{Abundance patterns of the integrated profiles for our sample of 11 DLAs} 
\label{general-trends}
\vspace{-0.3cm}
\begin{tabular}{l c c c c}
\hline \hline
\\[-0.3cm]
Element & \# DLAs & Mean $^{\rm a}$ & RMS $^{\rm b}$ & $\chi_{\nu}^2$ $^{\rm c}$ 
\smallskip 
\\ 
\hline
\multicolumn{5}{l}{\hspace{0.3cm} [X/Si]} \\
\hline
\\[-0.3cm]
N \dotfill  & \phantom{0}7 &	       $-$1.089{\scriptsize ($-$1.126)}           & 0.225{\scriptsize (0.195)} & \phantom{0}7.65{\scriptsize (7.772)\phantom{0}} \\ 
Mg \dotfill & \phantom{0}5 & \phantom{$-$}0.238{\scriptsize (0.197)\phantom{$-$}} & 0.447{\scriptsize (0.405)} &           38.37{\scriptsize (30.345)} \\
P \dotfill  & \phantom{0}4 &	       $-$0.127{\scriptsize ($-$0.173)}           & 0.193{\scriptsize (0.176)} & \phantom{0}6.44{\scriptsize (4.872)\phantom{0}} \\
S \dotfill  & \phantom{0}8 & \phantom{$-$}0.001{\scriptsize ($-$0.032)}           & 0.129{\scriptsize (0.121)} & \phantom{0}2.06{\scriptsize (1.785)\phantom{0}} \\
Ar \dotfill & \phantom{0}3 &	       $-$0.608{\scriptsize ($-$0.662)}           & 0.138{\scriptsize (0.200)} & \phantom{0}2.31{\scriptsize (2.666)\phantom{0}} \\  
Ti \dotfill & \phantom{0}4 &	       $-$0.147{\scriptsize ($-$0.295)}           & 0.327{\scriptsize (0.226)} &           10.28{\scriptsize (5.114)\phantom{0}} \\
Cr \dotfill & \phantom{0}9 &	       $-$0.314{\scriptsize ($-$0.359)}           & 0.148{\scriptsize (0.186)} & \phantom{0}6.01{\scriptsize (9.229)\phantom{0}} \\
Mn \dotfill & \phantom{0}9 &	       $-$0.656{\scriptsize ($-$0.695)}           & 0.102{\scriptsize (0.140)} & \phantom{0}3.57{\scriptsize (5.688)\phantom{0}} \\
Fe \dotfill &		10 &	       $-$0.436{\scriptsize ($-$0.436)}           & 0.115{\scriptsize (0.115)} & \phantom{0}6.27{\scriptsize (6.266)\phantom{0}} \\
Ni \dotfill & \phantom{0}7 &	       $-$0.404{\scriptsize ($-$0.417)}           & 0.109{\scriptsize (0.087)} & \phantom{0}4.07{\scriptsize (1.945)\phantom{0}} \\
Zn \dotfill & \phantom{0}8 & \phantom{$-$}0.007{\scriptsize ($-$0.042)}           & 0.174{\scriptsize (0.177)} & \phantom{0}9.73{\scriptsize (10.444)} \\
\hline
\multicolumn{5}{l}{\hspace{0.3cm} [X/Fe]} \\
\hline    
\\[-0.3cm]
N \dotfill  & \phantom{0}8 &           $-$0.678{\scriptsize ($-$0.706)}           & 0.278{\scriptsize (0.244)} &           12.96{\scriptsize (13.428)} \\
Mg \dotfill & \phantom{0}6 & \phantom{$-$}0.685{\scriptsize (0.645)\phantom{$-$}} & 0.376{\scriptsize (0.353)} &	   28.22{\scriptsize (23.584)} \\
Si \dotfill &		10 & \phantom{$-$}0.430{\scriptsize (0.430)\phantom{$-$}} & 0.114{\scriptsize (0.114)} & \phantom{0}6.63{\scriptsize (6.630)\phantom{0}} \\
P \dotfill  & \phantom{0}5 & \phantom{$-$}0.241{\scriptsize (0.193)\phantom{$-$}} & 0.277{\scriptsize (0.272)} & \phantom{0}9.62{\scriptsize (9.265)\phantom{0}} \\ 
S \dotfill  & \phantom{0}9 & \phantom{$-$}0.331{\scriptsize (0.303)\phantom{$-$}} & 0.195{\scriptsize (0.191)} & \phantom{0}4.22{\scriptsize (3.699)\phantom{0}} \\ 
Ar \dotfill & \phantom{0}4 &	       $-$0.092{\scriptsize ($-$0.140)}           & 0.234{\scriptsize (0.270)} & \phantom{0}4.64{\scriptsize (5.150)\phantom{0}} \\
Ti \dotfill & \phantom{0}4 & \phantom{$-$}0.261{\scriptsize (0.076)\phantom{$-$}} & 0.379{\scriptsize (0.233)} &	   15.59{\scriptsize (6.765)\phantom{0}} \\
Cr \dotfill &		10 & \phantom{$-$}0.131{\scriptsize (0.092)\phantom{$-$}} & 0.060{\scriptsize (0.107)} & \phantom{0}0.66{\scriptsize (3.582)\phantom{0}} \\
Mn \dotfill & \phantom{0}9 &	       $-$0.227{\scriptsize ($-$0.271)}           & 0.053{\scriptsize (0.077)} & \phantom{0}1.69{\scriptsize (2.515)\phantom{0}} \\
Ni \dotfill & \phantom{0}7 & \phantom{$-$}0.022{\scriptsize (0.007)\phantom{$-$}} & 0.085{\scriptsize (0.068)} & \phantom{0}2.42{\scriptsize (1.174)\phantom{0}} \\
Zn \dotfill & \phantom{0}9 & \phantom{$-$}0.396{\scriptsize (0.329)\phantom{$-$}} & 0.217{\scriptsize (0.199)} &	   26.51{\scriptsize (23.241)} \\
\hline
\end{tabular}
\begin{minipage}{83mm}
\smallskip
Note. The numbers in parentheses correspond to the abundance ratios [X/Si] and 
[X/Fe] computed by considering all of the gas in the \ion{Si}{ii} and \ion{Fe}{ii} 
profiles in comparison to the adopted method (see Sect.~\ref{DLA-trends}). \\
\\[-0.3cm]
$^{\rm a}$ Logarithmic weighted mean of [X/Si] and [X/Fe] using their \\
\phantom{$^{\rm a}$ }1\,$\sigma$ errors as weights. \\
$^{\rm b}$ Logarithmic RMS dispersion in [X/Si] and [X/Fe]. \\
$^{\rm c}$ Reduced $\chi^2$ of [X/Si] and [X/Fe] relative to their weighted mean.
\end{minipage}
\end{center}
\end{table}
%

Figure~\ref{general-trends-Fig} shows the nucleosynthetic abundance patterns as
a function of the atomic number of the seven DLAs studied in this paper 
(open circles) and the four DLAs studied in Paper~I (filled squares). We present 
both the [X/Si] and [X/Fe] abundance ratios relative to the solar values for 
the entire set of 22 detected elements X covering a range of atomic numbers 
from 5 to 36. This contrasts with the majority of DLAs for which only a handful 
of elements is usually detected \citep[e.g.][]{lu96,prochaska99,prochaska01}. 
We consider here the raw abundance ratios free from any correction (dust or
ionization). In this way we can try to identify whether the SFH, dust depletion, 
or ionization has a perceptible impact on some abundance ratios, and we can 
study the dispersion in the abundance patterns of the DLA galaxy population. In 
Table~\ref{general-trends} we present some statistical results on the observed 
abundance pattern trends. We computed the logarithmic weighted mean of the 
[X/Si] and [X/Fe] abundance ratios using their 1\,$\sigma$ errors as weights, 
the logarithmic RMS dispersion in [X/Si] and [X/Fe] ratios, and the reduced 
$\chi^2$ of [X/Si] and [X/Fe] relative to their weighted mean. The numbers 
in parentheses are given for information only. They correspond to the abundance 
ratios [X/Si] and [X/Fe] computed by considering all of the gas in the 
\ion{Si}{ii} and \ion{Fe}{ii} profiles, in comparison to the adopted method 
for the computation of the [X/Si] and [X/Fe] ratios (see the beginning of this 
section). To be conservative we do not take the limits in these calculations 
into account. We do note, however, that the limits would only lead to an 
increased dispersion in abundance ratios in which the detections already show a 
dispersion, e.g. P and Ti (see Fig.~\ref{general-trends-Fig}).

We underline that, although the abundance ratio measurements presented here are 
the raw ones (i.e. the directly observed gas-phase abundances), they show a 
remarkable uniformity\footnote{ This uniformity is even stronger when the 
abundance ratios [X/Si] and [X/Fe] are computed by considering all of the gas 
in the \ion{Si}{ii} and \ion{Fe}{ii} profiles (values in parenthesis in 
Table~\ref{general-trends}).}. Indeed, in the 11 DLA systems studied with 
redshifts between 1.7 and 2.5, \ion{H}{i} column densities covering one order 
of magnitude from $2\times 10^{20}$ to $4\times 10^{21}$ cm$^{-2}$, and 
metallicities from 1/55 to 1/5 solar, the abundance patterns show relatively 
low RMS dispersions, reaching only up to 2--3 times higher values than the 
statistical errors\footnote{The statistical errors are defined as the average 
of the 1\,$\sigma$ errors on measurements.} for the majority of elements. This 
suggests that the effects of nucleosynthesis enrichment, dust depletion, and 
ionization are negligible and that the abundance ratios of the integrated 
profiles for the DLA galaxy population are very uniform, as pointed out by 
\citet{molaro05}. In turn, this implies the respective star formation histories 
have conspired to yield one set of relative abundances. 
The time interval sampled by the DLA galaxies studied is of 1.8~Gyr for the 
adopted $H_0 = 65$ km s$^{-1}$ Mpc$^{-1}$, $\Omega_{\rm M} = 0.3$, and 
$\Omega_{\Lambda} = 0.7$ cosmology. 

We now discuss element per element the results derived for the [X/Si] and 
[X/Fe] ratios, along with their implications:

\noindent {\it N:}\hspace{0.2cm} The [N/Si] and [N/Fe] ratios both show a 
large dispersion, larger than 0.2~dex, with [N/Si] values ranging from $-0.8$ 
to $-1.5$. We expect that the large dispersion is due to different star 
formation histories (SFH) and ages of DLA galaxies, as discussed by a number of 
authors \citep[see][]{pettini02,prochaska02b,centurion03,chiappini03}. We do 
not wish to review the details here. 
While variations in the star formation histories are most likely the principal 
source of [N/Si] and [N/Fe] dispersions, ionization effects may also be 
important. Indeed, N$^0$ has a larger cross-section to photons with $h\nu > 2$ 
Ryd than H$^0$ \citep{sofia98}. Both in the DLA toward Q0450$-$13 and the DLA 
toward Q2343+12 with strong evidence for high ionization levels, we observe 
that a significant fraction of N is in the form of N$^+$ and not N$^0$ only. As 
a consequence N$^0$/(Si$^+$,Fe$^+$) will underestimate N/(Si,Fe) 
\citep{prochaska02b}.

\noindent {\it Mg:}\hspace{0.2cm} Aside from the measurement of 
\citet{srianand00}, the six Mg abundance measurements from our sample are the 
only Mg abundances derived in DLAs to date. The large dispersions observed in 
the [Mg/Si] and [Mg/Fe] ratios of +0.45 and +0.38~dex, respectively, are due to 
the upper open circle which corresponds to the DLA toward Q0450$-$13. This DLA 
shows several clues for a high ionization fraction (see Sect.~\ref{Q0450}). 
Hence, the measured abundance of Mg$^+$ is easily overestimated by several 
tenths of dex \citep[see Fig.~5 in][]{dessauges02}, which leads to the 
high [Mg/Si] and [Mg/Fe] ratios observed in this DLA. Mg is an $\alpha$-element 
and it should at least roughly trace Si as observed in Galactic metal-poor 
stars \citep{francois04}, although Si is probably produced in non-negligible 
amounts by Type~Ia supernovae (SNe) in addition to Type~II SNe. This is 
confirmed by our data, when excluding the DLA toward Q0450$-$13.

\noindent {\it Si:}\hspace{0.2cm} The low [Si/Fe] dispersion (0.11~dex,
$\chi_{\nu}^2 = 6.6$) indicates that the [Si/Fe] ratio remains nearly 
constant irrespective of the DLA galaxy and its dust depletion level. The mean 
[Si/Fe] value of +0.43~dex then suggests on average a high $\alpha$-element 
enhancement relative to iron-peak elements in all DLA systems, which is an 
enrichment by massive stars. Indeed, the $\alpha$-elements are produced in less 
than $2\times 10^7$ yrs by Type~II SNe resulting from massive stars, while the 
iron-peak elements are mainly produced by Type~Ia SNe on longer timescales 
between $3\times 10^7$ and $10^9$ yrs \citep{matteucci01}.

\noindent {\it P:}\hspace{0.2cm} P cannot be measured in Galactic halo
stars; thus the recent entry of P in the set of elements observed in DLAs 
offers a unique astronomical site where it can be measured at metallicities
significantly lower than solar. The [P/Si] and [P/Fe] ratios show a large
dispersion. The [P/Si] ratio is particularly interesting for observing the 
odd-even effect as it corresponds to an underabundance of odd-Z elements 
relative to even-Z elements of the same nucleosynthetic origin. The mean [P/Si] 
value of $-0.13$ shows evidence of a mild odd-even effect, lower than the 
expectations on the basis of yields by \citet{woosley95} and \citet{limongi00}. 
However, some DLAs show a strong odd-even effect, as the DLA at $z_{\rm abs} = 
2.375$ toward Q0841+129 with [P/Si] $= -0.34\pm 0.09$. This leads to a high 
[P/Si] dispersion due to various strengths of the odd-even effect from galaxy 
to galaxy. We emphasize that P is a non-refractory element and the variations 
are not very likely to be related to differential depletion.

\noindent {\it S:}\hspace{0.2cm} The [S/Si] ratio shows a relatively low
dispersion and a mean solar value. This implies that S traces Si very closely 
as in Galactic halo stars. Hence, the refractory $\alpha$-element Si appears to
be almost insensitive to dust depletion effects at the level they are observed 
in the DLAs studied, since it traces the volatile $\alpha$-element S whatever 
the dust depletion. The [S/Fe] ratio shows a large RMS dispersion of 
0.19~dex, larger than the one for the [Si/Fe] ratio. As a consequence, it might 
appear that the [S/Fe] ratio is more subject to variations due to dust 
depletion or nucleosynthesis enrichment than the [Si/Fe] ratio, but only in the 
order of the weak variations allowed by the [S/Si] ratios. The low reduced 
$\chi^2$ of the [S/Fe] ratio values of 4.2 confirms that the [S/Fe] intrinsic 
variations are in fact low and that [S/Fe] remains almost constant as [Si/Fe], 
whatever the DLA galaxy and its dust depletion level might be. The mean [S/Fe] 
value of +0.33~dex suggests, on average, an $\alpha$-element enhancement in all 
DLA systems as the [Si/Fe] ratio.

\noindent {\it Cl:}\hspace{0.2cm} We obtained a measurement and an upper 
limit to the column density of \ion{Cl}{i}. They both yield upper limits to the 
Cl abundance, since the dominant state of Cl in DLAs is Cl$^+$ and not Cl$^0$. 
Indeed, most of \ion{Cl}{i} is probably ionized, given that the ionization 
potential of Cl$^0$ is lower than 1~Ryd. Unfortunately, no general abundance 
pattern trend can be discussed for this element (see Paper~I for a description 
of its importance). 

\noindent {\it Ar:}\hspace{0.2cm} Ar is a typical product of Type~II SNe. 
It is presumed to track other $\alpha$-elements, although there is little
empirical evidence. No Ar abundance measurement exists in Galactic stars, and 
thus Ar measurements in DLAs are of high priority in the search to better 
understand the behavior of this element. The measured [Ar/Si] ratios have a low 
dispersion of 0.14~dex with a very low reduced $\chi^2$. Interestingly, all the 
obtained Ar measurements show significant underabundances relative to Si, the 
mean [Ar/Si] value being $-0.61$. No theoretical yields of Ar may explain such 
a large underabundance of Ar relative to Si, and the [Ar/Si] ratio is expected 
to be only weakly dependent on the SFH undergone by DLAs. Dust depletion 
effects are also not likely to explain these underabundances, since Ar is 
non-refractory. The only explanation can be found in the ionization effects. 
Indeed, \ion{Ar}{i} is very sensitive to ionization, in particular because its 
photoionization cross-section is one order of magnitude larger than for 
\ion{H}{i} \citep{sofia98}. Hence, Lyman continuum photons with energies $h \nu 
> 2$~Ryd are more efficient in ionizing \ion{Ar}{i} than \ion{H}{i}, if they 
are able to leak through the \ion{H}{i} layer \citep{vladilo03}. We wish to 
emphasize, however, that although all of our DLAs show a significant Ar 
underabundance, other ionization diagnostics imply the gas is predominantly 
neutral for the majority of our DLAs. Therefore, the [Ar/Si] ratio alone cannot 
characterize the ionization fraction of a DLA.

\noindent {\it Ti:}\hspace{0.2cm} Ti is generally accepted as an
$\alpha$-element, because it exhibits abundance patterns similar to other
$\alpha$-elements in Galactic stars \citep{edvardsson95,francois04}. It is a 
refractory element and has a high dust depletion level, even higher than Fe in 
Galactic ISM clouds \citep{savage96}. The dust depletion effects thus explain 
the large dispersion in the [Ti/Si] ratios of 0.33~dex, and the mean value of 
$-0.15$ is suggestive of some presence of dust in the majority of DLAs studied. 
The large dispersion in the [Ti/Fe] ratios of 0.38~dex is due both to dust 
depletion and nucleosynthesis enrichment effects. Positive departures of 
[Ti/Fe] from the solar value is evidence of an $\alpha$-enhancement, and 
negative [Ti/Fe] ratios provide evidence of dust depletion 
\citep[see][]{dessauges02}.

\noindent {\it Cr, Ni, Fe:}\hspace{0.2cm} Cr, Ni, and Fe are three iron-peak
elements with refractory properties. They trace each other in Galactic stars, 
but in gas phase the [Cr/Fe] and [Ni/Fe] ratios may show small differences
from the solar value, due to differential dust depletion. All the DLAs studied 
show an enhanced [Cr/Fe] ratio, the mean value being +0.13~dex in agreement 
with \citet{prochaska02c} findings. This is suggestive of dust depletion, since 
Galactic ISM lines of sight do exhibit a mild Cr overabundance relative to Fe. 
However, it is difficult to explain why every DLA shows enhanced [Cr/Fe] and 
why there is no trend with [Zn/Fe] (for the Zn discussion see below). The mean 
[Ni/Fe] value is solar, as it is observed in the Galactic ISM. These two 
abundance ratios show no dispersion, and they are uniform from one DLA to 
another with an RMS of 0.06 and 0.08~dex, respectively. This uniformity can be 
explained by the fact that Cr, Ni, and Fe have very similar dust depletion 
patterns in the Galactic ISM \citep{savage96}, and the DLAs do not have enough 
high dust depletion variations from one system to another to produce [Cr/Fe] 
and [Ni/Fe] variations.

\noindent {\it Mn:}\hspace{0.2cm} Mn is an iron-peak element, but it 
behaves differently from other iron-peak elements. The [Mn/Fe] ratios in 
Galactic stars are undersolar and show a decrease with metallicity. They 
illustrate a nice example of the odd-even effect for iron-nuclei. All the Mn 
abundance measurements obtained in the DLAs studied are also underabundant 
relative to Fe with a [Mn/Fe] mean value of $-0.23$. In addition, these 
measurements are very uniform with an RMS of 0.05~dex. This uniformity implies 
two important consequences. Firstly, the DLA [Mn/Fe] ratios reach a plateau at 
$-0.23$~dex with a small scatter for a metallicity of DLA systems between 
[Zn/H] $= -2$ and $-0.7$~dex. Secondly, the [Mn/Fe] ratios are similar 
irrespective of the dust depletion level of DLAs, as we have already argued in 
\citet{dessauges02}, since Mn has a very similar dust depletion level to that 
of Fe. In a future paper, we will consider if this result contradicts the 
conclusions of \cite{mcwilliam03} based on stellar abundances that Mn is a 
secondary element.

%

\begin{table*}[t]
\begin{center}
\caption{Abundance patterns and correlations between [X/Y] ratios and [Zn/Fe] for all clouds observed in our sample of 11 DLAs}
\label{allclouds-trends}
\begin{tabular}{l | c c c c | c c c c c}
\hline \hline
\\[-0.3cm]
$\lbrack$X/Y] & \# clouds $^{\rm 1}$ & Mean & RMS & $\chi_{\nu}^2$ & \# clouds $^{\rm 2}$ & $\tau$ $^{\rm a}$ & P($\tau$) $^{\rm b}$ & a $^{\rm c}$ & b $^{\rm d}$
\smallskip 
\\ 
\hline
\\[-0.3cm]
$\lbrack$Si/Fe] & 79 & \phantom{$-$}0.390 & 0.183 &           18.10 & 22 & \phantom{$-$}0.686 & 0.000 & $+0.241\pm 0.023$ & $+0.534\pm 0.044$ \\
$\lbrack$S/Fe]  & 34 & \phantom{$-$}0.376 & 0.251 & \phantom{0}9.85 & 18 & \phantom{$-$}0.757 & 0.000 & $+0.060\pm 0.038$ & $+0.961\pm 0.083$ \\
$\lbrack$Si/Zn] & 22 &  	 $-$0.033 & 0.173 &	      11.60 & 22 &	     $-$0.712 & 0.000 & $+0.246\pm 0.025$ & $-0.495\pm 0.041$ \\
$\lbrack$S/Zn]  & 18 & \phantom{$-$}0.032 & 0.101 & \phantom{0}1.62 & 18 &	     $-$0.343 & 0.047 & $+0.058\pm 0.032$ & $-0.059\pm 0.071$ \\
$\lbrack$S/Si]  & 32 & \phantom{$-$}0.047 & 0.213 & \phantom{0}5.03 & 16 & \phantom{$-$}0.561 & 0.002 & $-0.152\pm 0.037$ & $+0.416\pm 0.079$ \\
$\lbrack$N/Si]  & 36 &  	 $-$1.005 & 0.318 &	      17.04 & 12 & \phantom{$-$}0.412 & 0.062 & \multicolumn{2}{c}{no correlation} \\
$\lbrack$Mn/Fe] & 29 &  	 $-$0.221 & 0.108 & \phantom{0}5.37 & 22 & \phantom{$-$}0.264 & 0.086 & \multicolumn{2}{c}{no correlation} \\
$\lbrack$Cr/Fe] & 28 & \phantom{$-$}0.130 & 0.105 & \phantom{0}2.89 & 24 & \phantom{$-$}0.097 & 0.508 & \multicolumn{2}{c}{no correlation} \\
$\lbrack$Zn/Fe] & 25 & \phantom{$-$}0.427 & 0.263 &           30.01 &    &    &    & \\
\hline
\end{tabular}
\begin{minipage}{165mm}
\smallskip
Note. For the definition of a ``cloud'', see Sect.~\ref{indiv-clouds}. \\
\\[-0.3cm]
$^{\rm 1}$ Total number of clouds in our sample of 11 DLAs with a measurement of [X/Y]. The corresponding logarithmic weighted mean, \\
\phantom{$^{\rm 1}$} logarithmic RMS dispersion, and reduced $\chi^2$ are given in columns (3), (4), and (5), respectively. \\
$^{\rm 2}$ Number of clouds in our sample of 11 DLAs with both a measurement of [X/Y] and [Zn/Fe]. Data plotted in 
Fig.~\ref{allclouds-trends-Fig}. \\
$^{\rm a}$ Kendall correlation factor of [X/Y] versus [Zn/Fe]. A positive value of $\tau$ corresponds to a correlation and 
a negative value to an \\ \phantom{$^{\rm a}$} anti-correlation. \\
$^{\rm b}$ Probability under the null hypothesis of zero correlation from the Kendall test. A value $< 5$\% indicates a significant 
correlation. \\
$^{\rm c}$$^,$ $^{\rm d}$ Zero point and slope, respectively, and their 1\,$\sigma$ uncertainties, of the linear least-square 
regression [X/Y] $=$ a $+$ b $\times$ [Zn/Fe], \\ \phantom{$^{\rm c}$$^,$ $^{\rm d}$} computed by taking into account the errors 
on both [X/Y] and [Zn/Fe] data points.
\end{minipage}
\end{center}
\end{table*}
%

\noindent {\it Co:}\hspace{0.2cm} We obtained the third Co abundance
measurement in the DLA system toward Q1157+014. The first ones were obtained by 
\citet{ellison01} and \citet{rao05}. Co is an iron-peak element. Its dust 
depletion in the Galactic ISM is not well known; Co has probably a refractory 
nature similar to the one of Fe. In Galactic stars, [Co/Fe] shows a large 
scatter around the solar value. No general DLA abundance pattern trend can be 
discussed for this element yet. The three [Co/Fe] measurements are: [Co/Fe] 
$= +0.24\pm 0.13$ (our value), [Co/Fe] $= +0.31\pm 0.05$ \citep{ellison01}, and 
[Co/Fe] $= +0.05\pm 0.12$ \citep{rao05}. 

\noindent {\it Zn:}\hspace{0.2cm} Zn is an extremely important element in
DLA abundance studies. It is frequently considered as an iron-peak element 
because it traces the other iron-peak elements in Galactic stars, but Zn is 
probably produced through different nucleosynthetic processes 
\citep[e.g.][]{matteucci93}. Zn has a unique trump, which is that it is not 
readily incorporated into dust grains and thus is only very mildly refractory 
\citep{savage96}. As a consequence, the [Zn/Fe] ratio is widely used as a 
tracer of the dust depletion level in a DLA, with higher [Zn/Fe] values
implying higher depletion levels. Our measurements show a high dispersion of 
0.22~dex in the [Zn/Fe] ratios and a reduced $\chi^2$ of 26. The [Zn/Fe] values 
vary from 0 to +0.8, which indicates that the DLAs studied sample a relatively 
large range of dust depletion levels. The [Zn/Si] ratio is relatively free from 
dust depletion effects, since, as discussed above, the refractory 
$\alpha$-element Si traces the volatile element S well whatever the dust 
depletion level might be in the DLAs studied. Hence, the [Zn/Si] ratio is a good 
tracer of only nucleosynthetic enrichment. The [Zn/Si] dispersion of 0.17~dex 
with a $\chi_{\nu}^2$ of 10 is relatively large, and its most straightforward 
explanation can be found in different SFHs from one DLA to another. However, 
this dispersion, as well as the mean solar [Zn/Si] value, contradict the 
results obtained for [Si/Fe] and [S/Fe]. Indeed, as discussed above, 
the [Si/Fe] and [S/Fe] ratios seem to be uniform and enhanced relative to solar 
values. How can we reconcile the solar [Zn/Si] abundance pattern with the 
$\alpha$-enhancement observed in the [Si/Fe] and [S/Fe] ratios that cannot 
entirely be accounted for by dust depletion effects? The only solution is to 
assume that the [Zn/Fe] ratio can be larger than 0, independently from dust 
depletion effects. Recent measurements of Galactic metal-poor stars indeed 
suggest that Zn is overabundant relative to Fe in the range of 0 to +0.2~dex 
for metallicities between [Fe/H] $=-2$ and $-1$ 
\citep{prochaska00,mishenina02,nissen04,chen04,francois04}.

\noindent {\it Ge:}\hspace{0.2cm} Ge is the element with the highest
atomic number (Z\,=\,32) for which we obtained an abundance measurement. 
\citet{prochaska03b} were the first to draw attention to the prospect of
measuring Ge in DLAs. It is an element of the iron-peak, but the s-process in 
massive stars may also contribute to its production. It thus has a secondary 
origin, since it can only be produced when Fe is already present in stars. 
\citet{cowan05} have very recently shown that Ge abundances in Galactic 
metal-poor stars track the Fe abundances very well, but at a depressed level, 
$\langle$[Ge/Fe]$\rangle = -0.79\pm 0.04$. Hence, an explosive process on
iron-peak nuclei, rather than neutron capture, appears to be the dominant
synthesis mechanism for Ge at low metallicities. At higher metallicities, the
s-process production takes place, and it would be expected that the Ge 
abundances would increase with the Fe abundances. In DLAs, the Ge abundances do 
not seem to follow the same trend relative to the Fe abundances. Indeed, the 
two measured [Ge/Fe] abundance ratios are oversolar, [Ge/Fe] $= +0.41\pm 0.15$ 
(our measurement) and [Ge/Fe] $= +0.77$ \citep{prochaska03b}.

\noindent {\it B, As, Kr:}\hspace{0.2cm} We just obtained upper limits to
the abundances of these elements. They will be discussed in a future paper.

%

\section{Gas-phase abundance patterns of individual ``clouds''}
\label{indiv-clouds}

From the Voigt profile fitting of metal-lines, we obtained very accurate
component-to-component column density measurements. These components presumably 
correspond to interstellar medium clouds in the DLA galaxy on the QSO line of 
sight. We consider here these clouds as individual entities and study their
abundance pattern trends.

It is important, however, to define the concept of a cloud more strictly 
relative to the components defined in the Voigt profile fits. Indeed, to be 
able to consider the column density of a component as a physical property of an 
individual ISM cloud on a QSO line of sight, the component has to be 
independent, i.e. unblended with its neighbors. We assumed two components are 
independent if they satisfy the following criterion: the difference of position, 
$\Delta v$, in velocity space between two components has to be larger than the 
sum of their half line widths (i.e. their half $b$-values) within 5\,$\sigma$ 
of their respective errors. The results of this criterion agree well with an 
eye-ball determination of independent components from optical 
depth considerations. The components which do not satisfy this criterion are 
encompassed in Tables~\ref{Q0450-Ntable}--\ref{Q2348-Ntable} and together they 
form an independent cloud. Their column densities have to be summed and 
this sum corresponds to the column density of the cloud. The same criterion has 
been applied to the DLAs studied in Paper~I. In our sample of 11 DLA systems, 
we find a total of 84 clouds.

%

\begin{figure*}[!]
\centering
\includegraphics[width=17cm]{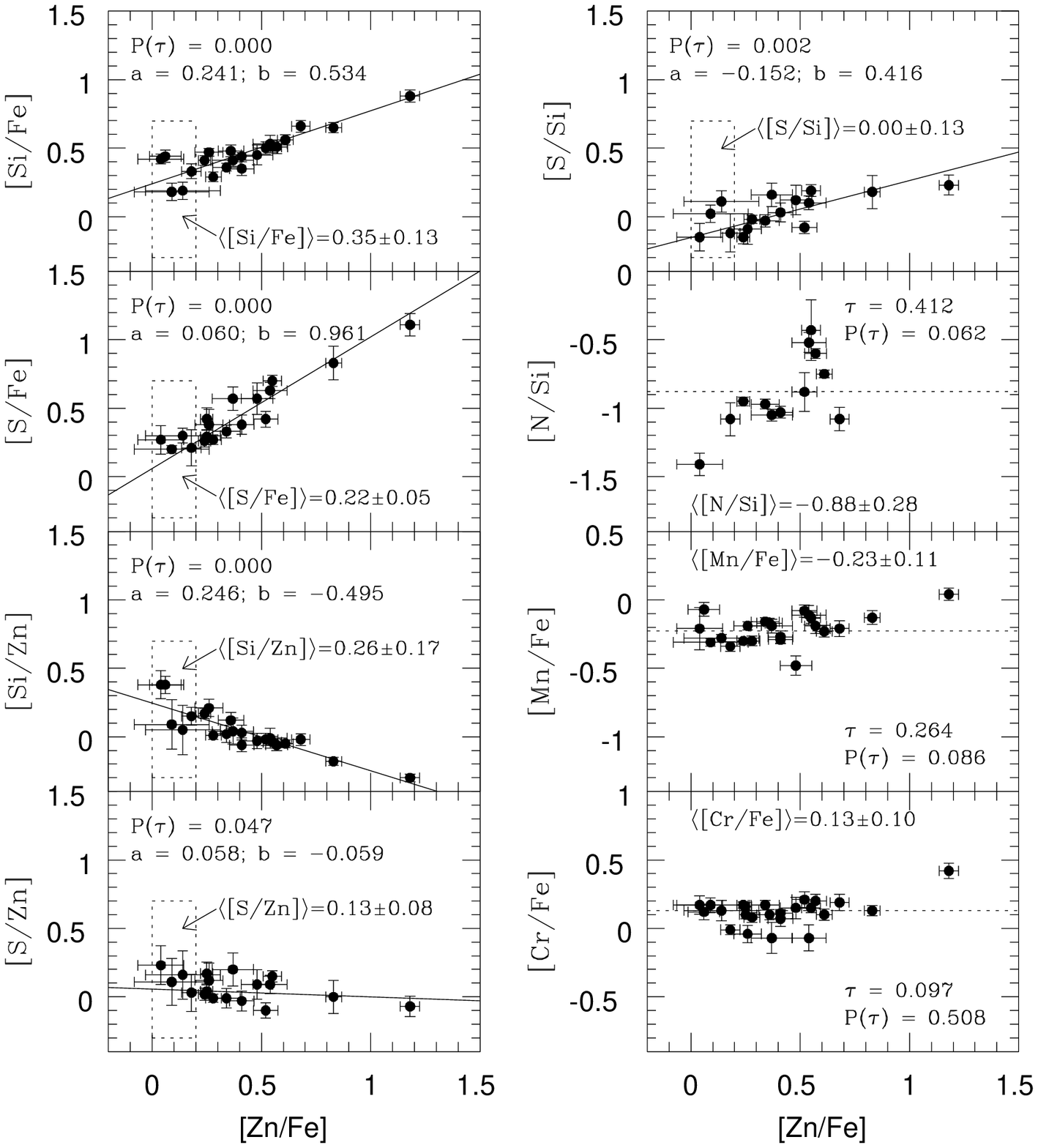}
\caption{Eight different abundance ratios [X/Y] versus [Zn/Fe] for the entire 
set of clouds observed in our sample of 11 DLAs. The Kendall test shows clear 
correlations between [Si/Fe], [S/Fe], [Si/Zn], [S/Zn], [S/Si], and [Zn/Fe] with 
a probability under the null hypothesis of zero correlation, P($\tau$), $< 5$\%. 
The derived linear least-square regressions, [X/Y] $=$ a $+$ b $\times$ [Zn/Fe], 
computed by taking into account the errors on both [X/Y] and [Zn/Fe] data 
points, are shown by the solid lines. The dotted horizontal lines correspond to 
the weighted mean of [X/Y] and are indicated when no correlation is observed. 
In panels 1-5 (from top to bottom and from left to right), we note the weighted 
mean of [X/Y] in the interval $0 <$ [Zn/Fe] $< 0.2$ (data points contained in 
the dotted boxes).}
\label{allclouds-trends-Fig}
\end{figure*}
%

In the first part of Table~\ref{allclouds-trends}, we present the dispersion 
properties for different abundance ratios --~[Si/Fe], [S/Fe], [Si/Zn], [S/Zn],
[S/Si], [N/Si], [Mn/Fe], [Cr/Fe], and [Zn/Fe]~-- of all the clouds observed in 
our sample of 11 DLA galaxies. We give the logarithmic weighted mean computed 
using the 1\,$\sigma$ errors as weights in column (3), the logarithmic RMS 
dispersion in column (4), and the reduced $\chi^2$ relative to the weighted 
mean in column (5). We notice that, in general, the mean abundance ratio
values of clouds are very similar to those of integrated profiles of DLAs, but 
the measured RMS dispersions and reduced $\chi^2$ are 2--3 times larger for 
individual clouds than the values obtained for the DLAs 
(see Table~\ref{general-trends}). This is partly due to the larger statistical 
errors in the cloud abundance ratio measurements. 

However, we do observe statistically significant dispersions of about 
5\,$\sigma$ in the [Zn/Fe] ratio and larger than 3\,$\sigma$ in the 
[$\alpha$/Fe,Zn] ratios, except for [S/Zn] which has a particularly low 
dispersion\footnote{The number of $\sigma$ is computed as the square root of 
the reduced $\chi^2$. This is a first order approximation of the probability to 
have a statistically significant dispersion assuming a Gaussian distribution.}. 
High dispersions are expected according to the observations made in the Milky 
Way and the Local Group galaxies, if these clouds probe different physical 
conditions and come from galaxies with various SFHs. Indeed, the observed 
gas-phase abundance ratios along a Galactic line of sight may vary by more than 
0.5~dex as the sightline penetrates clouds arising in various phases of the ISM 
\citep[e.g.][]{savage96}. Similarly, the LMC and SMC also exhibit large 
variations in their gas-phase abundance ratios \citep{welty99,welty01}. This is 
due to the fact that sightlines through the ISM probe gas with a range of 
physical conditions, e.g. various dust-to-gas ratios, volume densities, and 
ionization states. In addition, a range of at least 0.3~dex in the $\alpha$/Fe 
abundance ratios is observed both within a galaxy and from galaxy to galaxy 
when comparing the abundance measurements of stars in the Milky Way with those 
in the Magellanic Clouds and dwarf spheroidal galaxies 
\citep{venn99,shetrone03,tolstoy03}. Thus, the observed dispersion in the DLA 
cloud-to-cloud abundance ratios suggests at first glance that the individual 
clouds do not have a similar enrichment history or a uniform differential dust 
depletion. Moreover, the fact that this dispersion is higher than the one of 
global DLA abundance measurements indicates that the SFH, dust, and ionization 
variations are confined more to individual clouds rather than to the whole DLA 
galaxy. If confirmed, this will provide important constraints on the 
understanding of the ISM of high-redshift galaxies and the enrichment of gas in 
the early Universe. In Sect.~\ref{cloud-to-cloud} we study the cloud-to-cloud 
variations further within a given DLA system.


In the second part of Table~\ref{allclouds-trends}, we present the results of 
our study on different possible correlations between an abundance ratio [X/Y] 
and [Zn/Fe]. The [Zn/Fe] ratio is considered as a dust depletion indicator (see 
Sect.~\ref{DLA-trends}). A few of these correlations have already been explored 
by e.g. \citet{prochaska02c} for the abundance measurements obtained in DLAs, 
but this is the first time that we probe such properties for the individual 
clouds within DLAs. The trend discussed above that SFH/dust/ionization 
variations are likely to be more confined to individual clouds than to the 
global DLA galaxies suggests that if some correlations exist between some 
physical properties, they should be more easily identified in individual cloud 
studies. We searched for correlations between [Zn/Fe] and the nucleosynthesis 
indicators [Si/Fe], [S/Fe], [Si/Zn], [S/Zn], [N/Si], and [Mn/Fe], and between 
[Zn/Fe] and other dust depletion indicators [S/Si] and [Cr/Fe]. A Kendall test 
was performed to analyze these correlations, and in column (7) we give the 
Kendall correlation factor $\tau$ and in column (8) the probability P($\tau$) 
under the null hypothesis of zero correlation (values lower than 5\,\% indicate 
a significant correlation). In columns (9) and (10), we report the zero points 
and slopes with their 1\,$\sigma$ errors of the linear least-square regressions, 
[X/Y] = a~+~b\,$\times$\,[Zn/Fe], computed for the cases of significant 
correlations. The errors on both [X/Y] and [Zn/Fe] ratios were taken into 
account in this computation. We have not, however, accounted for the fact that 
Zn or Fe are generally present in the ratios along each axis (i.e. [Zn/Fe] and 
[X/Y]). Therefore, one should be more skeptical of correlations with less than 
99\,\% significance.

Figure~\ref{allclouds-trends-Fig} shows that there are clear correlations 
between [Si/Fe] and [S/Fe] versus [Zn/Fe] and clear anti-correlations between 
[Si/Zn] and [S/Zn] versus [Zn/Fe]. These trends are the result of a 
combination of dust depletion and nucleosynthesis enrichment effects. But, more 
precisely, the increase of the [$\alpha$/Fe] ratios and the decrease of the 
[Si/Zn] ratio with the dust depletion level can mainly be assigned to 
differential dust depletion effects. The nucleosynthesis enrichment effects 
contribute only negligibly to the strength of the evolution as a function of 
dust depletion, as illustrated by [S/Zn] versus [Zn/Fe]. Indeed, the [S/Zn] 
ratio, which is independent of dust depletion effects, is an intrinsic tracer 
of nucleosynthesis enrichment, and the correlation between [S/Zn] and [Zn/Fe] 
is observed at only 95\,\% confidence level, and the slope of its linear 
least-square regression is consistent with zero (${\rm b} = -0.059\pm 0.071$). 
This analysis, in addition, suggests that the high dispersion observed in the 
[Si/Fe], [S/Fe], and [Si/Zn] ratios (see the first part of 
Table~\ref{allclouds-trends}) is the result of only dust depletion effects. The 
nucleosynthesis enrichment effects are, indeed, negligible given the low 
dispersion measured in the [S/Zn] ratio and the low dispersion of [Si/Fe], 
[S/Fe], and [Si/Zn] along the linear least-square regressions (see 
Fig.~\ref{allclouds-trends-Fig}). Consequently, the individual clouds very 
likely have a similar enrichment history, but different dust depletion levels.

The signature of pure nucleosynthesis contribution (i.e. of the SFH) can be 
observed in the values of the [$\alpha$/Fe,Zn] ratios at [Zn/Fe] $\simeq 0$. 
Indeed, at [Zn/Fe] $\simeq 0$, the dust depletion level is low, and hence the 
$\alpha$-element over iron-peak element ratios are free from dust depletion 
effects. We computed the weighted means of data points contained in the 
interval [Zn/Fe] $= [0,+0.2]$ (see the dotted boxes in 
Fig.~\ref{allclouds-trends-Fig}), and interestingly we find enhanced 
[$\alpha$/Fe,Zn] ratios relative to solar in all cases: 
$\langle$[Si/Fe]$\rangle = +0.35\pm 0.13$, $\langle$[S/Fe]$\rangle = +0.22\pm 
0.05$, $\langle$[Si/Zn]$\rangle = +0.26\pm 0.17$, and $\langle$[S/Zn]$\rangle = 
+0.13\pm 0.08$. This suggests that the intrinsic abundance patterns of 
individual clouds within DLAs, when the nucleosynthesis enrichment can reliably 
be disentangled from dust depletion effects, show an $\alpha$-enhancement 
indicative of Type~II SNe. We would like, in addition, to underline the 
relatively important difference which exists between the mean [Si/Fe] ratio and 
the mean [S/Zn] ratio, yet measured in the same clouds free from dust depletion 
effects. The fact that [S/Zn] is almost solar, while [Si/Fe] shows a clear 
$\alpha$-enhancement, suggests that the [S/Zn] ratio, although independent from 
dust depletion effects, may not be a reliable tracer of nucleosynthesis 
enrichment as considered until now \citep[e.g.][]{centurion00}. Indeed, the S 
production is perhaps correlated in some way with the production of Zn 
\citep[see][]{fenner04}.

Some of the correlations and anti-correlations found in this analysis of clouds 
within DLAs have already been observed in DLAs themselves (e.g. [Si/Fe] versus 
[Zn/Fe]), while for some we observe only trends due to fewer data points. 
Similarly, the $\alpha$-enhanced abundance pattern unambiguously observed in 
individual clouds is still being debated for the global DLA abundance patterns 
\citep{prochaska02c,vladilo98,vladilo02,centurion00}. We stressed this issue in
Sect.~\ref{DLA-trends} with the mean DLA [Si,S/Fe] ratios showing an
$\alpha$-enhancement and the mean [Si/Zn] ratios showing solar values. 

We also explored the correlations of [Zn/Fe] with two other nucleosynthesis
enrichment indicators, [N/Si] and [Mn/Fe]. No clear correlation was observed for 
any of these ratios (see Fig.~\ref{allclouds-trends-Fig}). This is particularly 
interesting in the case of [Mn/Fe], since it shows that all the clouds have 
an undersolar [Mn/Fe] ratio, irrespective of dust depletion level, except for 
the dustiest cloud with [Zn/Fe] $> +1$. Hence, this underabundance is only a 
result of nucleosynthesis enrichment. Evidence of intrinsic subsolar [Mn/Fe] 
abundances in DLAs themselves were discussed in Sect.~\ref{DLA-trends}.

Finally, we found an interesting correlation between [Zn/Fe] and the ratio of 
two $\alpha$-elements: S, a volatile element, and Si, a refractory element (see
Fig.~\ref{allclouds-trends-Fig}). This correlation is a direct result of pure 
differential dust depletion effects observed in two $\alpha$-elements and two 
iron-peak elements. The mean value of the [S/Si] ratio in the interval [Zn/Fe] 
$= [0,+0.2]$ is solar, $\langle$[S/Si]$\rangle = 0.00\pm 0.13$, which shows that 
S traces Si in the presence of a weak dust depletion level.

%

\begin{table*}[!]
\begin{center}
\caption{Cloud-to-cloud chemical variation analysis in individual DLA systems}
\label{comp-analysis} 
\begin{tabular}{l c c c c c c c c}
\hline \hline
\\[-0.3cm]
$\lbrack$X/Y]& \# clouds $^{\rm a}$ & $\Delta v$ $^{\rm b}$ & Mean & RMS & $\chi_{\nu}^2$ & $\Delta_{sngl}$ & $\Delta_{all}$ & $\Delta_{best}$
\smallskip 
\\ 
\hline
\multicolumn{9}{l}{\hspace{0.3cm} DLA toward Q0450$-$13} \\
\hline    
\\[-0.3cm]   
$\lbrack$Si/Fe] &           11 &           232 & \phantom{$-$}0.313 & 0.151 & \phantom{0}13.66 & 0.69 & 0.34 & 0.21 \\
$\lbrack$O/Si]  & \phantom{0}7 &           232 &           $-$0.345 & 0.636 &           117.36 & 1.28 & 1.47 & 0.75 \\
$\lbrack$S/Si]  & \phantom{0}6 & \phantom{0}86 & \phantom{$-$}0.038 & 0.172 & \phantom{00}3.51 & 0.46 & 0.39 & 0.18 \\
$\lbrack$Mg/Si] & \phantom{0}6 & \phantom{0}86 & \phantom{$-$}1.103 & 0.227 & \phantom{00}6.57 & 0.54 & 0.51 & 0.27 \\
$\lbrack$N/Si]  & \phantom{0}7 &           138 &           $-$1.275 & 0.282 & \phantom{0}12.82 & 0.69 & 0.58 & 0.29 \\
$\lbrack$N/S]   & \phantom{0}6 & \phantom{0}86 &           $-$1.311 & 0.173 & \phantom{00}4.68 & 0.54 & 0.50 & 0.25 \\
\hline
\multicolumn{9}{l}{\hspace{0.3cm} DLA toward Q1157+014} \\
\hline
\\[-0.3cm]
$\lbrack$Si/Fe] & \phantom{0}4 &           144 & \phantom{$-$}0.448 & 0.026 & \phantom{00}0.26 & 0.07 & 0.06 & 0.00 \\
$\lbrack$Si/Zn] & \phantom{0}3 & \phantom{0}72 & \phantom{$-$}0.158 & 0.182 & \phantom{00}9.49 & 0.42 & 1.31 & 0.27 \\
$\lbrack$Mg/Si] & \phantom{0}4 &           144 &           $-$0.010 & 0.049 & \phantom{00}0.61 & 0.15 & 0.16 & 0.00 \\
$\lbrack$Mn/Fe] & \phantom{0}4 &           144 &           $-$0.214 & 0.095 & \phantom{00}5.74 & 0.31 & 0.44 & 0.14 \\
$\lbrack$Zn/Fe] & \phantom{0}3 & \phantom{0}72 & \phantom{$-$}0.311 & 0.201 & \phantom{00}8.26 & 0.41 & 1.21 & 0.26 \\
\hline
\multicolumn{9}{l}{\hspace{0.3cm} DLA toward Q1210+17} \\
\hline
\\[-0.3cm]
$\lbrack$Si/Fe] & \phantom{0}6 &           127 & \phantom{$-$}0.204 & 0.058 & \phantom{00}2.68 & 0.26 & 0.20 & 0.09 \\
$\lbrack$Si/Zn] & \phantom{0}3 & \phantom{0}51 & \phantom{$-$}0.013 & 0.060 & \phantom{00}0.12 & 0.00 & 0.01 & 0.07 \\
$\lbrack$S/Zn]  & \phantom{0}3 & \phantom{0}51 &           $-$0.003 & 0.139 & \phantom{00}0.67 & 0.29 & 0.37 & 0.07 \\
$\lbrack$S/Si]  & \phantom{0}3 & \phantom{0}51 &           $-$0.001 & 0.080 & \phantom{00}1.29 & 0.19 & 0.37 & 0.00 \\
$\lbrack$Mn/Fe] & \phantom{0}3 & \phantom{0}51 &           $-$0.296 & 0.015 & \phantom{00}0.29 & 0.07 & 0.10 & 0.03 \\
$\lbrack$Zn/Fe] & \phantom{0}3 & \phantom{0}51 & \phantom{$-$}0.267 & 0.156 & \phantom{00}0.88 & 0.33 & 0.45 & 0.00 \\
\hline
\multicolumn{9}{l}{\hspace{0.3cm} DLA toward 1331+17 $^{\rm 1}$} \\
\hline
\\[-0.3cm]
$\lbrack$Si/Fe] & \phantom{0}4 & \phantom{0}89 & \phantom{$-$}0.595 & 0.200 & \phantom{0}23.02 & 0.48 & 0.85 & 0.29 \\
$\lbrack$Si/Zn] & \phantom{0}4 & \phantom{0}89 &           $-$0.182 & 0.238 & \phantom{0}21.82 & 0.46 & 0.79 & 0.27 \\
$\lbrack$S/Zn]  & \phantom{0}4 & \phantom{0}89 & \phantom{$-$}0.005 & 0.090 & \phantom{00}0.85 & 0.32 & 0.40 & 0.02 \\
$\lbrack$S/Si]  & \phantom{0}4 & \phantom{0}89 & \phantom{$-$}0.139 & 0.149 & \phantom{00}2.00 & 0.44 & 0.59 & 0.18 \\
$\lbrack$Mg/Si] & \phantom{0}4 & \phantom{0}89 & \phantom{$-$}0.199 & 0.167 & \phantom{00}0.31 & 0.51 & 0.53 & 0.05 \\
$\lbrack$Mn/Fe] & \phantom{0}4 & \phantom{0}89 &           $-$0.154 & 0.226 & \phantom{0}13.99 & 0.43 & 0.77 & 0.22 \\
$\lbrack$Zn/Fe] & \phantom{0}4 & \phantom{0}89 & \phantom{$-$}0.810 & 0.476 & \phantom{0}60.47 & 0.81 & 1.45 & 0.59 \\
$\lbrack$Cr/Fe] & \phantom{0}4 & \phantom{0}89 & \phantom{$-$}0.164 & 0.192 & \phantom{0}10.94 & 0.46 & 0.82 & 0.26 \\
\hline
\multicolumn{9}{l}{\hspace{0.3cm} DLA toward Q2230+02} \\
\hline
\\[-0.3cm]
$\lbrack$Si/Fe] &           16 & 401 & \phantom{$-$}0.408 & 0.078 & \phantom{00}3.02 & 0.55 & 0.19 & 0.12 \\
$\lbrack$Si/Zn] & \phantom{0}4 & 144 &           $-$0.019 & 0.047 & \phantom{00}0.48 & 0.15 & 0.14 & 0.02 \\
$\lbrack$S/Zn]  & \phantom{0}4 & 144 &           $-$0.033 & 0.137 & \phantom{00}1.77 & 0.32 & 0.39 & 0.10 \\
$\lbrack$O/Si]  & \phantom{0}4 & 370 &           $-$0.593 & 0.303 & \phantom{0}10.27 & 0.72 & 1.31 & 0.43 \\
$\lbrack$S/Si]  & \phantom{0}6 & 144 &           $-$0.011 & 0.101 & \phantom{00}1.63 & 0.36 & 0.26 & 0.09 \\
$\lbrack$N/Si]  & \phantom{0}7 & 161 &           $-$1.003 & 0.122 & \phantom{00}1.75 & 0.33 & 0.20 & 0.07 \\
$\lbrack$N/S]   & \phantom{0}6 & 144 &           $-$1.021 & 0.202 & \phantom{00}2.50 & 0.55 & 0.44 & 0.18 \\
$\lbrack$Mn/Fe] & \phantom{0}6 & 161 &           $-$0.155 & 0.087 & \phantom{00}3.09 & 0.32 & 0.25 & 0.10 \\
$\lbrack$Zn/Fe] & \phantom{0}4 & 144 & \phantom{$-$}0.423 & 0.082 & \phantom{00}1.69 & 0.28 & 0.35 & 0.08 \\
\hline
\multicolumn{9}{l}{\hspace{0.3cm} DLA toward Q2231$-$00 $^{\rm 1}$} \\
\hline
\\[-0.3cm]
$\lbrack$Si/Fe] &           10 & 161 & \phantom{$-$}0.349 & 0.136 & \phantom{00}7.19 & 0.61 & 0.32 & 0.18 \\
$\lbrack$Si/Zn] & \phantom{0}4 & 123 & \phantom{$-$}0.015 & 0.093 & \phantom{00}2.16 & 0.36 & 0.41 & 0.10 \\
$\lbrack$S/Zn]  & \phantom{0}4 & 123 & \phantom{$-$}0.132 & 0.177 & \phantom{00}0.89 & 0.33 & 0.29 & 0.08 \\
$\lbrack$S/Si]  & \phantom{0}8 & 135 & \phantom{$-$}0.233 & 0.259 & \phantom{00}5.83 & 0.89 & 0.57 & 0.26 \\
$\lbrack$Mn/Fe] & \phantom{0}5 & 135 &           $-$0.216 & 0.097 & \phantom{00}5.02 & 0.38 & 0.41 & 0.16 \\ 
$\lbrack$Zn/Fe] & \phantom{0}4 & 123 & \phantom{$-$}0.391 & 0.231 & \phantom{0}14.45 & 0.62 & 1.04 & 0.30 \\
$\lbrack$Cr/Fe] & \phantom{0}5 & 135 & \phantom{$-$}0.066 & 0.110 & \phantom{00}3.43 & 0.39 & 0.39 & 0.12 \\
\hline
\multicolumn{9}{l}{\hspace{0.3cm} DLA toward Q2343+12 $^{\rm 1}$} \\
\hline
\\[-0.3cm]
$\lbrack$Si/Fe] &           19 &           355 & \phantom{$-$}0.515 & 0.208 & \phantom{0}24.37 & 1.09 & 0.45 & 0.33 \\ 
$\lbrack$Si/Zn] & \phantom{0}3 & \phantom{0}35 &           $-$0.045 & 0.021 & \phantom{00}0.25 & 0.08 & 0.12 & 0.00 \\
$\lbrack$O/Si]  &           17 &           305 &           $-$0.153 & 0.265 & \phantom{0}20.86 & 1.12 & 0.52 & 0.37 \\
$\lbrack$N/Si]  & \phantom{0}9 &           305 &           $-$0.810 & 0.465 & \phantom{0}18.54 & 1.00 & 0.77 & 0.36 \\
$\lbrack$Mn/Fe] & \phantom{0}3 & \phantom{0}35 &           $-$0.214 & 0.021 & \phantom{00}0.16 & 0.10 & 0.12 & 0.00 \\
$\lbrack$Zn/Fe] & \phantom{0}3 & \phantom{0}35 & \phantom{$-$}0.623 & 0.056 & \phantom{00}1.53 & 0.17 & 0.39 & 0.06 \\
\hline
\end{tabular}
\begin{minipage}{130mm}
\smallskip
$^{\rm 1}$ DLAs from our first sample, see their metal line profiles in Paper~I. \\
$^{\rm a}$ Number of clouds in a given DLA with a column density measurement for both the elements X and Y. \\
$^{\rm b}$ Velocity range in km~s$^{-1}$ covered by the clouds in the DLA galaxy for a given [X/Y] ratio. \\
\end{minipage}
\end{center}
\end{table*}
%

\section{Cloud-to-cloud chemical variations in individual DLA systems}
\label{cloud-to-cloud}

Having analyzed the cloud-to-cloud variations of the entire set of clouds 
observed in our sample of 11 DLA systems, we would now like to discuss the 
cloud-to-cloud chemical variations in individual DLAs in more detail. We apply 
the same definition of a cloud as in Sect.~\ref{indiv-clouds}. 
\citet{prochaska96} were the first to quantitatively investigate variations in 
the chemical abundances of a single DLA. They compared ionic column densities 
along the observed velocity profile using the apparent optical depth method 
\citep{savage91}. Then, \citet{lopez02} were the first to present a detailed 
cloud-to-cloud analysis for a $z>2$ damped system based on a Voigt profile 
analysis. And more recently, \citet{prochaska03a} performed the chemical 
abundance variation analysis along the sightlines of 13 DLAs. Together, these 
studies argue that the majority of DLAs have very uniform relative abundances. 
In Sect.~\ref{indiv-clouds} we have, however, shown that a high dispersion is
observed when all the clouds of different DLAs are studied together. 

There are three possible sources of abundance variations among clouds: 
(i)~different nucleosynthesis enrichments, i.e. different star formation 
histories from cloud-to-cloud; (ii)~different dust depletion levels, i.e. 
different dust amounts and/or physical conditions from cloud-to-cloud; and
(iii)~different ionization conditions, i.e. different densities (self-shielding) 
and/or ionizing fluxes from cloud-to-cloud. Each of these three sources can be
tested with specific abundance ratios (see Sect.~\ref{introduction}). 
Analyzing these specific abundance ratios in a given DLA galaxy, we tried to
determine the sources leading to cloud-to-cloud variations within a DLA galaxy.

In Table~\ref{comp-analysis} we describe the different abundance ratios 
analyzed and the number of clouds observed within each DLA. [Si/Fe] is the 
only abundance ratio for which a measurement is obtained in all the clouds 
along a given QSO sightline. In columns (4), (5), and (6) we give the first 
simple statistical results for the cloud-to-cloud abundance variations, namely 
the logarithmic weighted mean computed using the 1\,$\sigma$ errors as weights, 
the logarithmic RMS dispersion, and the reduced $\chi^2$ relative to the 
weighted mean, respectively. In addition, we performed a series of Monte-Carlo 
simulations, using the technique designed by \citet{prochaska03a}, to 
investigate the deviations allowed within each DLA and to derive quantitative 
limits and values of the cloud-to-cloud variations relative to the weighted 
mean. This technique consists of making three measures of the uniformity within 
individual DLAs. With one measure, we study an extreme scenario where all of 
the variation arises from a single cloud. In column (7) we give the values of 
$\Delta_{sngl}$, the minimum [X/Y] variation in a single cloud, which gives 
$\chi^2_{\rm M-C} > \chi^2_{\rm obs}$ in over 95\,\% of the 1000 trials in the 
Monte-Carlo analysis. The two other measures assume deviations in all the 
clouds with values drawn from a uniform distribution. In columns (8) and (9) we 
calculate, respectively, $\Delta_{all}$, the minimum variation which when 
applied to every cloud gives $\chi^2_{\rm M-C} > \chi^2_{\rm obs}$ in over 
95\,\% of the trials, and $\Delta_{best}$, the variation which when applied to 
each cloud has the highest probability of yielding $\chi^2 = 
\chi^2_{\rm obs}\pm 10$\,\%. The $\Delta_{all}$ values are the most realistic 
upper limits to abundance variations and $\Delta_{best}$ reflects the most 
likely value \citep[for more details, see][]{prochaska03a}.

%

\begin{figure}[t]
\centering
\includegraphics[width=9cm]{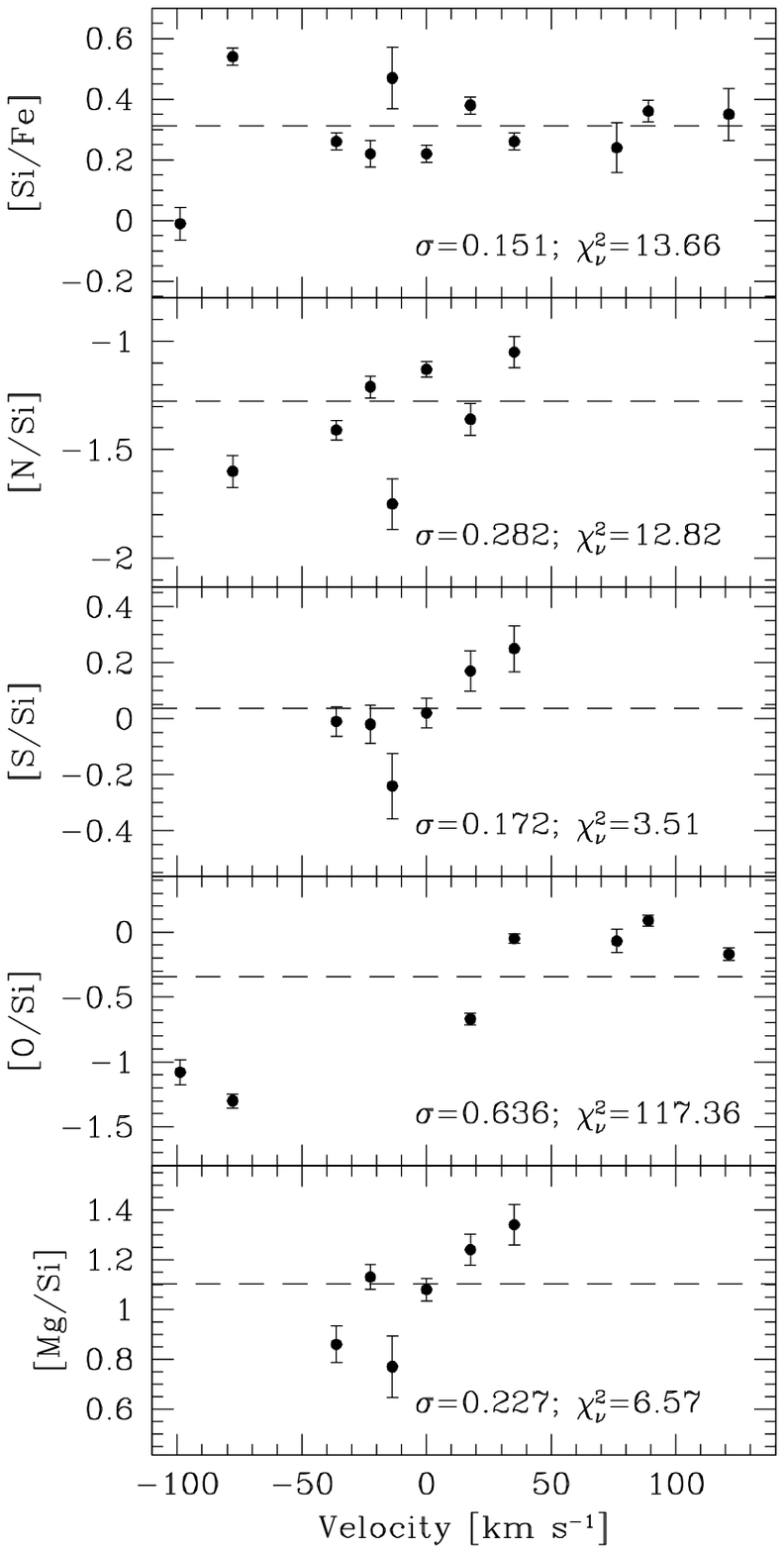}
\caption{Cloud-to-cloud [X/Y] abundance ratios in the DLA toward Q0450$-$13. 
A total number of 11 clouds covering 232 km~s$^{-1}$ in velocity space are
detected in this DLA along the QSO line of sight. The dashed line corresponds 
to the logarithmic weighted mean of the cloud-to-cloud [X/Y] ratios computed 
by using their 1\,$\sigma$ errors as weights. $\sigma$ indicates the 
logarithmic RMS dispersion in [X/Y], and $\chi_{\nu}^2$ the reduced $\chi^2$ 
relative to the weighted mean. In Table~\ref{comp-analysis} we give additional 
statistical information essential for an analysis of the amplitude of the 
cloud-to-cloud chemical variations in individual DLAs.}
\label{Q0450-comp}
\end{figure}
%

We performed the cloud-to-cloud chemical variation analysis only in the DLA
systems in which the majority of the abundance ratios considered can be 
measured in at least three clouds. For this reason, the DLAs toward Q0841+129,
Q2348$-$1444, and Q0100+13 were excluded. We now briefly discuss the
cloud-to-cloud abundance ratios for the seven remaining DLAs studied. For each 
analyzed DLA, we plot the cloud-to-cloud abundance ratios as a function of the 
velocity of the clouds.

%

\begin{figure}[t]
\centering
\includegraphics[width=9cm]{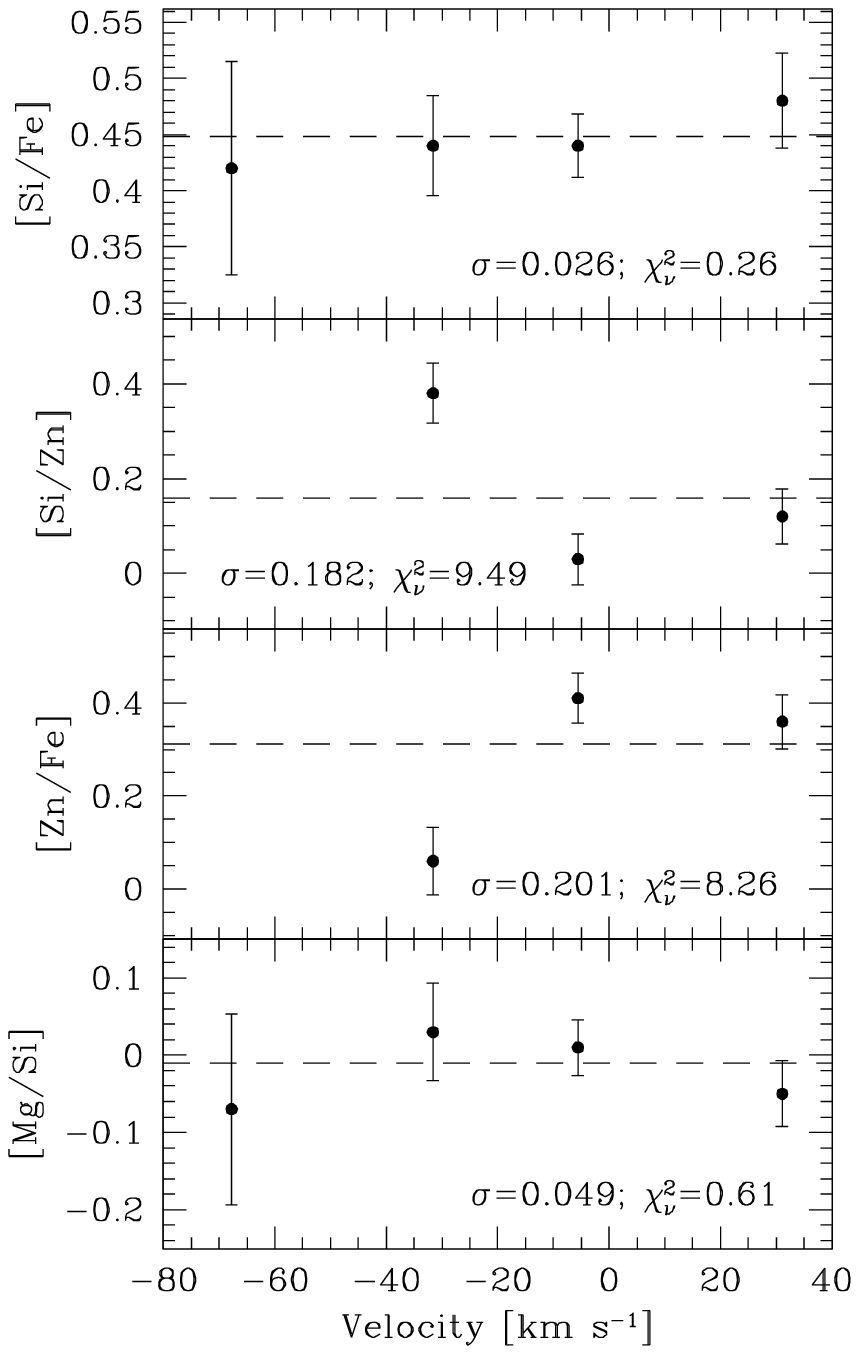}
\caption{Same as Fig.~\ref{Q0450-comp} for the DLA toward Q1157+014. A total
number of 4 clouds covering 144 km~s$^{-1}$ in velocity space are detected in
this DLA along the QSO line of sight.}
\label{Q1157-comp}
\end{figure}
%

\begin{figure}[t]
\centering
\includegraphics[width=9cm]{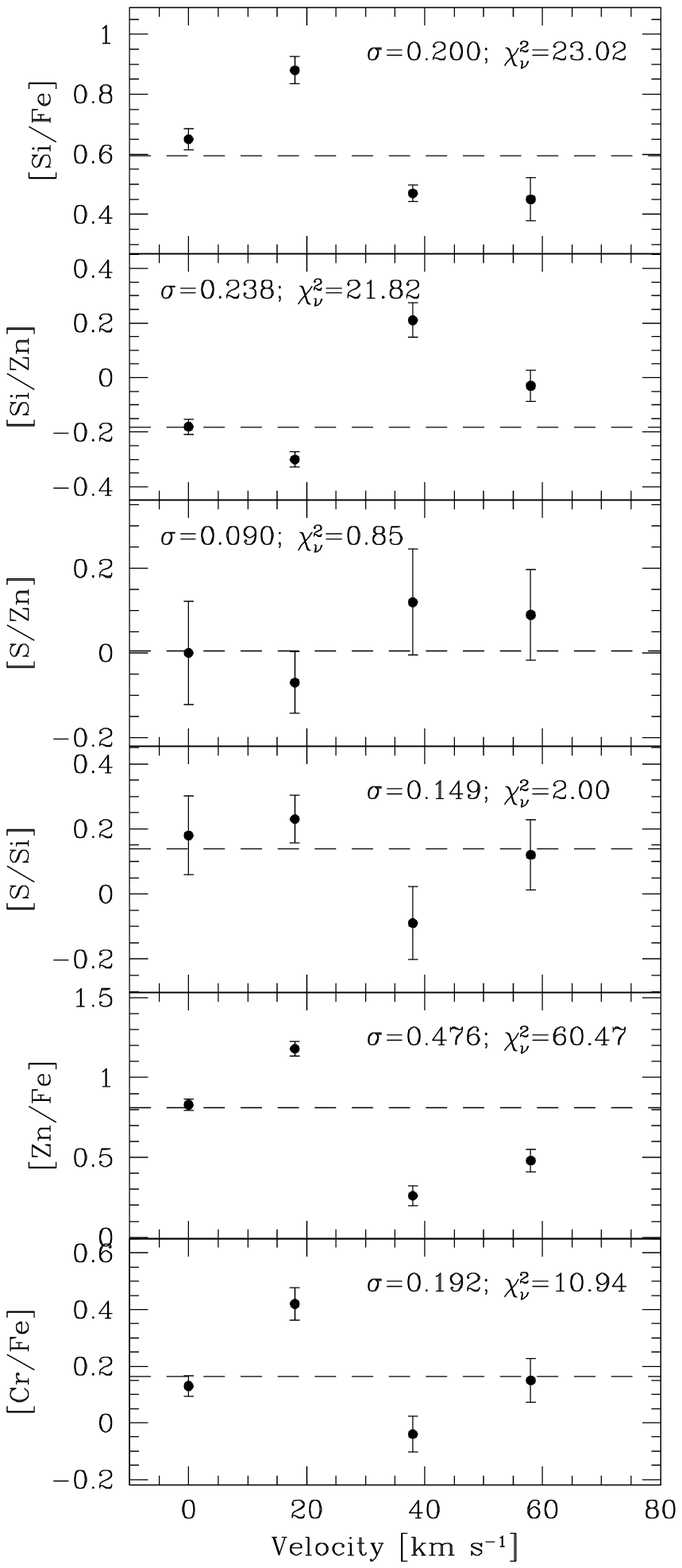}
\caption{Same as Fig.~\ref{Q0450-comp} for the DLA toward Q1331+17 from the
first sample of DLAs (see Paper~I). A total number of 4 clouds covering 89 
km~s$^{-1}$ in velocity space are detected in this DLA along the QSO line of 
sight.}
\label{Q1331-comp}
\end{figure}
%

\noindent {\it DLA toward Q0450$-$13.}\hspace{0.2cm} Figure~\ref{Q0450-comp} 
shows the cloud-to-cloud abundance ratios of this DLA. A large variation, larger 
than 0.2~dex, was observed in all the relative abundances analyzed in this DLA 
(see Table~\ref{comp-analysis}). The most impressive is the cloud-to-cloud 
variation of the [O/Si] ratio observed at 10\,$\sigma$ with $\Delta_{best}= 
0.75$~dex. It most likely results from a high variation in the ionization level 
from cloud-to-cloud. In the clouds at $v > +30$ km~s$^{-1}$, we observe solar 
[O/Si] ratios and in the clouds around $-90$ km~s$^{-1}$, the [O/Si] ratio is 
highly undersolar reaching $-1.5$~dex and indicating a high ionization level. 
The presence of strong ionization signatures in this DLA was already pointed 
out in Sect.~\ref{Q0450} with the detection of strong intermediate-ion 
lines, \ion{Fe}{iii} and \ion{N}{ii}. The deviation from uniformity observed in 
the other abundance ratios with $\Delta_{best} > 0.2$~dex is very interesting, 
all the more since a similar trend of increasing abundance ratios from bluer 
to redder clouds is observed. The fact that this trend is particularly 
pronounced in the [N/Si] ratio with a 3\,$\sigma$ variation confirms that it is
due to a variation in the ionization level from cloud-to-cloud, and it provides
constraints on the geometry of the ionizing flux in this DLA galaxy (see 
further details in Paper~IV).

%

\begin{figure}[t]
\centering
\includegraphics[width=9cm]{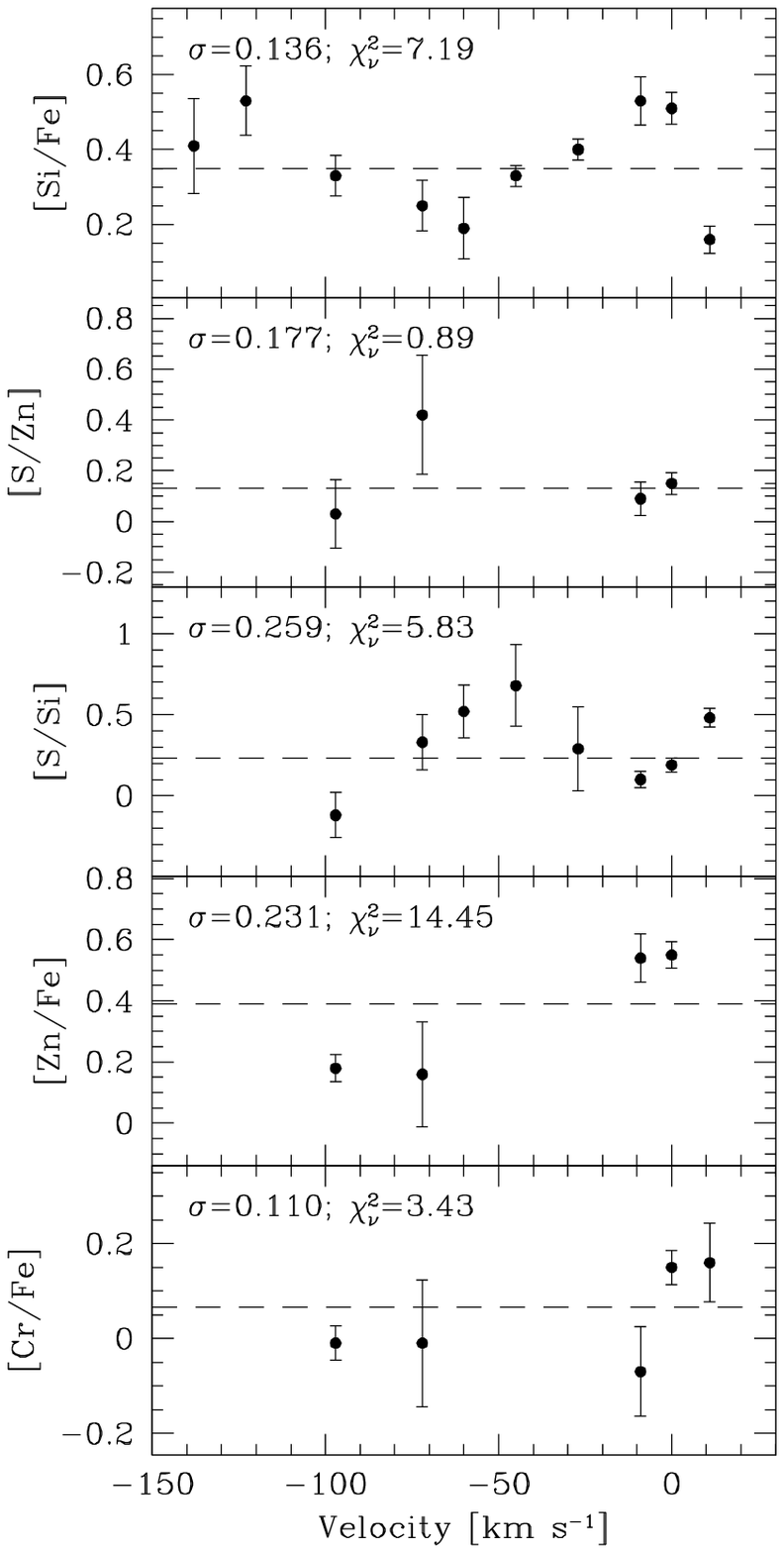}
\caption{Same as Fig.~\ref{Q0450-comp} for the DLA toward Q2231$-$00 from the
first sample of DLAs (see Paper~I). A total number of 10 clouds covering 161 
km~s$^{-1}$ in velocity space are detected in this DLA along the QSO line of 
sight.}
\label{Q2231-comp}
\end{figure}
%

\noindent {\it DLA toward Q1157+014.}\hspace{0.2cm} In Fig.~\ref{Q1157-comp}
we show the cloud-to-cloud abundance ratio plots for this DLA. The [Si/Fe] 
and [Mg/Si] ratios are perfectly uniform, while [Zn/Fe] and [Si/Zn] show 
variations at the 3\,$\sigma$ level (see Table~\ref{comp-analysis}). This 
variation is mainly due to one cloud at $-32$ km~s$^{-1}$ that has a lower 
[Zn/Fe] value and a higher [Si/Zn] value than the other two clouds with [Zn/Fe] 
and [Si/Zn] measurements. A careful examination of the \ion{Zn}{ii}, 
\ion{Fe}{ii}, and \ion{Si}{ii} profiles of this DLA already suggests some clues 
for variation. Indeed, the optical depth of component~3 (corresponding to 
the cloud at $-32$ km~s$^{-1}$) is much deeper in the \ion{Fe}{ii} and 
\ion{Si}{ii} lines than in the \ion{Zn}{ii} lines in comparison with the redder 
components. This variation can be assigned to a dust depletion variation from 
cloud-to-cloud. Indeed, the deeper \ion{Fe}{ii} and \ion{Si}{ii} optical depths 
in the cloud at $-32$ km~s$^{-1}$ indicate a lower dust amount. As a 
consequence, the abundances of Fe and Si, two refractory elements, are not 
depleted relative to Zn, a volatile element, in this cloud, while they are 
depleted in the clouds at $-5$ and $+31$ km~s$^{-1}$, in which we measure an 
enhanced [Zn/Fe] ratio and a low [Si/Zn] ratio.

%

\begin{figure}[t]
\centering
\includegraphics[width=9cm]{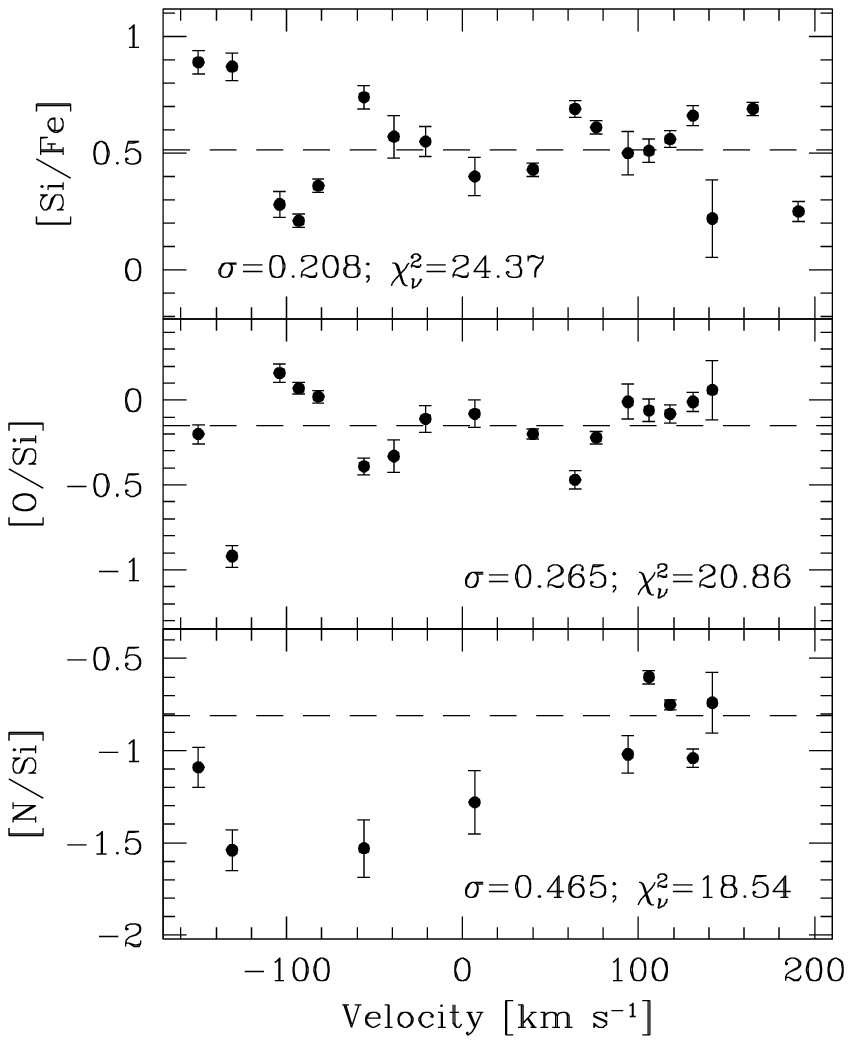}
\caption{Same as Fig.~\ref{Q0450-comp} for the DLA toward Q2343+12 from the
first sample of DLAs (see Paper~I). A total number of 19 clouds covering 355 
km~s$^{-1}$ in velocity space are detected in this DLA along the QSO line of 
sight.}
\label{Q2343-comp}
\end{figure}
%

\noindent {\it DLAs toward Q1210+17 and Q2230+02.}\hspace{0.2cm} For these 
DLAs, we do not show the cloud-to-cloud abundance ratio plots, because all their 
relative abundances are uniform to better than 0.1~dex (see 
Table~\ref{comp-analysis}). This reveals an important characteristic of these 
two high-redshift galaxies: the gas clouds which comprise these galaxies 
apparently have very similar physical properties (mainly when considering the 
differential dust depletion) and nucleosynthetic enrichment histories. Only the 
[O/Si] ratio in the DLA toward Q2230+02 seems to show a 3\,$\sigma$ departure 
from uniformity. This may indicate a variation in the ionization properties 
within this DLA galaxy, between the cloud observed at $-147$ km~s$^{-1}$ which 
appears to be more ionized, and the three clouds observed around $+180$ 
km~s$^{-1}$.

\noindent {\it DLA toward Q1331+17.}\hspace{0.2cm} In Fig.~\ref{Q1331-comp}
we show the cloud-to-cloud abundance ratio plots for this DLA. A high variation 
in the [Zn/Fe] ratio is observed at more than 7\,$\sigma$ with $\Delta_{best} = 
0.59$~dex. This variation is due to differential dust depletion variations from
cloud-to-cloud, with the clouds at $+38$ and $+58$ km~s$^{-1}$ being less
depleted (having lower [Zn/Fe] ratios) than the clouds at $-2$ and $+18$ 
km~s$^{-1}$. This DLA exhibits, in fact, one of the largest dust depletion level 
of any DLA. The presence of a high amount of dust in the clouds at $-2$ and 
$+18$ km~s$^{-1}$ is also favored by the detections of C$^0$ and Cl$^0$ in 
these clouds (see Fig.~4 in Paper~I). Indeed, these ions are usually associated 
with a dense, cold neutral medium, characteristic of highly depleted gas in the 
Milky Way. Furthermore, they generally suggest at least a modest molecular 
hydrogen fraction, which is indicative of dust. This strong cloud-to-cloud dust 
amount variation is also responsible for variations observed in: (i)~[Si/Fe], 
with Fe being more depleted than Si; (ii)~[Si/Zn], with Si being more 
depleted than the very mildly refractory Zn; (iii)~[Cr/Fe], with Fe being more
depleted than Cr; and (iv)~[S/Si], with Si being more depleted than the 
volatile S. Different SFHs from cloud-to-cloud do not seem to contribute to the 
[Si/Fe] and [Si/Zn] variations, since the [S/Zn] ratio is perfectly uniform. 
Interestingly, the [S/Zn] ratio is almost solar in all clouds, even though 
[Si/Fe] $\sim +0.4$~dex in clouds where [Zn/Fe] is low. This joins the comment 
made in Sect.~\ref{indiv-clouds} on the use of [S/Zn] as a nucleosynthesis 
enrichment indicator.

\noindent {\it DLA toward Q2231$-$00.}\hspace{0.2cm} In 
Fig.~\ref{Q2231-comp} we show the cloud-to-cloud abundance ratio plots for this
DLA. The departures from uniformity observed in this system are modest, lower 
than 0.2~dex (see Table~\ref{comp-analysis}). Only the [Zn/Fe] ratio seems to 
show a higher variation with $\Delta_{best} = 0.30$~dex and may reflect some 
cloud-to-cloud differential dust depletion variations. However, this does not 
affect the uniformity of the other abundance ratios strongly. We observe only 
a slight effect of a higher dust depletion level in the clouds at $-9$ and 
0 km~s$^{-1}$ compared to the clouds at $-97$ and $-72$ km~s$^{-1}$ on the 
[Si/Fe] and [S/Si] ratios. No sign of cloud-to-cloud SFH variations is detected, 
the [S/Zn] ratio is uniform.

\noindent {\it DLA toward Q2343+12.}\hspace{0.2cm} In Fig.~\ref{Q2343-comp}
we show the cloud-to-cloud abundance ratio plots for this DLA. The possible 
variations due to dust depletion and nucleosynthesis enrichment are difficult 
to highlight in this system, partly because many elements are detected in three 
clouds only (see Table~\ref{comp-analysis}). However, we observe high deviations 
from uniformity at more than 4\,$\sigma$ with $\Delta_{best} > 0.3$~dex in the 
cloud-to-cloud [Si/Fe], [O/Si], and [N/Si] ratios. They are likely to be 
dominated by ionization variations. Hints of a possible trend from 
cloud-to-cloud of higher [Si/Fe] ratios, when [O/Si] is lower resulting from 
ionization effects, can be suggested from our measurements. The presence of a 
high ionization level in this DLA was already pointed out in Paper~I with the 
detection of strong intermediate-ion lines, \ion{Fe}{iii}, \ion{N}{ii}, and 
\ion{S}{iii}, and will be discussed further in Paper~IV.

In summary, among the seven DLA systems for which we analyzed the cloud-to-cloud
chemical variations, five of them do show statistically significant 
cloud-to-cloud variations, namely higher than 0.2~dex (RMS $>0.2$~dex and 
$\Delta_{best} > 0.2$~dex) at more than 3\,$\sigma$, for at least two different 
abundance ratios. But, only two DLA systems toward Q0450$-$13 and Q1331+17 show
``extreme'' variations, that is, a higher dispersion than 0.3~dex at more than 
7\,$\sigma$. We were able to identify the sources of these variations thanks to 
the analysis of specific abundance ratios. These sources are either the 
differential dust depletion variations and/or the ionization condition 
variations from cloud-to-cloud. But, no evidence for variations due to 
different SFHs was highlighted. This suggests that the gas clouds within some 
DLA galaxies have different physical properties, namely different dust 
depletion and/or ionization levels, but they all seem to show a uniform 
nucleosynthetic enrichment history. 

In the \citet{prochaska03a} study based on a single abundance ratio, mainly 
[Si/Fe], only 2 out of 13 DLAs present cloud-to-cloud variations. Our work 
already shows a less uniform picture for the ISM of high-redshift DLA galaxies, 
since 5 out of 7 DLAs show chemical variations, and all DLAs except one have at 
least a high $\Delta_{sngl}$ ($> 0.2$~dex) for one abundance ratio indicating 
that there is at least a single cloud with an ``abnormal'' abundance. 
The main difference between these two studies comes from the fact that in 
our chemical variation analysis we considered other abundance ratios than 
[Si/Fe] alone. Given that the observed variations seem to mainly be due to dust 
depletion and ionization effects, as stated above, they are thus well-detected 
in the ratios tracing the dust depletion variations, i.e. [Zn/Fe], and the 
ionization variations. Consideration of the [Si/Fe] ratio alone would have 
shown variations in only 3 out of 7 DLAs, so that we would have missed 
variations in 2 DLA galaxies.

%

\section{Summary and concluding remarks}
\label{conclusions}

Analysis of our sample of damped Ly$\alpha$ systems has proved once again 
that these systems constitute the best laboratory for studying the chemical 
abundances and the interstellar medium properties of high redshift galaxies. We 
obtained new comprehensive sets of elemental abundances of seven DLAs in the 
redshift range $z_{\rm abs} = 1.8-2.5$ toward bright quasars. These were derived 
from UVES/VLT spectra combined with existing HIRES/Keck spectra. We detected 54
metal-line transitions, and obtained the column density measurements of 30 ions 
from 22 elements, $-$~B, C, N, O, Mg, Al, Si, P, S, Cl, Ar, Ti, Cr, Mn, Fe, Co,
Ni, Cu, Zn, Ge, As, Kr. Together with our first sample of four DLAs analyzed in
Paper~I, we have a sample of eleven DLA galaxies with uniquely comprehensive 
and homogeneous abundance measurements. 

In this paper (II in the series) we were able to study the abundance patterns 
and the chemical variations of a wide range of elements in the interstellar 
medium of galaxies outside the Local Group, all for the first time. Chemical 
variations from DLA galaxy to DLA galaxy and from cloud to cloud in the ISM 
within a galaxy are expected, if the line of sight to a distant QSO samples 
regions with different star formation histories and different ISM conditions. 
This is suggested by observations in the Milky Way, the Small and Large 
Magellanic Clouds, and dwarf spheroidal galaxies which show chemical variations 
from galaxy to galaxy and within the galaxy when comparing their stellar and 
gas-phase abundance patterns. We considered three potential sources of observed 
abundance variations: the star formation history (nucleosynthesis enrichment), 
differential dust depletion, and ionization.
 
\noindent Our main results can be summarized as follows:

1) The abundance patterns of the integrated profiles of DLAs show relatively low
RMS dispersions, reaching only 2--3 times higher values than the statistical 
errors, for the majority of elements. This uniformity is remarkable given that 
the quasar sightlines cross gaseous regions with \ion{H}{i} column densities 
covering an order of magnitude from $2\times 10^{20}$ to $4\times 10^{21}$ 
cm$^{-2}$ and with metallicities ranging from 1/55 to 1/5 solar. 
This implies that the respective star formation histories, if ever different, 
have conspired to yield one set of relative abundances and that the effects of
nucleosynthesis enrichment, dust depletion, and ionization are negligible. We 
discuss the implications of this uniformity element by element. The most 
interesting findings are: (i)~the [Si/Fe] and [S/Fe] abundance ratios show an 
$\alpha$-enhancement irrespective of the dust depletion level of the DLAs 
studied; to reconcile this $\alpha$-enhancement with the solar [Si/Zn] ratios, 
we suggest that the [Zn/Fe] ratios are intrinsically oversolar (independent of 
dust depletion effects) as indicated by the recent measurements in 
Galactic metal-poor stars; (ii)~all the Ar abundance measurements show a 
significant underabundance relative to Si; however, the [Ar/Si] ratio alone 
cannot characterize the ionization of a DLA, since DLAs with low [Ar/Si] ratios 
include cases where other ionization diagnostics imply the gas is predominantly 
neutral; (iii)~all the Mn abundance measurements are underabundant relative 
to Fe irrespective of the dust depletion of the DLAs studied; they reach a 
plateau at $-0.23$~dex; and (iv)~we obtained the second abundance measurement 
of Ge, an element beyond the iron-peak, in the metal-strong DLA toward 
Q1157+014, opening the way to investigation of s-process elements in DLAs.

2) From the Voigt profile fitting of metal-lines, we obtained very accurate 
component-to-component column density measurements. These components presumably 
correspond to interstellar medium clouds in the DLA galaxy on the QSO line of 
sight. By considering all the clouds of all the DLAs studied together, we 
see a statistically significant dispersion in several abundance ratios, for 
example about 5\,$\sigma$ in the [Zn/Fe] ratios and larger than 3\,$\sigma$ 
in the [$\alpha$/Fe,Zn] ratios. This indicates that the chemical variations are 
more confined to individual clouds within the DLA galaxies rather than to 
integrated profiles. If confirmed, this will provide important constraints on 
the understanding of the ISM of high-redshift galaxies and the enrichment of 
gas in the early Universe. We found unambiguous correlations between [Si/Fe], 
[S/Fe], and [S/Si] versus [Zn/Fe], and anti-correlations between [Si/Zn] and 
[S/Zn] versus [Zn/Fe]. These trends are primarily the result of differential 
dust depletion effects, which are also responsible for the high cloud-to-cloud 
abundance ratio dispersion. The signature of the pure nucleosynthesis 
enrichment contribution can be observed in the [$\alpha$/Fe,Zn] ratios at low 
dust depletion levels, [Zn/Fe] $\leq 0.2$. It is characterized by an 
$\alpha$-enhancement in the clouds. However, while the [S/Fe], [S/Fe], and 
[Si/Zn] ratios are highly $\alpha$-enhanced ($> 0.2$~dex), the [S/Zn] ratio 
remains almost solar, suggesting that [S/Zn] may not be a reliable tracer of
nucleosynthesis enrichment, the production of S being perhaps in some way 
correlated with the one of Zn.

3) We analyzed several specific abundance ratios in individual DLA galaxies to 
try to determine the origin of cloud-to-cloud chemical variations when observed 
in a given DLA galaxy. Study of the cloud-to-cloud chemical variations 
within seven individual DLA systems revealed that five of them show 
statistically significant variations, higher than 0.2~dex at more than 
3\,$\sigma$. Two of them show ``extreme'' variations with a dispersion higher 
than 0.3 dex at more than 7\,$\sigma$. The sources of these variations are 
either the differential dust depletion and/or the ionization effects, while 
there is no evidence of variations due to different star formation histories. 
This suggests that the gas clouds within DLA galaxies have different physical 
properties, but they all seem to show a uniform nucleosynthetic enrichment 
history. 

At lower redshift ($z_{\rm abs} < 1$), deep imaging shows that DLA galaxies
are a heterogeneous group that exhibits a variety of morphologies and surface 
brightnesses. If this is also the case at high redshift, we may have expected 
higher dispersions in the DLA abundance ratios. The uniformity observed in the 
global gas-phase abundance patterns of DLAs is thus even more surprising. 
Perhaps we are penalized by the small number statistics or, more important, 
there may be fundamental differences between high and low redshift DLAs. The 
results by \citet{kanekar03} may be relevant here. Indeed, they derived 
estimates of the spin temperature, $T_{\rm s}$, in 24 DLAs and found that all 
DLAs with high spin temperatures, $T_{\rm s} > 1000$~K, are identified with 
dwarf or low surface brightness galaxies, while DLAs with low $T_{\rm s}$ are 
associated with large, luminous galaxies. Interestingly, they observed that low 
redshift DLAs have both high and low values of $T_{\rm s}$, while high redshift 
DLAs ($z_{\rm abs} > 2$) have preferentially high $T_{\rm s}$. This result could 
help to understand the uniformity observed in the abundance ratios of DLAs in 
our sample. 

The uniformity in the nucleosynthesis enrichment observed in the clouds within 
DLAs is also surprising and poses important constraints on the formation of 
high-redshift galaxies. Indeed, one very promising scenario that allows us to 
explain the DLA kinematics is within the CDM hierarchical cosmology and 
describes a DLA as multiple merging ``clumps'' bound to individual dark matter 
halos \citep[e.g.][]{haehnelt98,maller01}. In that scenario, the uniformity of 
cloud-to-cloud abundance ratios within a DLA galaxy constrains the abundances 
of all of the protogalactic clumps making up a DLA system. In terms of
nucleosynthesis enrichment, this implies the clumps share a similar chemical 
enrichment pattern. This represents a challenge for the CDM simulations, since 
those protogalactic clumps which do not share a common gas reservoir and which 
merge over a large timescale would not be expected to necessarily have a 
unique enrichment history and be at the same stage of chemical evolution. It 
remains to be demonstrated whether these clumps really do express very similar 
nucleosynthetic enrichment patterns with, in addition, very different dust 
depletions in some cases. This also places strict constraints on the mixing
timescales of protogalaxies. Of course, access to the absolute values of 
metallicities of these clouds would help to confirm these statements. For this, 
however, we need to measure the \ion{H}{i} column densities of individual 
clouds within a DLA, which is not possible from the current observations.

%

\begin{acknowledgements}

The authors wish to thank everyone working at ESO/Paranal for the high quality 
of UVES spectra obtained in service mode. M.D.-Z. is supported by the Swiss 
National Funds and extends special thanks to Professor A.~Maeder for continuous 
encouragement. M.D.-Z. is grateful to the UCO/Lick Observatory for hosting her 
in Santa Cruz in September 2004, where the layout and main results of this 
paper were worked out. J.X.P. acknowledges support through the NSF grant 
AST~03-07824.

\end{acknowledgements}

%

\appendix

\section{An astrophysical determination of the 
{\rm\bf Ni}\,{\sc\bf II}\,\mathversion{bold}$\lambda$\mathversion{normal}\,1317 
oscillator strength} 
\label{appendix}
 
In the analysis of ionic column densities of damped Ly$\alpha$ systems, we 
noticed that we derive a systematically lower Ni$^+$ column density measurement
from the \ion{Ni}{ii}\,$\lambda$\,1317 line than from the \ion{Ni}{ii} lines at 
$\lambda_{\rm rest} = 1454$, 1709, 1741, and 1751 \AA, when all these lines are 
available in the same DLA system. The simplest way to explain this discrepancy 
is to call the reliability of the oscillator strength of the 
\ion{Ni}{ii}\,$\lambda$\,1317 line into question. The most recent atomic data 
published for the \ion{Ni}{ii} transitions are summarized in Table~\ref{NiII}. 
In this table we see that \citet{morton91} atomic data of the 
\ion{Ni}{ii}\,$\lambda$\,1454,\,1709,\,1741,\,1751 lines have been updated by 
\citet{fedchak00}, while no new measurement of the $f$-value exists for the 
\ion{Ni}{ii}\,$\lambda$\,1317 line.

Our UVES/VLT spectra, combined with HIRES/Keck spectra, allowed us to observe 
the complete sample of \ion{Ni}{ii} UV transitions in several DLAs with a 
resolution and a signal-to-noise ratio sufficient for accurate analysis. We 
therefore decided to use these data to derive an astrophysical oscillator 
strength of the \ion{Ni}{ii}\,$\lambda$\,1317 transition.

The adopted method to determine this $f$-value is relatively simple. We assumed
that the component-to-component column densities derived from the Voigt profile
fitting of the \ion{Ni}{ii} lines at $\lambda_{\rm rest} = 1454$, 1709, 1741, 
and 1751 \AA\ have to be the same as those derived from the 
\ion{Ni}{ii}\,$\lambda$\,1317 line, and we varied the 
\ion{Ni}{ii}\,$\lambda$\,1317 $f$-value by requiring that this line yield the 
same component-to-component column densities as the other \ion{Ni}{ii} lines.
More concretely, we determined the best value of the 
\ion{Ni}{ii}\,$\lambda$\,1317 oscillator strength via a $\chi^2$ minimization 
of theoretical Voigt profile fits to the observed absorption 
\ion{Ni}{ii}\,$\lambda$\,1317 line. We fixed the fitting parameters, the 
component-to-component $b$-values and redshifts, to the values of the 
best-fitting solution obtained for the low-ion lines, and the 
component-to-component Ni$^+$ column densities to the values obtained from the 
Voigt profile fitting of \ion{Ni}{ii} lines other than the 
\ion{Ni}{ii}\,$\lambda$\,1317 line (see 
Tables~\ref{Q0450-Ntable}--\ref{Q2348-Ntable}).

We performed this analysis in five DLA systems, in which we detect, with high
accuracy, at least two \ion{Ni}{ii} lines besides the 
\ion{Ni}{ii}\,$\lambda$\,1317 line. We do not consider the 
\ion{Ni}{ii}\,$\lambda$\,1370 line in this analysis, since it also seems 
to show some discrepancy to the best-fit solution obtained from the 
\ion{Ni}{ii}\,$\lambda$\,1454,\,1709,\,1741,\,1751 lines. The five selected DLAs
are the DLA at $z_{\rm abs} = 2.375$ toward Q0841+129, the DLA toward Q1157+014,
the DLA toward Q1210+17, the DLA toward Q2230+02, and the DLA toward Q2231$-$00
studied in our first sample of DLAs (see Paper~I). Their accessible 
\ion{Ni}{ii} lines, with their corresponding profiles, can be found in 
Figs.~\ref{Q0841-2p375-metals}, \ref{Q1157-metals}, \ref{Q1210-metals}, and
\ref{Q2230-metals}, and Fig.~6 of Paper~I, respectively. 
%

\begin{table}[t]
\begin{center}
\caption{Atomic data of the \ion{Ni}{ii} transitions}
\label{NiII} 
\begin{tabular}{l c c l}
\hline \hline
\\[-0.3cm]
Transition & $\lambda_{\rm rest}$ & $f$ & Reference
\smallskip 
\\ 
\hline\\[-0.3cm]
\ion{Ni}{ii}\,$\lambda$\,1317 & 1317.2170 & 0.07786 & \citet{morton91} \\
\ion{Ni}{ii}\,$\lambda$\,1370 & 1370.1310 & 0.07690 & \citet{fedchak99} \\
\ion{Ni}{ii}\,$\lambda$\,1454 & 1454.8420 & 0.03230 & \citet{fedchak00} \\
\ion{Ni}{ii}\,$\lambda$\,1709 & 1709.6042 & 0.03240 & \citet{fedchak00} \\
\ion{Ni}{ii}\,$\lambda$\,1741 & 1741.5531 & 0.04270 & \citet{fedchak00} \\
\ion{Ni}{ii}\,$\lambda$\,1751 & 1751.9156 & 0.02770 & \citet{fedchak00} \\
\hline
\end{tabular}
\end{center}
\end{table}
%

\begin{table}[t]
\begin{center}
\caption{The new \ion{Ni}{ii}\,$\lambda$\,1317 oscillator strength}
\label{NiII-values} 
\begin{tabular}{l c}
\hline \hline
\\[-0.3cm]
DLA system & New $f$ 
\smallskip 
\\ 
\hline\\[-0.3cm]
$z_{\rm abs} = 2.375$ toward Q0841+129  	    & $0.052\pm 0.010$ \\
$z_{\rm abs} = 1.944$ toward Q1157+014  	    & $0.064\pm 0.007$ \\
$z_{\rm abs} = 1.892$ toward Q1210+17		    & $0.052\pm 0.007$ \\
$z_{\rm abs} = 1.864$ toward Q2230+02		    & $0.054\pm 0.006$ \\
$z_{\rm abs} = 2.066$ toward Q2231$-$00 $^{\dagger}$& $0.064\pm 0.008$ \\
\hline\\[-0.3cm]
\multicolumn{2}{c}{$\langle$ $f_{\rm new}$(\ion{Ni}{ii}\,$\lambda$\,1317) $\rangle$ $= 0.057\pm 0.006$} \\
\\[-0.3cm]
\hline
\end{tabular}
\begin{minipage}{65mm}
\smallskip
$^{\dagger}$ DLA studied in Paper~I.
\end{minipage}
\end{center}
\end{table}
%

The new \ion{Ni}{ii}\,$\lambda$\,1317 oscillator strength results are 
illustrated in Fig.~\ref{NiII-plots} and summarized in Table~\ref{NiII-values}. 
Instead of showing the derived $f$-values through the best Voigt profile 
fitting solutions obtained for the \ion{Ni}{ii} lines, we present the results 
by plotting the apparent column density $N_{\rm a}(v)$ profiles for the 
analyzed \ion{Ni}{ii} lines. The apparent ionic column density for each pixel, 
$N_{\rm a}(v)$, is defined by:
$$N_{\rm a}(v) = 
\frac{m_{\rm e} c}{\pi {\rm e}^2} \frac{\tau_{\rm a}(v)}{f \lambda}~,$$
where $\tau_{\rm a}(v) = \ln[I_{\rm i}(v)/I_{\rm a}(v)]$, $f$ is the oscillator
strength, $\lambda$ the rest-wavelength, and $I_{\rm i}$ and $I_{\rm a}$ the 
incident intensity or, in other words, the continuum intensity ($I_{\rm i}(v) 
= 1$ in normalized spectra) and the measured intensity, respectively. For each 
of the five selected DLAs, we show the apparent column density $N_{\rm a}(v)$ 
profiles of the \ion{Ni}{ii}\,$\lambda$\,1709,\,1741,\,1751 lines overplotted 
by a shaded profile (see Fig.~\ref{NiII-plots}). This shaded profile corresponds 
to the \ion{Ni}{ii}\,$\lambda$\,1317 apparent column density profile which was 
computed with the best result derived in each DLA individually for the new 
\ion{Ni}{ii}\,$\lambda$\,1317 oscillator strength and its 1\,$\sigma$ error. 
We can observe that this shaded profile overlaps the apparent column density 
profiles of the \ion{Ni}{ii} lines at $\lambda_{\rm rest} = 1709$, 1741, and 
1751 \AA\ very well with exceptions for a few velocity components: the 
component at $v=-20$ km~s$^{-1}$ in Fig.~\ref{NiII-plots}\,b), the components 
at $v=+42$ and $+74$ km~s$^{-1}$ in Fig.~\ref{NiII-plots}\,d), and the component 
at $v=+11$ km~s$^{-1}$ in Fig.~\ref{NiII-plots}\,e). The thick solid profile 
corresponds to the \ion{Ni}{ii}\,$\lambda$\,1317 $N_{\rm a}(v)$ profile 
obtained with the old $f$-value from \citet{morton91}. It clearly shows the
 disagreement of this old $f$-value with the $N_{\rm a}(v)$ profiles of the 
\ion{Ni}{ii}\,$\lambda$\,1709,\,1741,\,1751 lines. 

The derived final new \ion{Ni}{ii}\,$\lambda$\,1317 oscillator strength is:
$$ f_{\rm new}(\textrm{Ni}\,\textsc{ii}\,\lambda\,1317) = 0.057\pm 0.006~.$$
It was obtained by computing a weighted mean of the five $f$-values derived in 
the five selected DLA systems (see Table~\ref{NiII-values}). The error 
corresponds to the RMS dispersion of the five $f$-values.
%

\begin{figure*}[!]
\centering
\includegraphics[width=12.7cm]{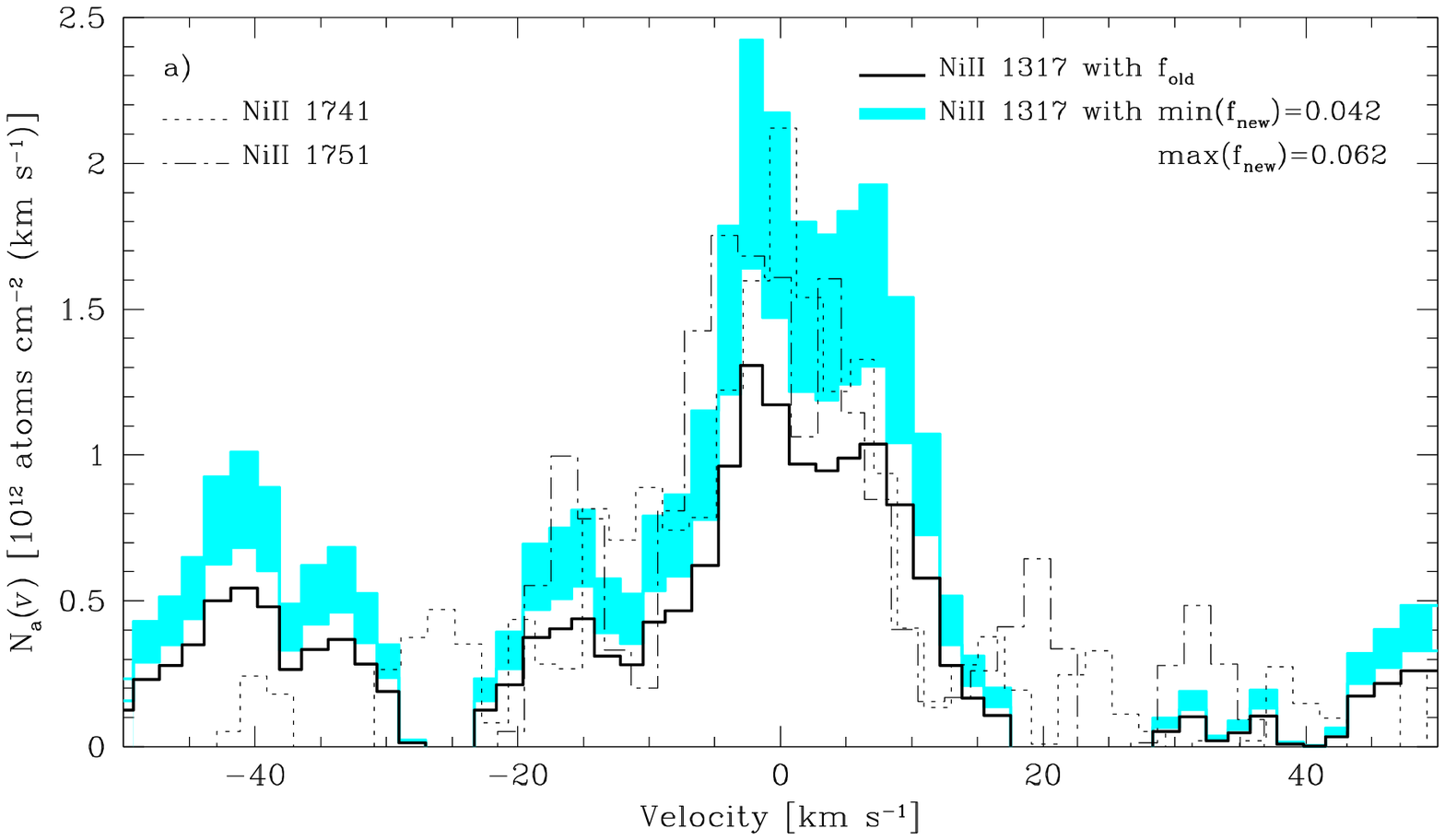}
\includegraphics[width=12.7cm]{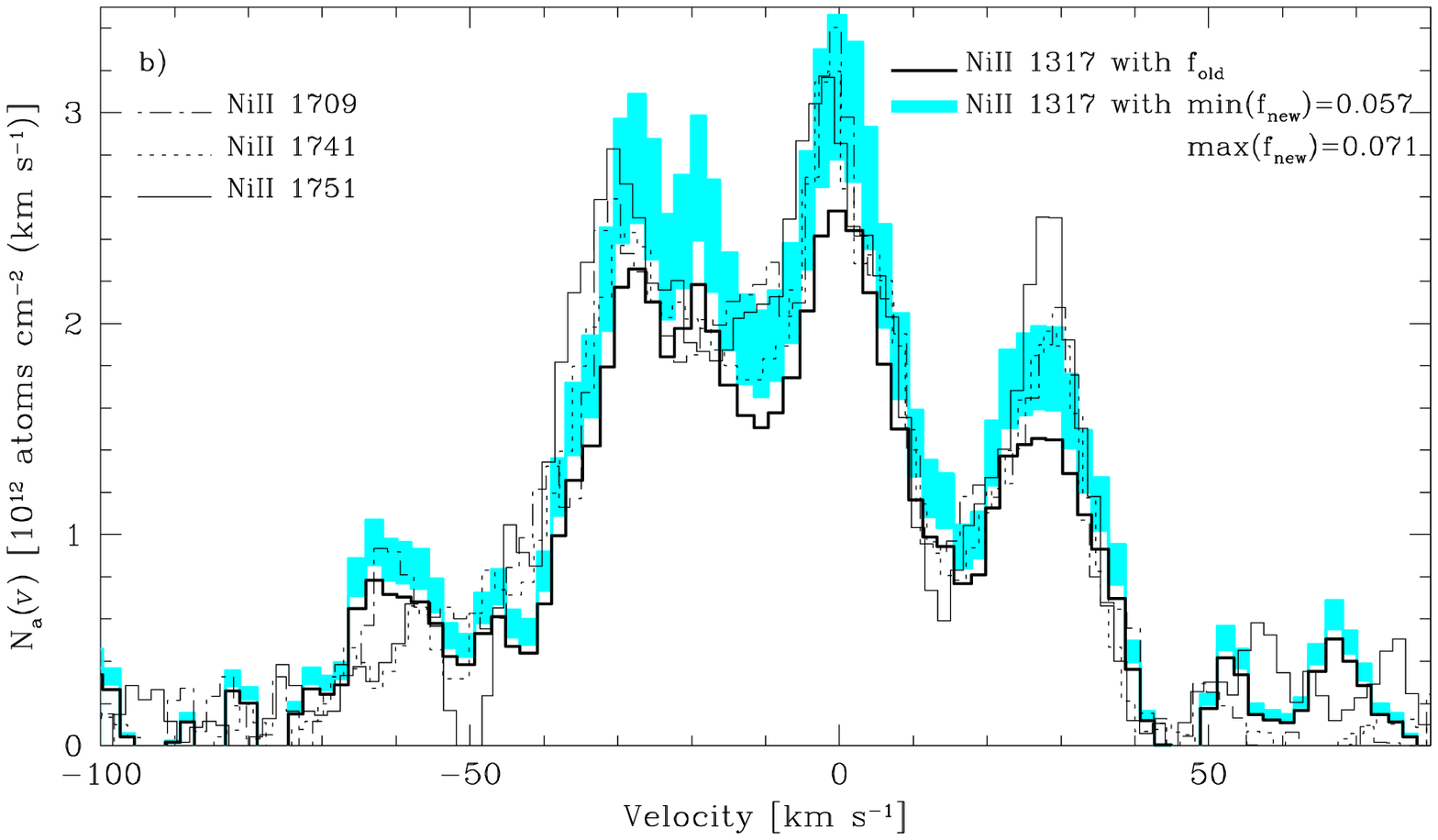}
\includegraphics[width=12.7cm]{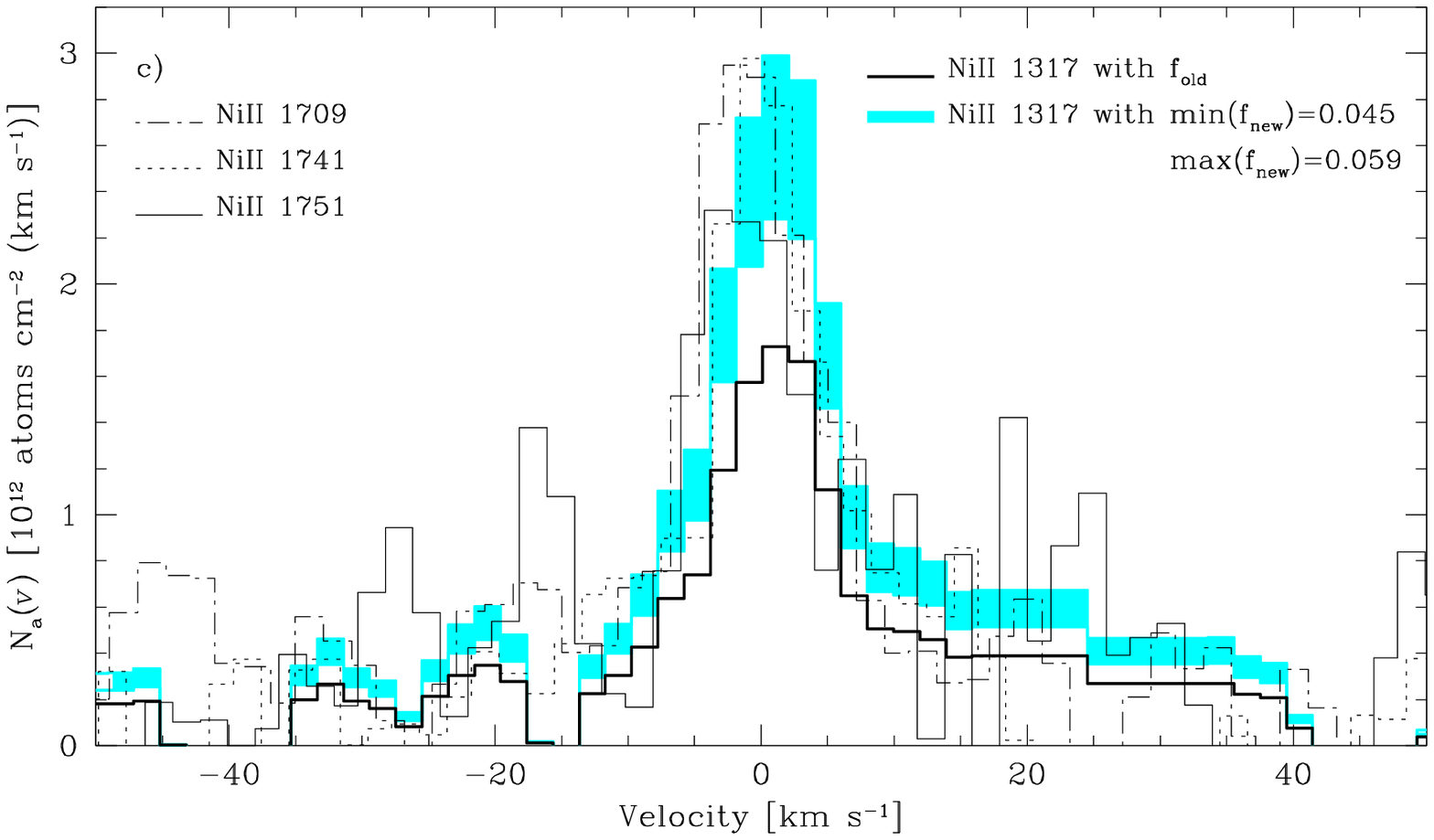}
\caption{Apparent column density $N_{\rm a}(v)$ profiles for the
\ion{Ni}{ii}\,$\lambda$\,1709,\,1741,\,1751 lines (thin solid, dotted, and
dashed-dotted lines). The shaded profile corresponds to the 
\ion{Ni}{ii}\,$\lambda$\,1317 apparent column density profile computed with the 
best result derived in each DLA individually for the new 
\ion{Ni}{ii}\,$\lambda$\,1317 oscillator strength and its 1\,$\sigma$ error. 
The thick solid profile corresponds to the \ion{Ni}{ii}\,$\lambda$\,1317 
$N_{\rm a}(v)$ profile obtained with the old $f$-value from \citet{morton91}. 
It is clearly lower than the apparent column density profiles for the 
\ion{Ni}{ii}\,$\lambda$\,1709,\,1741,\,1751 lines. The panels a), b), c), d), 
and e) correspond to the DLA at $z_{\rm abs} = 2.375$ toward Q0841+129, the DLA
toward Q1157+014, the DLA toward Q1210+17, the DLA toward Q2230+02, and the DLA
toward Q2231$-$00 studied in our first sample of DLAs (see Paper~I), 
respectively.} 
\label{NiII-plots}
\end{figure*}

\addtocounter{figure}{-1}
\begin{figure*}[!]
\centering
\includegraphics[width=12.7cm]{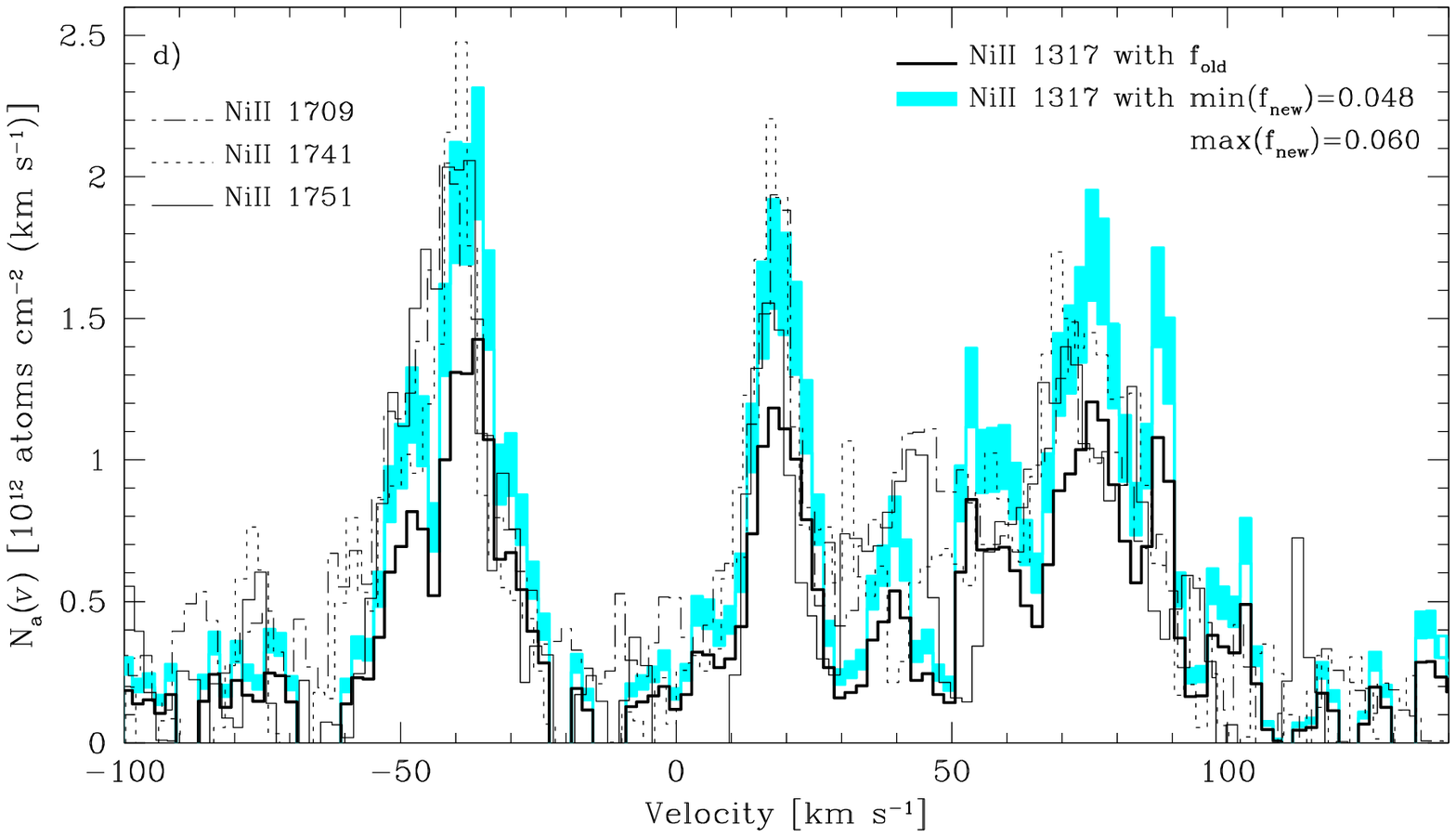}
\includegraphics[width=12.7cm]{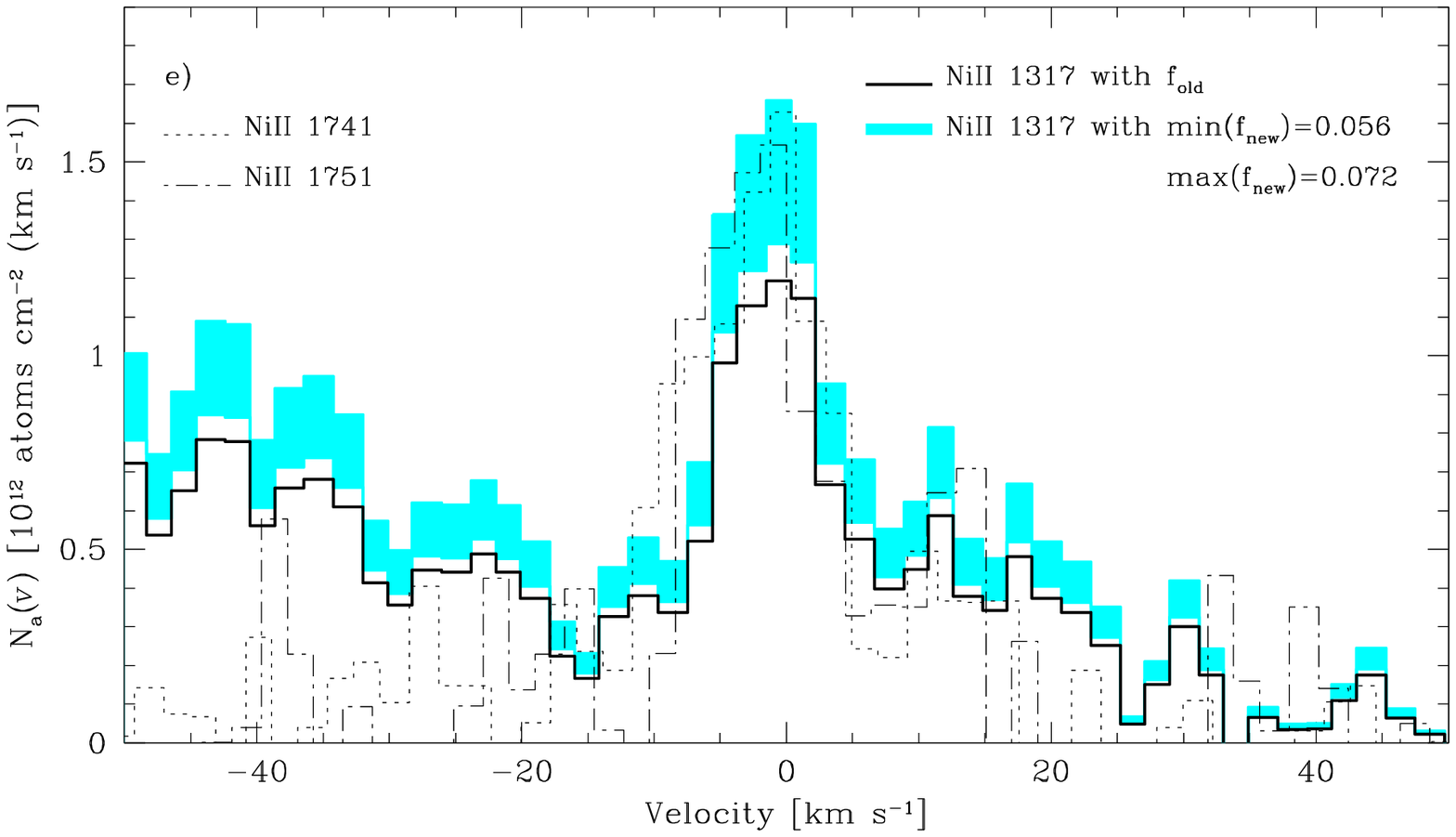}
\caption{{\em Continued}.}
\end{figure*}
%


\begin{thebibliography}{}

\bibitem[Ballester et al.(2000)]{ballester00}
Ballester, P., Modigliani, A., Boitquin, O., Cristiani, S., Hanuschik, R., 
Kaufer, A., \& Wolf, S. 2000, The Messenger, 101, 31

\bibitem[Calura et al.(2003)]{calura03}
Calura, F., Matteucci, F., \& Vladilo, G. 2003, MNRAS, 340, 59

\bibitem[Centuri\'on et al.(2000)]{centurion00}
Centuri\'on, M., Bonifacio, P., Molaro, P., \& Vladilo, G. 2000, ApJ, 536, 540

\bibitem[Centuri\'on et al.(2003)]{centurion03}
Centuri\'on, M., Molaro, P., Vladilo, G., P\'eroux, C., Levshakov, S. A., \& 
D'Odorico, V. 2003, A\&A, 403, 55

\bibitem[Chen et al.(2004)]{chen04}
Chen, Y. Q., Nissen, P. E., \& Zhao, G. 2004, A\&A, 425, 697

\bibitem[Chen \& Lanzetta(2003)]{chen03}
Chen, H.-W., \& Lanzetta, K. M. 2003, ApJ, 597, 706

\bibitem[Chiappini et al.(2003)]{chiappini03}
Chiappini, C., Matteucci, F., \& Meynet, G. 2003, A\&A, 410, 257

\bibitem[Cowan et al.(2005)]{cowan05}
Cowan, J. J., Sneden, C., Beers, T. C., et~al. 2005, ApJ, submitted

\bibitem[Dessauges-Zavadsky et al.(2004)]{dessauges04}
Dessauges-Zavadsky, M., Calura, F., Prochaska, J. X., D'Odorico, S., \& 
Matteucci, F. 2004, A\&A, 416, 79 [Paper~I]

\bibitem[Dessauges-Zavadsky et al.(2002)]{dessauges02}
Dessauges-Zavadsky, M., Prochaska, J. X., \& D'Odorico, S. 2002, A\&A, 391, 801

\bibitem[D'Odorico et al.(2000)]{dodorico00}
D'Odorico, S., Cristiani, S., Dekker, H., et~al. 2000, in SPIE 4005, 
Discoveries and Research Prospects from 8- to 10-Meter-Class Telescopes, 
ed. J. Bergeron, 121

\bibitem[Ellison et al.(2001)]{ellison01}
Ellison, S. L., Ryan, S. G., \& Prochaska, J. X. 2001, MNRAS, 326, 628

\bibitem[Edvardsson et al.(1995)]{edvardsson95}
Edvardsson, B., Pettersson, B., Kharrazi, M., \& Westerlund, B. 1995, A\&A, 
293, 75

\bibitem[Fedchak \& Lawler(1999)]{fedchak99}
Fedchak, J. A., \& Lawler, J. E. 1999, ApJ, 523, 734

\bibitem[Fedchak et al.(2000)]{fedchak00}
Fedchak, J. A., Wiese, L. M., \& Lawler, J. E. 2000, ApJ, 538, 773

\bibitem[Fenner et al.(2004)]{fenner04}
Fenner, Y., Prochaska, J. X., \& Gibson, B. K. 2004, ApJ, 606, 116

\bibitem[Fontana \& Ballester(1995)]{fontana95}
Fontana, A., \& Ballerster, P. 1995, The Messenger, 80, 37

\bibitem[Fran\c cois et al.(2004)]{francois04}
Fran\c cois, P., Matteucci, F., Cayrel, R., Spite, M., Spite, F., \& Chiappini, 
C. 2004, A\&A, 421, 613

\bibitem[Grevesse \& Sauval(1998)]{grevesse98}
Grevesse, N., \& Sauval, A. J. 1998, Space Science Reviews, 85, 161

\bibitem[Haehnelt et al.(1998)]{haehnelt98}
Haehnelt, M. G., Steinmetz, M., \& Rauch, M. 1998, ApJ, 495, 647

\bibitem[Jaunsen et al.(1995)]{jaunsen95}
Jaunsen, A. O., Jablonski, M., Pettersen, B. R., \& Stabell, R. 1995, A\&A,
300, 323

\bibitem[Kanekar \& Chengalur(2003)]{kanekar03}
Kanekar, N. \& Chengalur, J. N. 2003, A\&A, 399, 857

\bibitem[Le Brun et al.(1997)]{lebrun97}
Le Brun, V., Bergeron, J., Boiss\'e, P., \& Deharveng,  J. M. 1997, A\&A, 321, 
733


\bibitem[Ledoux et al.(2003)]{ledoux03}
Ledoux, C., Petitjean, P., \& Srianand, R. 2003, MNRAS, 346, 209

\bibitem[Limongi et al.(2000)]{limongi00}
Limongi, M., Straniero, O., \& Chieffi, A. 2000, ApJS, 129, 625

\bibitem[Lopez et al.(2002)]{lopez02}
Lopez, S., Reimers, D., D'Odorico, S., \& Prochaska, J. X. 2002, A\&A, 385, 778

\bibitem[Lu et al.(1996)]{lu96}
Lu, L., Sargent, W. L. W., Barlow, T. A., Churchill, C. W., \& Vogt, S. S. 
1996, ApJS, 107, 475

\bibitem[Maller et al.(2001)]{maller01}
Maller, A. H., Prochaska, J. X., Somerville, R. S., \& Primack, J. R. 2001,
MNRAS, 326, 1475

\bibitem[Matteucci et al.(1993)]{matteucci93}
Matteucci, F., Raiteri, C. M., Busson, M., Gallino, R., \& Gratton, R. 1993,
A\&A, 272, 421

\bibitem[Matteucci \& Recchi(2001)]{matteucci01}
Matteucci, F., \& Recchi, S. 2001, ApJ, 558, 351

\bibitem[McWilliam et al.(2003)]{mcwilliam03}
McWilliam, A., Rich, R. M., \& Smecker-Hane, T. A. 2003, ApJ, 592, L21

\bibitem[Mishenina et al.(2002)]{mishenina02}
Mishenina, T. V., Kovtyukh, V. V., Soubiran, C., Travaglio, C., \& Busso, M. 
2002, A\&A, 396, 189

\bibitem[Molaro(2005)]{molaro05}
Molaro, P. 2005, Chemical Abundances and Mixing in Stars in the Milky Way and 
its Satellites, Springer-Verlag Series, "ESO Astrophysics Symposia", Eds. 
L. Pasquini, S. Randich 

\bibitem[Morton(1991)]{morton91}
Morton, D. C. 1991, ApJS, 77, 119

\bibitem[Nestor et al.(2002)]{nestor02}
Nestor, D. B., Rao, S. M., Turnshek, D. A., Monier, E., Lane, W. M., \& 
Bergeron, J. 2002, Extragalactic Gas at Low Redshift, APS Conf. Ser. Vol. 254, 
Eds. J. S. Mulchaey, J. Stocke, 34

\bibitem[Nissen et al.(2004)]{nissen04}
Nissen, P. E., Chen, Y. Q., Asplund, M., \& Pettini, M. 2004, A\&A, 415, 993

\bibitem[Pettini et al.(1994)]{pettini94}
Pettini, M., Smith, L. J., Hunstead, R. W., \& King, D. L. 1994, ApJ, 426, 79

\bibitem[Pettini et al.(1995)]{pettini95}
Pettini, M., Lipman, K., \& Hunstead, R. W. 1995, ApJ, 451, 100

\bibitem[Pettini et al.(1997)]{pettini97}
Pettini, M., Smith, L. J., King, D. L., \& Hunstead, R. W. 1997, ApJ, 486, 665

\bibitem[Pettini et al.(2002)]{pettini02}
Pettini, M., Ellison, S. L., Bergeron, J., \& Petitjean, P. 2002, A\&A, 391, 21

\bibitem[Prochaska \& Wolfe(1996)]{prochaska96}
Prochaska, J. X., \& Wolfe, A. M. 1996, ApJ, 470, 403

\bibitem[Prochaska \& Wolfe(1999)]{prochaska99}
Prochaska, J. X., \& Wolfe, A. M. 1999, ApJS, 121, 369

\bibitem[Prochaska \& McWilliam(2000)]{prochaska00}
Prochaska, J. X., \& McWilliam, A. 2000, ApJ, 537, L57

\bibitem[Prochaska et al.(2001)]{prochaska01}
Prochaska, J. X., Wolfe, A. M., Tytler, D., et~al. 2001, ApJS, 137, 21

\bibitem[Prochaska et al.(2002a)]{prochaska02a} 
Prochaska, J. X., Howk, J. C., O'Meara, J. M., et~al. 2002a, ApJ, 571, 693

\bibitem[Prochaska et al.(2002b)]{prochaska02b}
Prochaska, J. X, Henry, R. B. C., O'Meara, J. M., et~al., 2002b, PASP, 114, 933

\bibitem[Prochaska \& Wolfe(2002)]{prochaska02c}
Prochaska, J. X., \& Wolfe, A. M. 2002, ApJ, 566, 68

\bibitem[Prochaska(2003)]{prochaska03a}
Prochaska, J. X. 2003, ApJ, 582, 49

\bibitem[Prochaska et al.(2003)]{prochaska03b}
Prochaska, J. X., Howk, J. C., \& Wolfe, A. M. 2003, Nature, 423, 57 

\bibitem[Rao et al.(2003)]{rao03}
Rao, S. M., Nestor, D. B., Turnshek, D. A., Lane, W. M., Monier, E. M., \& 
Bergeron, J. 2003, ApJ, 595, 94

\bibitem[Rao et al.(2005)]{rao05}
Rao, S. M., Prochaska, J. X., Howk, J. C., \& Wolfe, A. M. 2005, AJ, 129, 9

\bibitem[Savage \& Sembach(1991)]{savage91} 
Savage, B. D., \& Sembach, K. R. 1991, ApJ, 379, 245

\bibitem[Savage \& Sembach(1996)]{savage96} 
Savage, B. D., \& Sembach, K. R. 1996, ARA\&A, 34, 279

\bibitem[Shetrone et al.(2003)]{shetrone03}
Shetrone, M., Venn, K. A., Tolstoy, E., Primas, F., Hill, V., \& Kaufer A. 2003, 
AJ, 125, 684

\bibitem[Sofia \& Jenkins(1998)]{sofia98}
Sofia, U. J., \& Jenkins, E. B. 1998, ApJ, 499, 951

\bibitem[Srianand et al.(2000)]{srianand00}
Srianand, R., Petitjean, P., \& Ledoux, C. 2000, Nature, 408, 931

\bibitem[Storrie-Lombardi \& Wolfe(2000)]{storrie00}
Storrie-Lombardi, L. J., \& Wolfe, A. M. 2000, ApJ, 543, 552

\bibitem[Tolstoy et al.(2003)]{tolstoy03}
Tolstoy, E., Venn, K. A., Shetrone, M., Primas, F., Hill, V., Kaufer, A., \& 
Szeifert, T. 2003, AJ, 125, 707

\bibitem[Venn(1999)]{venn99}
Venn, K. A. 1999, ApJ, 518, 405

\bibitem[Viegas(1995)]{viegas95}
Viegas, S. M. 1995, MNRAS, 276, 268

\bibitem[Vladilo(1998)]{vladilo98}
Vladilo, G. 1998, ApJ, 493, 583

\bibitem[Vladilo et al.(2001)]{vladilo01}
Vladilo, G., Centuri\'on, M., Bonifacio, P., \& Howk, J. C. 2001, ApJ, 557, 
1007

\bibitem[Vladilo(2002)]{vladilo02}
Vladilo, G. 2002, A\&A, 391, 407

\bibitem[Vladilo et al.(2003)]{vladilo03}
Vladilo, G., Centuri\'on, M., D'Odorico, V., \& P\'eroux, C. 2003, A\&A, 402, 
487

\bibitem[Welty et al.(1999)]{welty99}
Welty, D. E., Frisch, P. C., Sonneborn, G., \& York, D. G. 1999, ApJ, 512, 636

\bibitem[Welty et al.(2001)]{welty01}
Welty, D. E., Lauroesch, J. T., Blades, J. C., Hobbs, L. M., \& York, D. G. 
2001, ApJ, 554, 75

\bibitem[Wolfe \& Briggs(1981)]{wolfe81}
Wolfe, A. M., \& Briggs, F. H. 1981, ApJ, 248, 460

\bibitem[Wolfe et al.(1986)]{wolfe86}
Wolfe, A. M., Turnshek, D. A., Smith, H. E., \& Cohen, R. D. 1986, ApJ, 61, 249

\bibitem[Wolfe et al.(1995)]{wolfe95}
Wolfe, A. M., Lanzetta, K. M., Foltz, C. B., \& Chaffee, F. H. 1995, ApJ, 454, 
698

\bibitem[Woosley \& Weaver(1995)]{woosley95}
Woosley, S. E., \& Weaver, T. A. 1995, ApJS, 101, 181

\end{thebibliography}
\end{document}